\documentclass[3p]{elsarticle}

\journal{J. Logic. Algebr. Program}

\usepackage{booktabs}
\usepackage{hyperref}

\usepackage{amssymb}
\usepackage{stmaryrd}
\let\lBrack=\llbracket
\let\rBrack=\rrbracket

\usepackage{amsmath}
\usepackage{amsthm}
\usepackage{calc}
\newtheorem{theorem}{Theorem}
\newtheorem{proposition}{Proposition}

\usepackage{tikz}
\usetikzlibrary{positioning,shapes.geometric,calc,arrows.meta}
\usepackage{contour}
\usepackage{syntax}
\usepackage{tabu}
\usepackage[noliterate]{maude}
\usepackage{xcolor}
\usepackage[capitalise,noabbrev]{cleveref}

\newsavebox{\sigue}
\savebox{\sigue}{\textcolor{red}{$\hookrightarrow$}\space}
\newcommand\coloneq{:=}

\usepackage{newunicodechar}
\newunicodechar{ρ}{$\rho$}
\AtBeginDocument{\renewcommand\setminus{\;\hbox{\textbackslash}\;}}

\lstnewenvironment{maudexec}[1][]{\lstset{language={}, #1}}{}

\newcommand*\lststrat[1][]{\ifmmode\hbox\bgroup\fi\lstinline[basicstyle=\ttfamily,literate={=>}{{=>}}2{-->}{{-->}}3{->}{{->}}2{<-}{{<-\;}}2
{alpha}{$\alpha$}1{beta}{$\beta$}1{gamma}{$\gamma$}1,#1]}

\newcommand\idle{\skywd{idle}}
\newcommand\fail{\skywd{fail}}
\newcommand\seq{\texttt;}
\newcommand\disj{\texttt|}
\newcommand\ifthel[3]{#1 \texttt? #2 \texttt: #3}

\newcommand\skywd[1]{\texttt{\color{darkgray}\bfseries #1}}
\newcommand\kywd[1]{\texttt{\bfseries #1}}

\newcommand\ssem[1]{\lBrack #1 \rBrack_\Delta}
\newcommand\venv{\ensuremath{\mathrm{VEnv}}}
\newcommand\sfun{\ensuremath{\mathrm{SFun}}}
\newcommand\N{\ensuremath{\mathbb{N}}}

\newenvironment{mgrammar}{\vspace*{-1.5em}\begin{quotation}\hspace*{-\grammarindent}\begin{grammar}}{\end{grammar}\end{quotation}}

\let\varc=\theta

\makeatletter
\def\ps@pprintTitle{\let\@oddhead\@empty
     \let\@evenhead\@empty
     \def\@oddfoot
       {\hbox to \textwidth {\ifnopreprintline\relax\else
        \@myfooterfont \ifx\@elsarticlemyfooteralign\@elsarticlemyfooteraligncenter \hfil\@elsarticlemyfooter\hfil \else \ifx\@elsarticlemyfooteralign\@elsarticlemyfooteralignleft \@elsarticlemyfooter\hfill{}\else \ifx\@elsarticlemyfooteralign\@elsarticlemyfooteralignright {}\hfill\@elsarticlemyfooter \else \normalshape\hfill\begin{tikzpicture}
			\node at (0, 2em) {};
			\node[draw=black!70, fill=black!5, inner sep=5pt, text width=.705\linewidth]{
				Accepted authors' manuscript of the article published in \hfill \@journal\ 134 \\
				DOI: \href{https://doi.org/10.1016/j.jlamp.2023.100887}{10.1016/j.jlamp.2023.100887} \hfill License: CC-BY-NC-ND
			};
		\end{tikzpicture} \hfill\fi \fi \fi \fi }
       }\let\@evenfoot\@oddfoot}
\makeatother

\begin{document}
\begin{frontmatter}

\address[sri]{SRI International, Menlo Park, CA, USA}
\address[ucm]{Facultad de Informática, Universidad Complutense de Madrid, Spain}
\address[uiuc]{Univerity of Illinois at Urbana-Champaign, IL, USA}

\cortext[cor1]{Corresponding author}

\title{The Maude strategy language}
\author[sri]{Steven Eker}
\ead{eker@csl.sri.com}
\author[ucm]{Narciso Martí-Oliet}
\ead{narciso@ucm.es}
\author[uiuc]{José Meseguer}
\ead{meseguer@illinois.edu}
\author[ucm]{Rubén Rubio\corref{cor1}}
\ead{rubenrub@ucm.es}
\author[ucm]{Alberto Verdejo}
\ead{jalberto@ucm.es}

\begin{abstract}
	Rewriting logic is a natural and expressive framework for the specification of concurrent systems and logics. The Maude specification language provides an implementation of this formalism that allows executing, verifying, and analyzing the represented systems. These specifications declare their objects by means of terms and equations, and provide rewriting rules to represent potentially non-deterministic local transformations on the state. Sometimes a controlled application of these rules is required to reduce non-determinism, to capture global, goal-oriented or efficiency concerns, or to select specific executions for their analysis. That is what we call a strategy. In order to express them, respecting the separation of concerns principle, a Maude strategy language was proposed and developed. The first implementation of the strategy language was done in Maude itself using its reflective features. After ample experimentation, some more features have been added and, for greater efficiency, the strategy language has been implemented in C++ as an integral part of the Maude system.
This paper describes the Maude strategy language along with its semantics, its implementation decisions, and several application examples from various fields.
\end{abstract}


\end{frontmatter}

\section{Introduction} \label{sec:intro}

	Rewriting logic specifications describe computational systems as collections of rewriting rules that are applied to the terms of an equational specification. A great generality is achieved by allowing the rules not to be confluent or terminating. However, and specially when the specifications become executable, it may be convenient to control this non-determinism in order to avoid undesired evolutions. This control is realized by \emph{strategies}, whose need has been identified since the first developments of rewriting logic and its implementations~\cite{rewritingLogic}.

	Maude~\cite{allmaude} is a specification language based on rewriting logic as well as an interpreter with several formal analysis features and a formal tool environment \cite{mfe} that allows executing, verifying, and analyzing the specified systems. The non-deterministic rewriting process can be explored in different ways in Maude. In some cases, it is enough to see where a single path of exhaustive rewriting leads to, and this is possible with the \texttt{rewrite} and \texttt{frewrite} commands that select the next rewrite by some fixed criteria ensuring some fairness properties of the chosen path. Sometimes, all possible execution paths need to be explored using the \texttt{search} command; for example, to search for a violation of an invariant. However, to observe paths satisfying some user-chosen restrictions, strategies are required. Due to the reflective properties of rewriting logic, and using some Maude \emph{metalevel} functions that reproduce the rule application operations, strategies have been traditionally expressed as metacomputations inside the logic~\cite{clavel96}. This mechanism is complete and powerful, but it is sometimes cumbersome for users that are not familiar with the metalevel, because of its conceptual complexity and the verbosity of its notation.
Hence, following the influence of previous implementations of strategic rewriting such as ELAN~\cite{elan}, Stratego~\cite{stratego} and TOM~\cite{tom}, a simple but expressive object-level strategy language has been proposed for Maude. Its design is based on the previous Maude experience with strategies at the metalevel and on its predecessors, but it deviates from the treatment of strategies in ELAN in a remarkable aspect. In ELAN, rules and strategies are tightly coupled and strategies can be used in the definition of rules. The Maude language advocates a clear separation between them, making strategies a separate layer above rules. This is enforced and emphasized by banning strategies from \emph{system modules}, i.e., from rewrite theory specifications. Instead, different kinds of strategies can be specified separately for the same system module in different \emph{strategy modules}, so that the same rewrite rules can be controlled with different strategies by means of the chosen strategy module. Of course, for simplicity and ease of use, each strategy language design favors some features and may not directly support others.
Nevertheless, besides the fact that Maude's strategy language supports user-definable recursive strategies, which may be used to specify various additional features, Maude itself, including its strategy language, is \emph{user-extensible} thanks to its reflective nature, supported by Maude's \texttt{META-LEVEL} module, so that new features can be added by reflective extension.

	Strategies are a useful specification and execution tool~\cite{barendregt}. Its origins date back to combinatory logic and the $\lambda$-calculus, where choosing the next reduction position by a fixed structural criterion ensures finding a normal form in case it exists. In this vein, strategies can be used to make systems confluent or terminating by imposing additional restrictions. Moreover, they can provide specifications with a notion of global control over the eminently local meaning of rules in various ways: as simple as choosing a rule precedence, or arbitrarily complex by depending on the execution history. The \emph{Rule + Strategies} approach~\cite{pettorossi,lescanneOrme}, as an evolution of the Kowalski’s motto \emph{Algorithm = Logic + Control}~\cite{kowalski}, can be exploited to build specifications with a clear \emph{separation of concerns}. Typically, it is easier to prove that the logical part given by rules ensures correct deductions or computations, while leaving to the strategies the responsibilities on efficiency and goal directedness (as shown in \cref{sec:completion}). In some other cases, strategies are actually needed to ensure that the system behaves as intended. In addition, strategies can be used to analyze plain rewriting logic specifications too, by the observation of selected rewriting evolutions. Many examples from different fields have already been specified using the Maude strategy language: it has been applied to express deduction procedures as strategies controlling a declarative inference system~\cite{completion, strategies06}, to specify aspects of the semantics of programming languages~\cite{eden,operational}, of the ambient calculus~\cite{ambientCalculus}, Milner's CCS~\cite{ccs}, process scheduling policies~\cite{smcJournal},\footnote{Process scheduling policies, implemented by operating systems to distribute the processor time among multiple simultaneous processes, are an archetypical example of strategies programmed in real software using standard programming languages. Specifying these strategies at a higher level in Maude can be useful for better reasoning about them.} membrane systems~\cite{membrane,memstratmc-jlamp}, etc.

	The first prototype of the strategy language was written at the metalevel as an extension of Full Maude~\cite{allmaude}. Now, the complete language is available and efficiently supported at the Maude interpreter level, implemented in C++, from its version 3.0 onwards~\cite{maude,maude30}. The most recent version of Maude can be downloaded from \texttt{\href{http://maude.cs.illinois.edu}{maude.cs.illinois.edu}}, and the examples described in this article and many more are available at \texttt{\href{https://maude.ucm.es/strategies}{maude.ucm.es/strategies}} and \texttt{\href{https://github.com/fadoss/strat-examples}{github.com/fadoss/strat-examples}}.

	This article extends, updates, and integrates the information already presented in some workshop and conference talks~\cite{towardsStrategy, strategies06, rewSemantics, pssm,ccs}, and a PhD thesis~\cite{mitesis}. For the first time, this article describes strategy modules and the meta-representation of the strategy language, and provides a complete and formal denotational semantics of the language with strategy calls. More details are given on the strategy language constructors, and design and implementation decisions are discussed. New and significantly improved examples are included, as well as a comparison with the similar strategy languages of ELAN, Stratego, TOM, and ρLog.

\section{Rewriting logic and Maude}

	Rewriting logic~\cite{rewritingLogic,20years} is a computational logic for expressing both concurrent computation and logical deduction in a general and natural way. Maude~\cite{maude,allmaude} is a specification language whose programs are exactly rewrite theories. These can be executed in the language interpreter, and analyzed and verified by means of various formal analysis features in Maude itself, as well as by several formal analysis tools in Maude's formal tool environment~\cite{mfe}. Typically, those formal tools are themselves implemented in Maude through reflection. Maude has already been applied to the specification and verification of many logics, models of concurrency, programming languages, hardware and software modeling languages, distributed algorithms, network protocols, cryptographic protocols, real-time and cyber-physical systems, and biological systems (see \cite{20years} for a survey of rewriting logic, Maude, and its applications).

\subsection{Abstract reduction} \label{sec:ars}	

	First, some standard notation and terminology will be reviewed. Given a set $S$ of states and a binary relation $(\to) \in S \times S$, we refer to $s \to s'$ as a reduction or execution step, and $s_0 \to \cdots \to s_n$ as a derivation or execution of length $n$. Non-terminating executions of the form $s_0 \to s_1 \to \cdots$ are also considered. We say that $s \in S$ is \emph{irreducible} if no $s' \in S$ makes $s \to s'$ hold. We write $s \to^n s'$ if there is a derivation of length $n$ from $s$ to $s'$, $\to^+$ for the transitive closure of $\to$ defined as $\cup_{n \geq 1} (\to^n)$, and $\to^*$ for the transitive and reflexive closure of $\to$. We also write $s \to^! s'$ if $s \to^* s'$ and $s'$ is irreducible, and say that $s'$ is a \emph{normal form} of $s$.

	The pair $(S, \to)$ is an \emph{abstract reduction system}. It is \emph{confluent} if for all $s, s_1, s_2 \in S$ such that $s \to^* s_1$ and $s \to^* s_2$, there is a $s'$ such that $s_1 \to^* s'$ and $s_2 \to^* s'$. It is \emph{terminating} if every execution from any state $s$ is finite. In a confluent and terminating reduction system, also known as \emph{convergent}, every state has a single normal form. Moreover, given another relation $\to'$ on $S$, we say $\to'$ is \emph{coherent}~\cite{crc-jlap} with $\to$ if for all $s, u, s_1, s_2 \in S$ such that $s \to^! u$, $s \to' s_1$, and $u \to' s_2$, there is a $u'$ such that $s_1 \to^! u'$ and $s_2 \to^! u'$. In simpler words, $\to'$ is coherent with $\to$ if $\to'$ can be applied on the canonical forms of $\to$ without loss of generality.

\subsection{Membership equational logic}

	A \emph{signature} in membership equational logic~\cite{spmel} is given by a set of \emph{sorts} $S$ and a collection $\Sigma$ of operators $f : s_1 \cdots s_n \to s$ from which terms are constructed. Sorts are related by a partial order $s_1 < s_2$ representing subsort inclusion. Each connected component in the sort relation is called a \emph{kind}, and the kind of a sort $s$ is denoted by $[s]$. Each operator $f : s_1 ... s_n \to s$ is also lifted to the kind level as $f : [s_1] \ldots [s_n] \to [s]$. Kinds are interpreted as sets containing all the well-formed terms of the related sorts, as well as some error elements that may arise from partially-defined functions or type errors. Sorts and kinds are uniformly referred to as \emph{types}. The set of terms of a given type $s$ over some variables $X$ is written $T_{\Sigma, s}(X)$ and the full set of terms is written $T_{\Sigma}(X)$. Terms without variables are called \emph{ground terms}. A \emph{substitution} is a type-preserving function $\sigma : X \to T_\Sigma(X)$ that assigns terms to variables. It can be extended to a function $\overline\sigma : T_\Sigma(X) \to T_\Sigma(X)$ that replaces the occurrences of the (typed) variables in a term inductively. For any pair of substitutions $\sigma_1, \sigma_2$, we define their composition $\sigma_2 \circ \sigma_1$ by the equality $(\sigma_2 \circ \sigma_1)(x) \coloneq \overline{\sigma_2}(\sigma_1(x))$. It satisfies $\overline{\sigma_2 \circ \sigma_1} = \overline{\sigma_2} \circ \overline{\sigma_1}$ in the usual functional sense. The line over the extension is usually omitted.

	Membership equational logic for a given theory $(\Sigma, E)$ has two kinds of atomic sentences, \emph{equations} and sort \emph{membership axioms}, on top of the above signature. They can be conditioned by premises in the form of other equations and sort membership axioms to yield Horn clauses of the form:

	\[ 	t = s \qquad \text{if } \bigwedge_i u_i = u'_i \wedge \bigwedge_j v_j : s_j \qquad\text{and}\qquad
		t : s \qquad \text{if } \bigwedge_i u_i = u'_i \wedge \bigwedge_j v_j : s_j \]
where $t : s$ is the membership axiom stating that $t$ has sort $s$. Apart from equations and membership axioms, operators can be annotated with \emph{structural axioms} like associativity, commutativity, and identity. Of course, such structural axioms are a special case of equations. However, in Maude they are specified together with their corresponding operators because they are used in a built-in way to efficiently support deduction \emph{modulo} such structural axioms.

	These (possibly conditional) equations and memberships $E$, and structural axioms $B$ induce an equality relation $=_{E \cup B}$ that identifies different terms up to provable equality with $E \cup B$. The \emph{initial term algebra} $T_{\Sigma/E \cup B}$ is the quotient of the ground terms $T_\Sigma(\emptyset)$ modulo this equality relation. Its elements $[t]$ are equivalence classes modulo $=_{E\cup B}$, but we will usually write simply $t$ when dealing with well-defined characteristics and operations that are independent of the class representatives.

\subsection{Rewriting logic} \label{sec:rwlog}

	A \emph{rewrite theory} $\mathcal{R} = (\Sigma, E \cup B, R)$ is a membership equational theory $(\Sigma, E \cup B)$ together with a set $R$ of rewriting rules, which are interpreted as non-equational transitions. A possibly conditional rewriting rule has the form:
\[ l \Rightarrow r \qquad \text{if } \bigwedge_i u_i = u'_i \wedge \bigwedge_j v_j : s_j \wedge \bigwedge_k w_k \Rightarrow w'_k \]
The application of a rule to a term $t$ is the replacement of an instance (modulo $B$) of $l$ in some position $p$ of $t$ by $r$ at that same position $p$ instantiated accordingly if the condition holds. Conditions of the third type are named \emph{rewriting conditions}. They are satisfied if the instance of each $w_k$ can be rewritten in zero or more steps to match an instance (modulo $E \cup B$) of $w_k'$. Formally, a rule as above rewrites a term $t$ to $t'$, $t \to_{\mathcal{R}} t'$, if there is a position $p$ in $t$ and a substitution $\sigma : X \to T_{\Sigma}(X)$ such that $t|_p =_{E \cup B} \sigma(l)$ and $t' =_{E \cup B} t[\sigma(r)]_p$, and $\sigma(u_i) =_{E \cup B} \sigma(u'_i)$ for all $i$, $\sigma(v_j) \in T_{\Sigma, s_j}(\emptyset)$ for all $j$, and $\sigma(w_k) \to^*_{\mathcal{R}} \sigma(w'_k)$ for all $k$.

	For execution purposes, since the equality relation $=_{E \cup B}$ can be undecidable, equations are handled as oriented rewrite rules $\to_{E/B}$ modulo structural axioms. Moreover, some additional executability requirements are given on $\mathcal{R}$, namely (1) $(\Sigma, \to_{E/B})$ is assumed \emph{convergent} modulo $B$; and (2) the rules $R$ are assumed \emph{coherent} modulo $B$ with respect to the (oriented) equations $E$. This second requirement means that the relation $\to_{R/B}$, consisting in $\to_{\mathrm{R}}$ with $=_{E \cup B}$ replaced by $=_B$, is coherent with $\to_{E/B}$ as defined in \cref{sec:ars}. The effect of assumptions (1)-(2) is that $\to_{\mathcal{R}}$ can be computed in terms of the much simpler and decidable equality relation $=_B$, instead of requiring the complex relation $=_{E \cup B}$, as $\to_{R/B} \circ \to_{E/B}^!$. These executability requirements are expressed at the Maude level by the notion of \emph{admissible} Maude modules, as explained in \cref{sec:maude}.

\subsection{Maude} \label{sec:maude}

	Maude is a specification language~\cite{maude} whose programs are a straight translation to ASCII of the previous mathematical notation for equations, membership axioms, and rules. Specifications are organized in modules that can include or extend other modules. Some useful modules are included in the Maude \emph{prelude} specifying natural and floating-point numbers, lists, strings, sets, reflective operations, and so on. Pure equational theories are specified in \emph{functional modules}, introduced by the \texttt{fmod} keyword.

\begin{figure}[ht]\centering
\newcommand\innersep{0.025}
\newcounter{tmp}
\hfill
\begin{tikzpicture}
	\draw[thick] (-2*\innersep, -2*\innersep) rectangle (4+2*\innersep, 4+2*\innersep);

	\foreach \i in {0, ..., 3}
	{
		\foreach \j in {1, ..., 3}
		{
			\draw (\i + \innersep, \j + \innersep)  rectangle +(1 - 2 * \innersep, 1 - 2 * \innersep);
			\node at (\i + 0.5, \j + 0.5) {\setcounter{tmp}{1+\i+12-4*\j}\thetmp};
		}
	}

	\foreach \i in {0, ..., 2}
	{
		\draw (\i + \innersep, \innersep)  rectangle +(1 - 2 * \innersep, 1 - 2 * \innersep);
		\node at (\i + 0.5, 0.5) {\setcounter{tmp}{13+\i}\thetmp};
	}
\end{tikzpicture}
\hfill\hfill
\begin{tikzpicture}
	\draw[thick] (-2*\innersep, -2*\innersep) rectangle (4+2*\innersep, 4+2*\innersep);

	\foreach \i/\j/\v in {0/3/5, 1/3/1, 2/3/4, 3/3/8, 0/2/2, 1/2/14, 2/2/15,
		3/2/3, 0/1/9, 1/1/7, 2/1/6, 3/1/11, 0/0/13, 1/0/10, 3/0/12}
	{
		\draw (\i + \innersep, \j + \innersep)  rectangle +(1 - 2 * \innersep, 1 - 2 * \innersep);
		\node at (\i + 0.5, \j + 0.5) {\v};
	}
\end{tikzpicture}
\hfill
\hphantom{}
\caption{The 15-puzzle in its goal position and in a shuffled one.} \label{fig:15puzzle}
\end{figure}

The following functional module specifies the board of the 15-puzzle game, which consists of 15 sliding numbered square tiles lying in a $4 \times 4$ frame, as depicted in~\cref{fig:15puzzle}. The puzzle has been represented as a semicolon-separated list of rows, each a list of tiles, which are in turn either a natural number or a blank \texttt{b}. Declarations and statements can be annotated with attributes between brackets. The \texttt{ctor} attribute for operator declarations states that the operator is intended to be a data constructor of the range sort. Structural axioms of commutativity, associativity, and identity element $e$ are respectively written as operator attributes \texttt{comm}, \texttt{assoc}, and \texttt{id:} $e$. The syntax for conditions and membership axioms follows directly from the mathematical notation of the previous section, except that
a special form of equality condition \lstinline|l := r|, known as \emph{matching condition}, allows free variables in its lefthand side to be bound by matching. Apart from this exception, all variables in the condition and the righthand side of an equation should appear in the lefthand side of that equation.

\begin{lstlisting}[escapechar=^,xrightmargin=2ex]
fmod 15PUZZLE-BOARD is
	protecting NAT . ^\hfill^ *** module inclusion

	sorts Tile Row Puzzle .              ^\hfill^ *** sort declarations
	subsorts Nat < Tile < Row < Puzzle . ^\hfill^ *** subsorts         ^\hphantom{}^

	op b : -> Tile [ctor] .	^\hfill^ *** operator declarations
	op nil : -> Row [ctor] .
	op __  : Row Row -> Row [ctor assoc id: nil prec 25] .
	op _;_ : Puzzle Puzzle -> Puzzle [ctor assoc] .

	var T : Tile . ^\hfill^ *** variables            ^\hphantom{}^
	var R : Row .  ^\hfill^ *** (local to the module)

	op size : Row -> Nat .
	eq size(nil) = 0 . ^\hfill^ *** equations
	eq size(T R) = size(R) + 1 .
endfm
\end{lstlisting}

	Not all functional modules are \emph{admissible}, as explained in \cref{sec:rwlog}. Maude uses equations as left-to-right simplification rules, which are applied exhaustively to obtain canonical forms of the terms modulo the equations and axioms. For this procedure to be sound, the simplification relation $\to_{E/B}$ should be confluent and terminating modulo the axioms $B$ (in this example, the specified associativity and identity axioms). Maude does not check this automatically, since it is generally undecidable, but the Maude Formal Environment \cite{mfe} includes tools for checking these properties.

	\emph{System modules} specify rewrite theories by adding rules. In this case, the rule application relation $\to_{\mathcal{R}}$ need not be confluent nor terminating. However, Maude works efficiently by reducing a term to normal form before applying a rule, so that an $\mathcal{R}$-rewriting step is achieved by the much simpler relation $\to_{R/B} \circ \to_{E/B}^!$. This assumes that the rules $R$ are coherent with the equations $E$ modulo structural axioms $B$, i.e., the second requirement in the last paragraph of \cref{sec:rwlog} must be satisfied. 

	For example, rules can be added to the functional module \texttt{15PUZZLE-BOARD} above. The rules \texttt{left}, \texttt{right}, \texttt{down}, and \texttt{up} displace the tile at each side of the blank towards it. In other words, they move the blank towards such direction. Using these operations, the goal of the game is to arrive at the configuration of the first board of~\cref{fig:15puzzle} from any shuffled board like the second board.
\begin{lstlisting}
mod 15PUZZLE is
	protecting 15PUZZLE-BOARD .

	var T : Tile . vars LU RU LD RD : Row . var P : Puzzle .

	 rl [left]  : T b => b T .
	 rl [right] : b T => T b .
	crl [down]  : (LU b RU) ; (LD T RD)
	           => (LU T RU) ; (LD b RD) if size(LU) = size(LD) .
	crl [up]    : (LU T RU) ; (LD b RD)
	           => (LU b RU) ; (LD T RD) if size(LU) = size(LD) .
endm
\end{lstlisting}
Rules may be named by means of an optional \emph{label} between brackets. This will be very helpful for the specification of strategies in the strategy language. Conditions for rules are equational conditions plus possible rewriting condition fragments \lstinline[keepspaces]|l => r|, which may contain free variables in its righthand side \texttt{r} to be instantiated by matching.

	The Maude interpreter provides various commands to execute its programs. The \texttt{reduce} command simplifies a given term to its normal form with the equations and memberships $E$ modulo the structural axioms $B$.
\begin{maudexec}
Maude> reduce size(1 b 2 3) .
rewrites: 14
result NzNat: 4
\end{maudexec}
The \texttt{rewrite} and \texttt{frewrite} commands~\cite[\S 5.4]{maude} rewrite a term using all the rewriting rules in the module (as well as the
equations and memberships $E$, all applied modulo the axioms $B$, as explained above). An optional bound on the number of rewriting steps can be given between brackets.
\begin{maudexec}
Maude> rewrite [21] 1 b 2 3 .
rewrites: 21
result Row: b 1 2 3
\end{maudexec}
Moreover, the \texttt{search} command lets the user find all terms reachable by rewriting that match a pattern and satisfy a specified condition. The rewriting paths that lead to the found terms can also be inspected.
\begin{maudexec}
Maude> search 1 b 2 3 =>* 1 2 R .

Solution 1 (state 2)
states: 3  rewrites: 2
R:Row --> b 3

Solution 2 (state 3)
states: 4  rewrites: 5
R:Row --> 3 b

No more solutions.
states: 4  rewrites: 6
\end{maudexec}
Further details about the language and the interpreter can be found in the Maude manual~\cite{maude}.

\subsection{Strategies}

	In a broad sense, strategies are seen as recipes to control a process in which multiple decisions are possible. Several formal definitions of strategy have been given for reduction and rewriting systems~\cite{extstrat,rewritingStrategies,visser05}. In the $\lambda$-calculus~\cite{barendregt}, a strategy is usually formalized by a function from $\lambda$-terms to $\lambda$-terms, whose output can be derived from the input by some reductions. They typically apply a rule in a position determined by a fixed criterion, like a $\beta$-reduction on the leftmost-innermost redex. More general notions of strategy let the selection of the next rewrite be non-deterministic and may depend on the derivation history~\cite{terese}.

	Considering rule labels, a rewriting system can be seen as a labeled graph $(T_\Sigma, A)$ whose states are terms and its transitions $(t, l, t') \in A$ are rule applications $t \to_R t'$ with a rule of label $l$. Thus, an execution is a path within the graph, either finite or infinite. In this context, we can see strategies from different points of view:
\begin{itemize}
	\item Seeing them as a choice of allowed executions, an \emph{abstract strategy} is formally defined as a subset $\mathcal{S}$ of the paths in the graph.
	\item Focusing on the results of these computations, strategies are seen as functions from an initial term to the set of its results
\[ \mathcal S(t) \coloneq \left\{ t' \in T_\Sigma \text{ s.t. there is a path } \; t \stackrel{l_1}\to_R t_1 \stackrel{l_2}\to_R \cdots \stackrel{l_n}\to_R t' \text{ in } \mathcal S \right\} \]
Notice that abstract strategies may also select infinite paths, which are useful in the specification of reactive systems, as described in~\cite{fscd,smcJournal,smcJournal-btime}.
	\item Looking at the allowed next steps, \emph{intensional strategies} are defined as partial functions that map an execution in progress to the admitted next steps. This formalization is often used in games and concurrent agent systems. 

	\item Regarding syntax, we can contemplate strategies as expressions in some strategy specification language, whose semantics can be given in any of the more abstract forms above.
\end{itemize}
Although the expressive power of abstract strategies is greater than that of intensional ones~\cite{extstrat}, and sets of results lack information about the intermediate states, all these descriptions may be useful in different situations. In practice, our strategies will be specified explicitly with a specific syntax: the Maude strategy language that we describe in this paper.

\section{The Maude strategy language} \label{sec:language}

	The design of the Maude strategy language is highly influenced by the traditional reflective specification of strategies as metalevel programs in Maude~\cite[\S 14]{allmaude, strategiesClavel}, while avoiding their conceptual difficulties and verbosity, and by other strategy languages like ELAN~\cite{elan}, Stratego~\cite{stratego}, and Tom~\cite{tom}, as already discussed in \cref{sec:intro}. In this section, we describe the syntax and informal semantics of the language, step by step and with examples.

	The basic instruction of the Maude strategy language is the selective rule application, and other operators are available to combine them and build more complex strategic programs. These \emph{combinators} include the typical programming constructs (sequential composition, tests, conditions, loops, recursive definitions), nondeterministic choices, and others that specifically exploit the structure of terms. For executing a strategy expression $\alpha$ on a term $t$, the Maude interpreter provides a command \texttt{srewrite $t$ using $\alpha$}, thoroughly described in~\cref{sec:commands}. At the moment, it is enough to know that it enumerates all terms that can be obtained by rewriting with the strategy from the given term, each called a \emph{result} or \emph{solution} of the strategy. The term to which the strategy is applied will be called the \emph{subject term}. Since strategies may be non-deterministic, these results can be zero, one, or even an infinite number. Looking at the results as a whole, strategies can be seen as transformations from a term to a set of terms. \emph{Strategy modules}, the modules where strategies can be named and recursive strategies can be defined, are described in \cref{sec:modules}. Formal semantics of what is here informally described can be found in \cref{sec:semantics}.

	To illustrate each strategy constructor at work, let us go back to the example specified in the modules \texttt{15PUZZLE-BOARD} and \texttt{15PUZZLE} of \cref{sec:maude}. The data representation of the square is not intended to be efficient, as shown by the conditional rules for \texttt{down} and \texttt{up}, but just to be easily readable and illustrative for the strategy examples in this section. The game will be discussed in more detail in~\cref{sec:15puzzle}.

\subsection{Idle and fail}

	The simplest strategies are the constants \lststrat|idle| and \lststrat|fail|.
\begin{mgrammar}
	<Strat>		::= "idle" | "fail"
\end{mgrammar}
The \lststrat|fail| strategy always \emph{fails} for any given state, i.e., it does not produce any result.

\begin{maudexec}
Maude> srewrite 1 b 2 using fail .

No solution.
rewrites: 0
\end{maudexec}
In a broader sense, we say that a strategy $\alpha$ \emph{fails} in a state $t$ if it does not produce any result. On the contrary, the \lstinline|idle| constant always succeeds, and returns the given state unaltered as its only result.
\begin{maudexec}
Maude> srewrite 1 b 2 using idle .

Solution 1
rewrites: 0
result Row: 1 b 2

No more solutions.
rewrites: 0
\end{maudexec}

These strategy constants are useful in combination with the conditional operator (see \cref{sec:conditional}), the rule application operator with rewriting conditions (see \cref{sec:ruleapp}), and strategy definitions (see \cref{sec:parameterization,sec:15puzzle,sec:RIP} for examples where \skywd{idle} is used to fill the base case of an inductive strategy).

\subsection{Rule application} \label{sec:ruleapp}

	Since strategies are used to control rewriting, the cornerstone of the language is the application of rules. Rules are usually selected by their labels, but a finer control on how they are applied is possible using an initial ground substitution, which can optionally be provided between brackets.
Moreover, in order to apply a conditional rule with rewriting conditions, strategies must be provided to control them.
\begin{mgrammar}
	<Strat>		::=	<RuleApp>
				\alt "top(" <RuleApp> ")"

	<RuleApp>	::=	<Label> [ "[" <Substitution> "]" ] [ "{" <StratList> "}" ]
				\alt "all"

	<Substitution> ::=	<Variable> "<-" <Term>
				\alt <Substitution> "," <Substitution>

	<StratList> ::=		<Strat>
				\alt <Strat> "," <StratList>
\end{mgrammar}
The strategy expression $\mathit{label}\,[x_1 \,\texttt{<-}\, t_1\texttt, \ldots\texttt, x_n \,\texttt{<-}\, t_n]\{\alpha_1\texttt, \ldots\texttt, \alpha_m\}$ denotes the application anywhere within the subject term of any rule labeled $\mathit{label}$ in the current module having exactly $m$ rewriting condition fragments.

\begin{maudexec}
Maude> srewrite 1 b 2 ; 3 b 4 using right .

Solution 1
rewrites: 1
result Puzzle: 1 2 b ; 3 b 4

Solution 2
rewrites: 2
result Puzzle: 1 b 2 ; 3 4 b

No more solutions
rewrites: 2
\end{maudexec}
Moreover, the variables in both rule sides and in its condition are previously instantiated by the substitution that maps $x_i$ to $t_i$ for all $1 \leq i \leq n$.
\begin{maudexec}
Maude> srewrite 1 b 2 ; 3 b 4 using left[T <- 1] .

Solution 1
rewrites: 1
result Puzzle: b 1 2 ; 3 b 4

No more solutions
rewrites: 1
\end{maudexec}

	When strategies are specified within curly brackets, each $\alpha_i$ controls the rewriting of the $i$-th rewriting condition fragment \lstinline[mathescape,keepspaces]|$l_i$ => $r_i$| of the selected rule. As usual when evaluating rule conditions, both sides $l_i$ and $r_i$ are instantiated with the substitution determined by the matching of the lefthand side and the previous condition fragments. However, the instance of $l_i$ is rewritten according to the strategy $\alpha_i$, and only its solutions are matched against $r_i$. As usual again, the evaluation of the next rewriting condition fragments continues with the partial substitution obtained by each result of the match modulo $B$, yielding potentially different one-step rewrites.
We now present a simple module illustrating the application of rules with rewriting fragments. The \texttt{PuzzleLog} pair maintains a history of applied moves in addition to the puzzle being solved, and both are updated simultaneously with the \texttt{move} rule. This rule is marked with the \texttt{nonexec}(utable) attribute, which causes the Maude rewrite commands to ignore it, because it contains a free variable \texttt{M} in the right-hand side. However, it can be executed with a rule application strategy that instantiates such a variable.
\begin{lstlisting}[escapeinside=^]
mod 15PUZZLE-LOG is
	protecting 15PUZZLE .
	protecting LIST{Qid} .

	sort PuzzleLog .
	op <_|_> : List{Qid} Puzzle -> PuzzleLog [ctor] .

	var M : Qid . var L : List{Qid} . vars P P' : Puzzle .

	crl [move] : < L | P > => < L M | P' > if P => P' [nonexec] .
endm
\end{lstlisting}
The imported module \texttt{LIST\{Qid\}} is an instance of the parameterized module \texttt{LIST} (described in \cref{sec:parameterization}) with quoted identifiers as elements. These identifiers are arbitrary words prefixed by a single quote, like \texttt{'left}, and they are appended to the list by juxtaposition.
\begin{maudexec}
Maude> srewrite < nil | 1 b 2 > using move[M <- 'left]{left} .

Solution 1
rewrites: 3
result PuzzleLog: < 'left | b 1 2 >

No more solutions.
rewrites: 3
\end{maudexec}

	A special strategy for rule application is the constant \skywd{all} that executes any rule in the module, labeled or not. Rewriting conditions are evaluated without restrictions, as in the usual \texttt{rewrite} command.
However, rules marked with \texttt{nonexec} and the implicit rules that handle external objects~\cite[\S 9]{maude} are excluded.
\begin{maudexec}
Maude> srewrite < nil | 1 b 2 > using all .

Solution 1
rewrites: 3
result PuzzleLog: < nil | b 1 2 >

Solution 2
rewrites: 4
result PuzzleLog: < nil | 1 2 b >

No more solutions.
rewrites: 4
\end{maudexec}

	Rules are usually applied anywhere in the subject term by default; but rule application expressions can be prefixed by the \lststrat|top| modifier to restrict matching to the top of the subject term. In combination with the subterm rewriting operator, to be explained in~\cref{sec:matchrew}, this allows restricting rewriting to specific positions within the term structure. Nevertheless, \lststrat|top(alpha)| is not necessarily deterministic, because matching of the rule lefthand side takes place modulo structural axioms $B$ such as associativity and commutativity. Additionally, rules can contain matching and rewriting condition fragments that may produce multiple substitutions. For example, if we had a rule \texttt{multimv} as \lstinline[keepspaces]{LU b RU => LU RU b} that jumps multiple positions at once, we would obtain:

\begin{maudexec}
Maude> srewrite 1 b 2 b 3 using top(multimv) .

Solution 1
rewrites: 1
result Row: 1 2 b 3 b

Solution 2
rewrites: 2
result Row: 1 b 2 3 b

No more solutions.
rewrites: 2
\end{maudexec}

\subsection{Tests} \label{sec:tests}

	Strategies often need to check some property of the subject term to decide whether to continue rewriting it or not, or whether it can be given as a solution. Test can be used to do these checks and abandon those rewriting paths in which they are not satisfied. Since matching is one of the most basic features of rewriting, tests are based on matching and on evaluation of equational conditions.
\begin{mgrammar}
	<EqCondition> ::=	<BoolTerm>
				\alt <Term> "=" <Term>
				\alt <Term> ":=" <Term>
				\alt <EqCondition> "/\textbackslash" <EqCondition>

	<Test>	::=		"amatch" <Pattern> [ "s.t." <EqCondition> ]
				\alt "match" <Pattern> [ "s.t." <EqCondition> ]
				\alt "xmatch" <Pattern> [ "s.t." <EqCondition> ]
\end{mgrammar}
Three variants of tests are available regarding where matching takes place: \skywd{match} tries to match at the top of the subject term, \skywd{amatch} matches anywhere, and \skywd{xmatch} matches only at the top, but \emph{with extension} modulo the structural axioms of the top symbol.\footnote{Suppose $+$ is an associative and commutative (AC) symbol, $f$ is a unary symbol, and $a$, $b$, and $c$ are constants. Consider the rule $f(x) + a \Rightarrow f(x) + b$. This rule cannot be applied at the top to the term $(f(x) + c) + a$. However, it can be applied \emph{with extension} modulo AC if we add the AC-extension rule $(f(x) + a) + y \Rightarrow (f(x) + b) + y$. See~\cite[\S 4.8]{allmaude} for a detailed explanation with examples of extension rules.} Their behavior is equivalent to \skywd{idle} when the test passes, and to \fail{} when it does not. In other words, the test strategy applied to a term $t$ will evaluate to the set of results $\{ t \}$ if the matching and condition evaluation succeeds, and to $\emptyset$ if they fail.

\begin{maudexec}
Maude> srewrite 1 b 2 using xmatch b N s.t. N =/= 1 .

Solution 1
rewrites: 1
result Row: 1 b 2

No more solutions.
rewrites: 1
\end{maudexec}

	If the pattern and the condition contain variables that are bound in the current scope, they will be instantiated before matching. The condition is then evaluated like an equational condition \cite[\S 4.3]{maude}.

\subsection{Regular expressions}

	The first strategy combinators to describe execution paths are the typical regular expression constructors.

\begin{mgrammar}
	<Strat> ::=		<Strat> ";" <Strat>
				\alt <Strat> "|" <Strat>
				\alt <Strat> "*"
				\alt <Strat> "+"
\end{mgrammar}
The \emph{sequential composition} or \emph{concatenation} $\alpha \,\seq\, \beta$ rewrites the subject term with $\alpha$ and then with $\beta$. Hence, every result of $\alpha \,\seq\, \beta$ is a result of applying $\beta$ to a term obtained from $\alpha$. Of course, if $\alpha$ fails, no result will be obtained regardless of $\beta$.
The \emph{nondeterministic choice} or \emph{alternation} combinator $\alpha \disj \beta$ rewrites the subject term using either $\alpha$ or $\beta$, chosen nondeterministically. The solutions of the alternation is the union of the solutions of both strategies.

\begin{maudexec}
Maude> srewrite 1 2 ; 3 b using left ; up | up ; left .

Solution 1
rewrites: 24
result Puzzle: b 2 ; 1 3

Solution 2
rewrites: 32
result Puzzle: b 1 ; 3 2

No more solutions.
rewrites: 32
\end{maudexec}
Notice that the previous strategy is parenthesized as \texttt{(left ; up) | (up ; left)}, with the usual precedence of regular expressions.
The \emph{iteration} \lststrat|alpha *| executes $\alpha$ zero or more times consecutively. It can be described recursively as \lststrat[mathescape]{alpha * $\equiv$ idle | alpha ; alpha *}.
\begin{maudexec}
Maude> srewrite 1 b 2 3 using right * .

Solution 1
rewrites: 0
result Row: 1 b 2 3

Solution 2
rewrites: 1
result Row: 1 2 b 3

Solution 3
rewrites: 2
result Row: 1 2 3 b

No more solutions.
rewrites: 2
\end{maudexec}
Iterations are the simplest form of a loop in the strategy language, although they finish nondeterministically after each iteration. The typical loop \texttt{while $P$ do $\alpha$ end} where $P$ is a predicate on the subject term can be encoded as \lststrat[mathescape]|(match $X$ s.t. $P(X)$ ; $\alpha$) * ; match $X$ s.t. not $P(X)$| for an appropriate variable $X$. They can also be useful to expand the transitive closure of a strategy, like in the example above, where we obtain all states reachable by successive applications of the \texttt{right} rule.

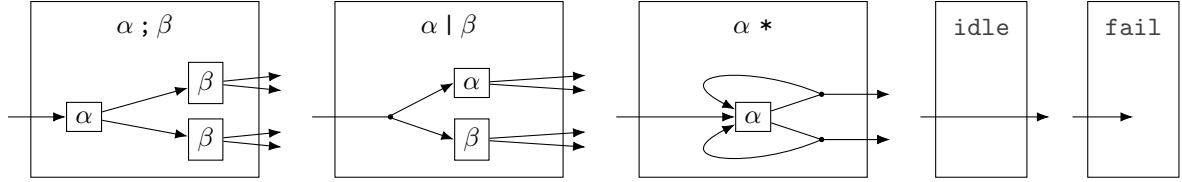
\begin{figure}[t]\centering
\begin{tikzpicture}[sbox/.style={draw}, >=LaTeX]
\begin{scope}
\draw (-1.5, -2) rectangle (1.5, .33);
\node at (0, 0) (S) {$\alpha\, \seq\, \beta$};
\node[sbox] at (-.8, -1.2) (SA) {$\alpha$};
\node[sbox] at (.8, -.75) (SB1) {$\beta$};
\node[sbox] at (.8, -1.5) (SB2) {$\beta$};
\draw[->] (SA) -- (SB1);
\draw[->] (SA) -- (SB2);
\draw[->] ($(SA)+(-1, 0)$) -- (SA);
\draw[->] (SB1) -- ($(SB1)+(1, .1)$);
\draw[->] (SB1) -- ($(SB1)+(1, -.1)$);
\draw[->] (SB2) -- ($(SB2)+(1, .1)$);
\draw[->] (SB2) -- ($(SB2)+(1, -.1)$);
\end{scope}

\begin{scope}[shift={(4, 0)}]
\draw (-1.5, -2) rectangle (1.5, .33);
\node at (0, 0) (S) {$\alpha\, \disj\, \beta$};
\node[sbox] at (.3, -1.5) (DA) {$\beta$};
\node[sbox] at (.3, -.75) (DB) {$\alpha$};
\draw[->] (DA) -- ($(DA)+(1.5, .1)$);
\draw[->] (DA) -- ($(DA)+(1.5, -.1)$);
\draw[->] (DB) -- ($(DB)+(1.5, .1)$);
\draw[->] (DB) -- ($(DB)+(1.5, -.1)$);
\draw (-1.8, -1.2) -- (-.8, -1.2);
\fill (-.77, -1.2) circle (1pt);
\draw[->] (-.8, -1.2) -- (DA.west);
\draw[->] (-.8, -1.2) -- (DB.west);
\end{scope}

\begin{scope}[shift={(8, 0)}]
\draw (-1.5, -2) rectangle (1.5, .33);
\node at (0, 0) (L) {$\alpha$ \texttt{*}};
\node[sbox] at (0, -1.2) (LA) {$\alpha$};
\draw[->] (-1.8, -1.2) -- (LA);
\draw (LA) -- (.9, -.9);
\draw (LA) -- (.9, -1.5);
\fill (.9, -.9) circle (1pt);
\fill (.9, -1.5) circle (1pt);
\draw[->] (.9, -.9) -- (1.8, -.9);
\draw[->] (.9, -1.5) -- (1.8, -1.5);
\draw[->] (.9, -.9) .. controls (0, -.5) and (-1.2, -.6) .. ($(LA.north west)!0.5!(LA.west)$);
\draw[->] (.9, -1.5) -- (1.8, -1.5);
\draw[->] (.9, -1.5) .. controls (0, -1.9) and (-1.2, -1.8) .. ($(LA.south west)!0.5!(LA.west)$);
\end{scope}

\begin{scope}[shift={(11, 0)}]
\draw (-.6, -2) rectangle (.6, .33);
\node at (0, 0) {\skywd{idle}};
\draw[->] (-.8, -1.2) -- (.9, -1.2);
\end{scope}

\begin{scope}[shift={(13, 0)}]
\draw (-.6, -2) rectangle (.6, .33);
\node at (0, 0) {\skywd{fail}};
\draw[->] (-.8, -1.2) -- (0, -1.2);
\end{scope}
\end{tikzpicture}
\caption{Schema for the operators of the regular expressions sublanguage.} \label{fig:regex}
\end{figure}

	The already seen \lststrat|idle| and \lststrat|fail| constants complete the regular expressions sublanguage. The usual notation for the non-empty iteration \lststrat|alpha +| is also available in the language as a derived combinator \lststrat[mathescape]{alpha + $\equiv$ alpha ; alpha *}. \cref{fig:regex} summarizes in a diagram the meaning of the main operators presented in this section.

\subsection{Conditional and derived operators} \label{sec:conditional}

	Conditional statements and expressions are an elementary control mechanism in any programming language. The next strategy combinator is the typical \emph{if-then-else} construct. However, its condition is a strategy too, which is considered satisfied if it provides at least one result. This idea has been borrowed from Stratego~\cite{stratego} and ELAN~\cite{elan}.
\begin{mgrammar}
	<Strat>	::=	<Strat> "?" <Strat> ":" <Strat>
\end{mgrammar}
More precisely, the evaluation of the expression \lststrat|alpha ? beta : gamma| starts by evaluating the condition $\alpha$. If $\alpha$ succeeds, its results are continued with the positive branch $\beta$ just as if $\alpha \seq \beta$ were run. But if $\alpha$ fails, the negative branch $\gamma$ is evaluated instead on the initial term. In case $\alpha$ is just a test, the conditional behaves like a typical Boolean conditional.

\begin{maudexec}
Maude> srewrite 1 b 2 3 4 using right * ; (right ? fail : idle) .

Solution 1
rewrites: 6
result Row: 1 2 3 4 b

No more solutions.
rewrites: 6
\end{maudexec}

	Using the conditional strategy, multiple useful derived combinators can be defined and are available in the language:
\begin{mgrammar}
	<Strat>	::=	<Strat> "or-else" <Strat>
			\alt "not(" <Strat> ")"
			\alt <Strat> "!"
			\alt "try(" <Strat> ")"
			\alt "test(" <Strat> ")"
\end{mgrammar}
The combinator \lststrat|alpha or-else beta| evaluates to the result of $\alpha$ unless $\alpha$ fails; in that case, it evaluates to the result of $\beta$. Consequently, it is equivalent to
\[ \hbox{\lststrat|alpha or-else beta| $\;\equiv\;$ \lststrat|alpha ? idle : beta|} \]
This is one of the most used combinators of the strategy language because rule precedences appear in many situations. The negation combinator \lststrat|not(alpha)| is defined as
\[ \hbox{\lststrat|not(alpha)| $\equiv$ \lststrat|alpha ? fail : idle|,} \]
thus reversing the binary result of evaluating $\alpha$. More precisely, the initial term is obtained when $\alpha$ fails, and \texttt{not} fails when $\alpha$ succeeds.  The \emph{normalization} operator \lststrat|alpha !| evaluates a strategy repeatedly until just before it fails.
\[ \hbox{\lststrat[mathescape]{alpha ! $\equiv$ alpha * ; not(alpha)}} \]
Observe that the strategy in the \texttt{srewrite} example above can now be simply written as \texttt{right !}. The strategy \lststrat|try(alpha)| works as if $\alpha$ were evaluated, but when $\alpha$ fails it results in $\{t\}$ instead, where $t$ was the initial subject term.
\[ \hbox{\lststrat|try(alpha)| $\equiv$ \lststrat|alpha ? idle : idle|} \]
Finally, \lststrat|test(alpha)| checks the success/failure result of $\alpha$, but it does not change the subject term.
\[ \hbox{\lststrat|test(alpha)| $\equiv$ \lststrat|not(alpha) ? fail : idle|} \]
Notice that this is the same as \lststrat|not(not(alpha))|.

\subsection{Rewriting of subterms} \label{sec:matchrew}

	As seen in~\cref{sec:ruleapp}, rules are applied anywhere within the subject term by default, but we may be interested in concentrating their application into selected subterms. The following family of combinators allows rewriting multiple selected subterms using some given strategies. Additionally, it can be used to extract information from the subject term and use it for the control of the strategy execution.
\begin{mgrammar}
	<Strat> ::=		"amatchrew" <Pattern> [ "s.t." <EqCondition> ] "by" <VarStratList>
				\alt "matchrew" <Pattern> [ "s.t." <EqCondition> ] "by" <VarStratList>
				\alt "xmatchrew" <Pattern> [ "s.t." <EqCondition> ] "by" <VarStratList>

	<VarStratList> ::=	<VarId> "using" <Strat>
				\alt <VarStratList> "," <VarStratList>
\end{mgrammar}
The subterms to be rewritten are selected by matching against a pattern $P$. Some of the variables in the pattern $x_1, \ldots, x_n$ will match selected subterms $u_1, \ldots, u_n$ to which the strategies $\alpha_1, \ldots, \alpha_n$ are respectively applied.
\[ \hbox{\lstinline[mathescape]|matchrew $P[x_1, \ldots, x_n, x_{n+1}, \ldots, x_m]$ s.t. $C$ by $x_1$ using $\alpha_1$, ..., $x_n$ using $\alpha_n$|} \]
These variables must be distinct, but they may appear more than once in the pattern. There may be additional variables $x_{n+1}, \ldots, x_m$ that do not designate subterms to be rewritten.

\begin{figure}[t]\centering
\begin{tikzpicture}
	\node (TP) {\ttfamily f(... g(...) ...) --> f(... g(...) ...)};
	\node[below=1.5em of TP] (T) {\ttfamily g(...$u_i$...$u_j$...) --> g(...$u_i'$...$u_j'$...)};

	\draw (-2, -1.4) edge[bend right,->] (1.3, -1.4);
	\draw (-1.35, -1.4) edge[bend right=25,->] (2.2, -1.4);
	\draw (-1.9, -0.3) edge[bend right=10,->] (-1.7, -0.8);
	\draw (2, -0.8) edge[bend right=10,->] (2.3, -0.3);

	\node at (-3.1, -0.55) {matching};
	\node at (3.7, -0.55) {substitution};
	\node at (3.9, -1.7) {rewriting};
	\node at (-2.1, -1.7) {$\alpha_i$};
	\node at (2.3, -1.8) {$\alpha_j$};
\end{tikzpicture}

\caption{Behavior of the \texttt{amatchrew} combinator} \label{fig:matchrew}
\end{figure}
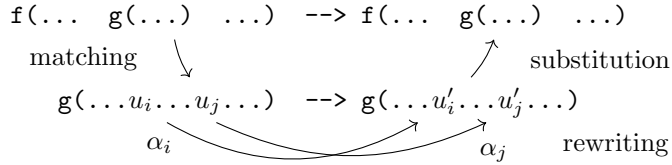

The strategy explores all possible matches of the pattern $P$ for the subject term that satisfy the condition $C$. For each, it extracts the subterms $u_i$ to $u_n$ that match the variables $x_1$ to $x_n$, and rewrites them separately in parallel using their corresponding strategies $\alpha_1$ to $\alpha_n$. Every combination of their results is reassembled in the original term in place of the original subterms. Like for the tests, three variants can be selected by changing the initial keyword: \skywd{matchrew} for matching on top, \skywd{xmatchrew} for doing so with extension, and \skywd{amatchrew} for matching anywhere. The behavior of the \skywd{amatchrew} operator is illustrated in \cref{fig:matchrew}.
\begin{maudexec}
Maude> srewrite 1 b 2 ; 3 b 4 using matchrew RU ; RD
                                    by RU using left, RD using right .

Solution 1
rewrites: 2
result Puzzle: b 1 2 ; 3 4 b

No more solutions.
rewrites: 2
\end{maudexec}
The following simple example shows that patterns may be non-linear and that multiple matches are possible.
\begin{maudexec}
Maude> srewrite 1 b ; 1 b ; 2 b ; 2 b using xmatchrew R ; R by R using left .

Solution 1
rewrites: 1
result Puzzle: b 1 ; b 1 ; 2 b ; 2 b

Solution 2
rewrites: 2
result Puzzle: 1 b ; 1 b ; b 2 ; b 2

No more solutions.
rewrites: 2
\end{maudexec}

	Variables bound in the outer scope are instantiated in the pattern $P$ and condition $C$, like in tests. Moreover, the variable scope of the substrategies $\alpha_1, \ldots, \alpha_n$ is extended with the variables in $P$ and $C$. Notice that \lststrat|matchrew| and its variants are the only strategies that define static variable scopes, and the only ones that bind variables along with strategy calls. 
In fact, \skywd{matchrew} operators are often used for binding variables and introduce their values in the strategy control flow, as shown in almost all examples in \cref{sec:examples}, even with patterns as simple as a single variable. A common idiom is \lststrat[mathescape]{matchrew S s.t. $P$ := S by S using $\alpha$} where $\alpha$ needs to be applied to the whole term, but depends on information that can be obtained by pattern matching from some part of it.\footnote{In a draft version of the language, arbitrary subpatterns could be used to select the subterm to be rewritten, like in \texttt{\skywd{matchrew} f(X, g(Y)) \skywd{by} g(Y) \skywd{using} $\alpha$}. However, this option was removed for simplicity and to avoid some pathological cases with overlapping patterns.}

\subsection{Pruning of alternative solutions} \label{sec:one}

	The strategy language combinators are nondeterministic in different ways: rule applications may produce multiple rewrites, the alternation operator $\alpha \disj \beta$ could choose $\alpha$ or $\beta$, the iteration \lststrat|alpha *| may execute $\alpha$ any number of times, and \lststrat|matchrew| could start with different matches. The default behavior of the strategy rewriting engine is to perform an exhaustive search on all the allowed rewriting paths, and show all reachable results. However, the user may be interested in picking any solution without distinction, or may know that a single solution exists. In these cases and for efficiency, the \texttt{one} combinator is available.
\begin{mgrammar}
	<Strat> ::= "one(" <Strat> ")"
\end{mgrammar}
\lststrat|one(alpha)| evaluates $\alpha$ on the given term. If $\alpha$ fails, so \lststrat|one(alpha)| does. Otherwise, a solution for $\alpha$ is chosen nondeterministically as its single result. However, since the purpose of introducing this strategy is efficiency, the chosen solution will actually be the first that is found.

\begin{maudexec}
Maude> srewrite 1 b 2 3 using one(right +) .

Solution 1
rewrites: 1
result Row: 1 2 b 3

No more solutions.
rewrites: 1
\end{maudexec}
Notice that all deterministic choices will still be explored until a solution is found. This means that \skywd{one} will be more effective the fewer steps are required to execute its argument. Another consequence is that \skywd{one} cannot be used to limit the search of a \skywd{matchrew} to a single match, which may be desired in some situations, and multiple matches may be tried until a solution of the whole \skywd{matchrew} is encountered.

	More meaningful usage examples of \texttt{one} can be found in \cref{sec:RIP}. ELAN \cite{elan} counts with similar constructs \texttt{dc one} and \texttt{first one}, whose behaviors coincide with our \texttt{one} when a single strategy argument is provided (see~\cref{sec:comparison}).

\subsection{Strategy calls} \label{sec:calls}

	Instead of writing huge monolithic strategy expressions and copy them verbatim to be executed, complex strategies are better split into several subexpressions referred to by meaningful names. These strategies are defined in strategy modules, which will be explained in \cref{sec:modules}, but also extend the language of expressions with a \emph{strategy call} constructor.
\begin{mgrammar}
	<Strat> ::= <StratCall>

	<StratCall> ::= <StratId> "(" <TermList> ")"
	\alt <StratId>
\end{mgrammar}
Strategies are invoked by a strategy label, and may receive input arguments of any sort in the module, according to their declarations. These parameters are written in a comma-separated list between parentheses. For strategies without parameters, the parentheses can be omitted, except when there is a rule with the same label in the module.

For example, suppose we have defined a strategy \texttt{move} that moves the blank as many horizontal and vertical positions as indicated by its two arguments (this strategy will actually be defined in \cref{sec:modules}). The following command executes it with an offset of two horizontal positions:
\begin{maudexec}
Maude> srewrite 1 b 2 3 4 using move(2, 0) .

Solution 1
rewrites: 2
result Puzzle: 1 2 3 b 4

No more solutions.
rewrites: 2
\end{maudexec}

\subsection{Strategy commands} \label{sec:commands}

	Strategies are evaluated in the interpreter using the \texttt{srewrite} and \texttt{dsrewrite} commands, which look for solutions of the strategy and show all those they find.
\begin{mgrammar}
	<Commands>	::= "srewrite" [ "[" <Nat> "]" ] [ "in" <ModId> ] <Term> "using" <Strat> "."
			\alt "dsrewrite" [ "[" <Nat> "]" ] [ "in" <ModId> ] <Term> "using" <Strat> "."
\end{mgrammar}
Following the usual convention of Maude commands, an optional bound on the number of solutions to be calculated can be specified between brackets after the command keyword. Nevertheless, more solutions can be requested afterwards using the \texttt{continue} command. By default, the command will be executed in the current Maude module, but a different module could be specified preceded by \texttt{in}. The \texttt{srewrite} keyword can be abbreviated to \texttt{srew}, and \texttt{dsrewrite} to \texttt{dsrew}.

	The \texttt{srewrite} command explores the rewriting tree using a \emph{fair} policy that ensures that all solutions are eventually found if there is enough memory. Not being completely a breadth-first search, it may explore multiple execution paths in parallel. More details can be found in \cref{sec:implementation}.
\begin{maudexec}
Maude> srew [2] b 1 2 ; b 3 using right + .

Solution 1
rewrites: 2
result Puzzle: 1 b 2 ; b 3

Solution 2
rewrites: 3
result Puzzle: b 1 2 ; 3 b
\end{maudexec}
On the contrary, the \texttt{dsrewrite} command performs a depth-first exploration of the tree. It is usually faster and uses less memory, but some solutions may not be reached because of nonterminating execution branches.
\begin{maudexec}
Maude> dsrew [2] b 1 2 ; b 3 using right + .

Solution 1
rewrites: 1
result Puzzle: 1 b 2 ; b 3

Solution 2
rewrites: 2
result Puzzle: 1 2 b ; 3 b
\end{maudexec}
Notice that the order in which solutions are obtained may differ depending on the type of search, as they do in the execution above. The displayed rewrite count reflects all the equational and rule rewrites that have been applied until the solution was found, but its origin could be in other execution branches not yet completed, or abandoned because they do not lead to a solution.

	The search conducted by the \texttt{srewrite} and \texttt{dsrewrite} commands theoretically explores the subtree of the rewriting tree pruned by the restrictions of the strategy, but their search space is actually a graph. The execution engine is able to detect already visited execution states, thus preventing the redundant evaluation of the same strategy on the same term. Consequently, the strategy evaluation may finish in situations where, operationally, nonterminating executions are involved.
\begin{maudexec}
Maude> srew 1 b using (left | right) * .

Solution 1
result Row: 1 b

Solution 2
result Row: b 1

No more solutions.
\end{maudexec}
However, strategy evaluation is not always terminating, since the underlying rewriting system may have infinitely many states. In addition, the cycle detector does not operate for strategy calls, unless they are tail recursive and do not have parameters.

\subsection{Strategy modules and recursion} \label{sec:modules}

	As anticipated in \cref{sec:calls}, callable strategies can be declared and defined in strategy modules. Apart from the benefits already described, this increases the expressiveness of the strategy language via recursive and mutually recursive strategies, which can also keep a control state in their parameters.

	Strategy modules are a third level of Maude modules, devoted to represent the control of rewriting systems by means of the strategy language, as the classical functional and system modules were dedicated to represent equational and rewrite theories, respectively. They are introduced by the \kywd{smod} keyword and closed by \kywd{endsm}.
\begin{mgrammar}
	<Module>	::= "smod" <ModId> [ <ParameterList> ] "is" <SmodElt>* "endsm"

	<SmodElt>	::= <ModElt> | <StratDecl> | <StratDef>
\end{mgrammar}
Strategy modules are extensions of system modules in a similar way as system modules are extensions of functional modules. Therefore, they may include any declaration or statement that is allowed in these lower-level modules. However, to promote a clean separation between the rewriting theory specification and its control, we encourage including only strategy-related statements in strategy modules, apart from importations and variable declarations. Only strategy modules are able to import other strategy modules, but they can import modules of any kind using the usual statements: \texttt{including}, \texttt{extending}, \texttt{generated-by}, and \texttt{protecting}. The semantic difference between these importation modes is described in~\cite[\S 10.2.1]{maude}.

	The strategy-related statements are strategy declarations and strategy definitions.
\begin{mgrammar}
	<StratDecl>	::= "strat" <SLabel>+ [ ":" <Type>* ] "@" <Type> "."

	<StratDef>	::= "sd" <StratCall> ":=" <Strat> "."
			\alt "csd" <StratCall> ":=" <Strat> "if" <Condition> "."
\end{mgrammar}
The following line declares a strategy $\mathit{slabel}$ that receives $n$ parameters of sorts $s_1, \ldots, s_n$, and that is intended to control rewriting of terms of sort $s$.
\begin{lstlisting}[mathescape]
strat $\mathit{slabel}$ : $s_1$ $\ldots$ $s_n$ @ $s$ .
\end{lstlisting}
This latter sort $s$ is understood as a mere comment and ignored by Maude.
Many strategies with a common signature can be defined in the same declaration, by writing multiple identifiers, in which case the plural keyword \lststrat|strats| is preferred. The input parameter sorts and the colon are omitted if the strategy has no parameters.

	Strategies are defined by means of conditional or unconditional strategy definitions.
\begin{lstlisting}[mathescape, keepspaces]
sd $\mathit{slabel}$($p_1$, $\ldots$, $p_n$) := $\alpha$ .
csd $\mathit{slabel}$($p_1$, $\ldots$, $p_n$) := $\alpha$ if $C$ .
\end{lstlisting}
These definitions associate the strategy name $\mathit{slabel}$ to an expression $\alpha$ whenever the input parameters match the patterns $p_1, \ldots, p_n$.
The syntax of conditions is the same as that of equations and tests, explained in~\cref{sec:tests}. The lefthand sides of the definitions are strategy calls as described in~\cref{sec:calls}. Variables in the pattern and the condition may appear in the strategy expression $\alpha$. When a strategy is called, all strategy definitions that match the input arguments will be executed, and the union of all their results is the result of the call. Hence, strategies without any definition at all behave like \fail.

	The following is an example of strategy module that imports the system module \texttt{15PUZZLE}, declares two strategies, \texttt{loop} and \texttt{move}, and gives definitions for them.\footnote{The \texttt{15PUZZLE-STRATS} module only includes unconditional strategy definitions. For an example using conditional definitions, see the \texttt{solveLoop} strategy in \cref{sec:15puzzle}.}
\begin{lstlisting}
smod 15PUZZLE-STRATS is
	protecting 15PUZZLE .
	protecting INT .

	strat loop           @ Puzzle .
	strat move : Int Int @ Puzzle .

	var N : Nat . var M : Int .

	sd loop := left ; up ; right ; down .

	sd move(0, 0)      := idle .
	sd move(s(N), M)   := right ; move(N, M) .
	sd move(- s(N), M) := left  ; move(- N, M) .
	sd move(0, s(N))   := down  ; move(0, N) .
	sd move(0, - s(N)) := up    ; move(0, - N) .
endsm
\end{lstlisting}
While \texttt{loop} simply displaces the blank in a fixed loop, \texttt{move} moves it some offset in the board. Its five definitions have disjoint patterns, so that only a single definition will be executable for any given input. Therefore, the strategy is deterministic. If \texttt{M} were written instead of \texttt{0} in the last two definitions, multiple definitions could be activated for the same call, and both vertical and horizontal displacements would be mixed nondeterministically to bridge the distance.

\begin{maudexec}
Maude> srewrite 1 2 3 ; 4 5 6 ; 7 b 8 using move(1, -2) .

Solution 1
rewrites: 70
result Puzzle: 1 2 b ;
               4 5 3 ;
               7 8 6

No more solutions.
rewrites: 70
\end{maudexec}

\subsection{Parameterization} \label{sec:parameterization}

	Maude's support for \emph{parameterized programming}~\cite[\S 6.3]{parameterized,maude} in its functional and system modules has also been extended to strategy modules. To describe parameterization for strategy modules, we will first introduce the basic and common building blocks: theories, parameterized modules, and views. \emph{Theories} are used to express the interface of parameterized modules, by describing the requirements that any actual parameter must satisfy. Their syntax is almost identical to that of functional and system modules, but they are delimited by keywords \lstinline|fth| and \lstinline|endfth|, for functional theories, and \lstinline|th| and \lstinline|endth| for system theories. However, their declarations are understood as formal objects, and the executability requirements that a module must obey are not required for theories. The simplest one, although extensively used, is the following theory \texttt{TRIV} specifying a single parameter sort:
\begin{lstlisting}
fth TRIV is
	sort Elt .
endfth
\end{lstlisting}

\begin{figure}[t]\centering
\begin{tikzpicture}
	\node (TH1) {\texttt{TH1}};
	\node[below=of TH1] (THn) {\texttt{THn}};
	\node[below=of THn] (PM) {\texttt{PM\{X1 :: TH1, \ldots, Xn :: THn\}}};
	\node[right=3cm of PM] (IN) {\texttt{PM\{View1, \ldots, Viewn\}}};
	\node[above=of IN] (Mn) {\texttt{Mn}};
	\node[above=of Mn] (M1) {\texttt{M1}};

	\node[above=.25 of THn] {$\vdots$};
	\node[above=.25 of Mn] {$\vdots$};

	\draw[->] (TH1) -- node[above] {\texttt{View1}} (M1);
	\draw[->] (THn) -- node[above] {\texttt{Viewn}} (Mn);
	\draw[->] (PM) -- (IN);

	\draw[->] (THn) -- (PM);
	\draw[->] (Mn) -- (IN);
\end{tikzpicture}
\caption{Diagram of a parameterized module instantiation.} \label{fig:instantiation}
\end{figure}

	A \emph{parameterized module} includes in its header \texttt{PM\{X1 :: TH1, \ldots, Xn :: THn\}} a list of one or more formal parameters, each bound to a theory and identified by a name. In its body, the parameter sorts are referred to by their formal names prefixed by the parameter name and a dollar sign \lstinline|$|, while the formal operators are referred to by their original names. For example, a parameterized module for lists can be specified as follows:
\begin{lstlisting}
fmod LIST{X :: TRIV} is
	sorts NeList{X} List{X} .
	subsort X$Elt < List{X} .

	op nil : -> List{X} .
	op __ : List{X} List{X} -> List{X} [comm assoc id: nil] .

	*** (more declarations and equations)
endfm
\end{lstlisting}

	How a module satisfies the requirements of a theory\footnote{That the requirements of theory $T$ are satisfied by view $V$ is a \emph{proof obligation} that is not checked by Maude. It can instead be discharged with the help of tools in the Maude Formal Environment~\cite{mfe}.} is specified using a \emph{view}, which maps the formal objects in the theory to the actual objects in the chosen target module where the theory is interpreted. Views are then used to instantiate the parameterized modules as depicted in~\cref{fig:instantiation}.
For example, the following view interprets \texttt{NAT} as a \texttt{TRIV} by mapping the formal sort \texttt{Elt} to the actual sort \texttt{Nat} of natural numbers.
\begin{lstlisting}
view Nat from TRIV to NAT is
	sort Elt to Nat .
endv
\end{lstlisting}
Then, the parameterized module \lstinline|LIST{X :: TRIV}| is instantiated by the view \texttt{Nat} to produce the module \lstinline|LIST{Nat}| of lists of natural numbers. Such module instantiation expressions can appear in importation statements to be used in the importing module.
\begin{lstlisting}
fmod COUNTDOWN is
	protecting LIST{Nat} .
	op countdown : -> List{Nat} .
	eq countdown = 5 4 3 2 1 0 .
endm
\end{lstlisting}

	Strategy modules can be parameterized by functional and system theories too, but also with formal strategies expressed as \emph{strategy theories}.
\begin{mgrammar}
	<Theory>	::= "sth" <ModId> "is" <SmodElt>* "endth"
\end{mgrammar}
Strategy theories are syntactically identical to strategy modules. They should only contain strategy declarations and definitions, but they can also include functional or system theories using the \lstinline{including} mode (or modules using any importation mode). The formal strategies declared in the theory should be realized by the actual target modules.
Modules other than strategy modules cannot be parameterized by strategy theories. A simple example of strategy theory is the following \texttt{STRIV} that declares a single strategy without arguments operating on the formal sort of the \texttt{TRIV} theory.
\begin{lstlisting}
sth STRIV is
	including TRIV .
	strat st @ Elt .
endsth
\end{lstlisting}
Parameterized by this theory, we can then specify the following module \texttt{REPEAT}, which defines a strategy \texttt{repeat($n$)} that applies $n$ times the parameter strategy \texttt{st}.
\begin{lstlisting}
smod REPEAT{X :: STRIV} is
	protecting NAT .

	strat repeat : Nat @ X$Elt .

	var N : Nat .

	sd repeat(0)    := idle .
	sd repeat(s(N)) := st ; repeat(N) .
endsm
\end{lstlisting}
Again, views are required to interpret strategy theories and instantiate strategy modules. Thus, the syntax of views is extended to support strategy mappings:
\begin{mgrammar}
	<ViewElt>	::= "strat" <StratId> "to" <StratId> "."
			\alt "strat" <StratId> [ : <Type> * ] "@" <Type> "to" <StratId> "."
			\alt "strat" <StratCall> "to expr" <Strat> "."
\end{mgrammar}
Three different forms of mapping are offered with the same structure as operator mappings~\cite[\S 6.3.2]{maude}: using \texttt{\skywd{strat} $\mathit{formalName}$ : $s_1$ ... $s_n$ @ $s$ \skywd{to} $\mathit{actualName}$}, the formal strategy with the given name $\mathit{formalName}$ and signature is mapped to an actual strategy in the target module whose name is $\mathit{actualName}$ and whose input arguments' sorts are the translation of the formal signature according to the sort mappings of the view. To map all overloaded strategies with a given name at once, another mapping \texttt{\skywd{strat} $\mathit{formalName}$ to $\mathit{actualName}$} is available. Finally, \texttt{\skywd{strat} $\mathit{slabel}(x_1, \ldots, x_n)$ \skywd{to expr} $\alpha$} maps the strategy $\mathit{slabel}$ whose input types are those of $x_1, \ldots, x_n$ to the strategy expression $\alpha$ that may depend on these variables. Only variables are allowed as arguments in the lefthand side.

	For instance, we can repeatedly apply the \texttt{loop} strategy defined in the strategy module \texttt{15PUZZLE-STRATS} at the beginning of this section, by instantiating the \texttt{REPEAT} module with the following view from \texttt{STRIV} to \texttt{15PUZZLE-STRATS}.
\begin{lstlisting}
view Loop from STRIV to 15PUZZLE-STRATS is
	sort Elt to Puzzle .
	strat st to loop .
endv
\end{lstlisting}
\texttt{Loop} maps the \texttt{Elt} sort to the \texttt{Puzzle} sort, and the formal strategy \texttt{st} to the \texttt{loop} strategy in \texttt{15PUZZLE-STRAT}. Instantiating \lstinline|REPEAT{X :: STRIV}| with the view \texttt{Loop} by writing \lstinline|REPEAT{Loop}|, we can now apply \texttt{repeat} for \texttt{loop}.
\begin{maudexec}
Maude> srew 1 2 ; 3 b using repeat(2) .

Solution 1
rewrites: 64
result Puzzle: 3	1 ;
               2	b

No more solutions.
rewrites: 64
\end{maudexec}

	Typically, the target of a view from a strategy theory is a strategy module but, using the \texttt{to expr} mapping, views can be directly specified for system modules. For example, the following view \texttt{Right} to the system module \texttt{15PUZZLE} maps \texttt{st} to the rule application expression \texttt{right}.
\begin{lstlisting}[escapechar=^]
view Right from STRIV to 15PUZZLE is
	sort Elt to Puzzle .
	strat st ^\textbf{to expr}^ right .
endv
\end{lstlisting}
Then, in a module including \lstinline|REPEAT{Right}|, we can execute:
\begin{maudexec}
Maude> srew b 1 2 3 4 using repeat(3) .

Solution 1
rewrites: 3
result NeRow: 1 2 3 b 4

No more solutions.
rewrites: 3
\end{maudexec}
Other combinations of initial theories and target modules or theories are possible, including views from functional or system theories to strategy modules or theories. These possibilities and their implications are discussed in \cite[\S 6.3.2]{maude}.
Several other examples of parameterized strategy modules are available in~\cite{pssm}.

\subsection{Reflecting strategies at the metalevel} \label{sec:metalevel}

	As pointed out in the introduction, Maude is a reflective language. This means that functional and system modules (and, as we shall see below, also strategy modules) can be treated as \emph{data} and can then be transformed and manipulated in very powerful \emph{meta-programming}, reflective ways within Maude. For example, virtually all formal verification tools, which of course must inspect, manipulate, and sometimes transform Maude modules as data, are built this way.

	In fact, all new features added to Maude are routinely reflected at the metalevel to make Maude metaprogramming even more powerful. For strategies, as for any other Maude feature, what this allows us to do is to make Maude's object language \emph{user-extensible}. This is related to our remark in the introduction that Maude's strategy language design, as any other such design, does not try to support \emph{all} conceivable features, but tries to make some judicious design decisions about what features to directly support. In a non-reflective setting, such language design decisions are somewhat dramatic, since if a desirable feature is not supported we may be out of luck.  But this is not so with reflection, since the user can easily extend the given object language with new features implemented in Maude at the metalevel. Furthermore, reflecting strategies at the metalevel is also useful to be able to access the new strategic functionality for meta-programming purposes, and so that interactive applications and specific frameworks built on top of Maude can benefit from those, and potential formal tools can reason about strategies.

	The Maude metalevel is a hierarchy of modules specifying the different Maude entities and operations~\cite[\S 16]{maude}. Terms are metarepresented in the \texttt{META-TERM} module as terms of sort \texttt{Term}, modules are defined in \texttt{META-MODULE} as terms of sort \texttt{Module} along with its statements, views are represented in \texttt{META-VIEW} as terms of sort \texttt{View}, and \texttt{META-LEVEL} represents operations like reduction, rule application, rewriting, etc., as \emph{descent functions}. Thus, to reflect the strategy language, we have specified it in a new module \texttt{META-STRATEGY}, extended the \texttt{META-MODULE} and \texttt{META-VIEW} modules with strategy modules and views, and incorporated the strategy-rewriting operations into \texttt{META-LEVEL}. First, the strategy language constructs have been defined as follows:
\begin{lstlisting}[moredelim={[is][]{\#}{\#}}, escapechar=^]
sorts RuleApplication CallStrategy Strategy StrategyList .
subsorts RuleApplication CallStrategy < Strategy < StrategyList .

ops #fail# #idle# : -> Strategy [ctor] .
op #all# : -> RuleApplication [ctor] .
op _[_]{_} : Qid Substitution StrategyList -> RuleApplication [ctor ...] .
op #top# : RuleApplication -> Strategy [ctor] .
op match_s.t._ : Term Condition -> Strategy [ctor ...] .
op _|_ : Strategy Strategy -> Strategy [ctor assoc comm id: #fail# ...] .
op _;_ : Strategy Strategy -> Strategy [ctor assoc id: #idle# ...] .
op _or-else_ : Strategy Strategy -> Strategy [ctor assoc ...] .
op _+ : Strategy -> Strategy [ctor] .
op _?_:_ : Strategy Strategy Strategy -> Strategy [ctor ...] .
op matchrew_s.t._by_ : Term Condition UsingPairSet -> Strategy [ctor] .
op _[[_]] : Qid TermList -> CallStrategy [ctor prec 21] .
op #one# : Strategy -> Strategy [ctor] .
*** and others (see ^\textit{\cite[\S 16.3]{maude}}^ or the Maude distribution)
\end{lstlisting}
Using this metarepresentation of strategy expressions, strategy declarations and definitions are represented as operators in \texttt{META-MODULE}:
\begin{lstlisting}
sorts StratDecl StratDefinition .
op strat_:_@_[_]. : Qid TypeList Type AttrSet -> StratDecl [ctor ...] .
op sd_:=_[_].     : CallStrategy Strategy AttrSet -> StratDefinition [ctor ...] .
op csd_:=_if_[_]. : CallStrategy Strategy EqCondition AttrSet
                      -> StratDefinition [ctor ...] .
\end{lstlisting}
And its \texttt{Module} sort has also been extended with new symbols for strategy modules and theories with slots to hold these new module items.
\begin{lstlisting}
op smod_is_sorts_._______endsm : Header ImportList SortSet
  SubsortDeclSet OpDeclSet MembAxSet EquationSet RuleSet
  StratDeclSet StratDefSet -> StratModule [ctor ...] .
op sth_is_sorts_._______endsth : Header ImportList SortSet
  SubsortDeclSet OpDeclSet MembAxSet EquationSet RuleSet
  StratDeclSet StratDefSet -> StratTheory [ctor ...] .
\end{lstlisting}
Similarly, the view symbol of \texttt{META-VIEW} has been extended with a new entry for strategy bindings.

	Finally, the functionality of the commands \texttt{srewrite} and \texttt{dsrewrite} is meta-represented in a descent function defined in the \texttt{META-LEVEL} module.
\begin{lstlisting}
op metaSrewrite : Module Term Strategy SrewriteOption Nat
                  ~> ResultPair? [special (..)] .

sort SrewriteOption .
ops breadthFirst depthFirst : -> SrewriteOption [ctor] .
\end{lstlisting}
\texttt{metaSrewrite} receives the meta-representations of a module,\footnote{\texttt{upModule('$\mathit{name}$, false)} can be used to obtain the meta-representation of the module $\mathit{name}$, where \texttt{false} can be replaced by \texttt{true} to obtain a flattened version where importations are resolved. Similarly, \texttt{upTerm($t$)} can be used to obtain the meta-representation of a term $t$.} a term, a strategy, and the search type for the results of rewriting the term according to the strategy in that module. As expected from \cref{sec:commands}, \texttt{breadthFirst} corresponds to \texttt{srewrite} and \texttt{depthFirst} to \texttt{dsrewrite}. Since this may lead to multiple solutions, the last parameter is used to enumerate them in increasing order until the \texttt{failure} constant is returned. For example, \texttt{srewrite 0 b 0 using right *} at the metalevel is:
\begin{maudexec}
Maude> red in META-LEVEL : metaSrewrite(upModule('15PUZZLE, false),
 '__['0.Zero, 'b.Tile, '0.Zero], ('right[none]{empty}) *, breadthFirst, 0) .
rewrites: 2
result ResultPair: {'__['0.Zero,'b.Tile,'0.Zero],'Row}

Maude> red metaSrewrite(upModule('15PUZZLE, false),
 '__['0.Zero, 'b.Tile, '0.Zero], ('right[none]{empty}) *, breadthFirst, 1) .
rewrites: 2
result ResultPair: {'__['0.Zero,'0.Zero,'b.Tile],'Row}

Maude> red metaSrewrite(upModule('15PUZZLE, false),
 '__['0.Zero, 'b.Tile, '0.Zero], ('right[none]{empty}) *, breadthFirst, 2) .
rewrites: 2
result ResultPair: (failure).ResultPair?
\end{maudexec}

	Using its metarepresentation, the strategy language can be extended in different ways.
In~\cite{metatrans-jlamp}, a general schema for extending the language at the metalevel is presented and applied to some constructs of other languages that are directly available in Maude, as discussed in~\cref{sec:comparison}. They are the \emph{congruence operators} from ELAN/Stratego and \emph{generic traversals} from Stratego. These and other extensions are available in the Maude strategy language webpage~\cite{stratweb}.
 
\section{Examples} \label{sec:examples}

	The strategy language has already been applied to several examples related to semantics of programming languages \cite{eden,operational,ccs}, proof systems \cite{membrane,memstratmc-jlamp,completion}, the ambient calculus \cite{ambientCalculus}, neural networks \cite{neuralNetworks}, membrane computing~\cite{membrane,memstratmc-jlamp}, games~\cite{metatrans-jlamp}, a Sudoku solver \cite{sudoku}, etc. In this section we show the features of the strategy language in action with examples from different fields:
\begin{enumerate}
	\item Two simple introductory examples in \cref{sec:exampleintro}. They illustrate several aspects of the language including the \skywd{matchrew} combinator, strategy modules, recursive strategies, and parameterization.
	\item A strategy for the running example of Section 3, the 15-puzzle, that solves the game. This section shows how a rather complex algorithm can be implemented by a strategy in a modular way. The very useful mutually recursive strategies, conditional strategy definitions, and strategy arguments are exemplified here. Moreover, this covers games, one of the most relevant application areas for strategies.
	\item The RIP protocol in \cref{sec:RIP} demonstrates the application of strategies to specify and simulate realistic examples like a communication protocol, which can later be used for verification~\cite{mitesis}. In this example, we also care about the performance of the simulation and make use of the \skywd{one} combinator for that purpose. The controlled delivery of messages in this object-oriented model can also be translated to other specifications of this kind.
	\item Finally, the Knuth-Bendix is a paradigmatic example of the advocated separation of concerns between rules and strategies, because it specified several completion algorithm in separate strategy modules that operate on exactly the same deduction system. Moreover, it is a classical example in field of automated deduction, where strategies are also quite relevant.
\end{enumerate}

\subsection{Two simple introductory examples} \label{sec:exampleintro}

	We start the section with two simple examples proposed in the beginnings of the strategy language~\cite{towardsStrategy}: a blackboard game and a generic backtracking scheme. The following system module defines a blackboard as a multiset of natural numbers along with a rule \texttt{play} that replaces two of them by their arithmetic mean.

\begin{lstlisting}
mod BLACKBOARD is
	protecting NAT .

	sort Blackboard .
	subsort Nat < Blackboard .

	op __ : Blackboard Blackboard -> Blackboard [assoc comm] .

	vars M N : Nat .

	rl [play] : M N => (M + N) quo 2 .
endm
\end{lstlisting}
The goal of the game is to obtain the greatest number after reducing to whole blackboard to a single figure, since the order in which means are calculated affects the result, as we can see by running \texttt{play} exhaustively.
\begin{maudexec}[escapechar=^]
Maude> srew 8 7 4 3 2 1 using play ! .

Solution 1
rewrites: 1407
result NzNat: 6

^\itshape{[...] (the omitted solutions are 5, 4, 3)}^

Solution 5
rewrites: 13077
result NzNat: 2

No more solutions.
rewrites: 24510
\end{maudexec}

	Since choosing the next two numbers at random is not suited to win the game, a player may try some strategies: combining first the two greatest numbers, the two lowest, the greatest and the lowest, etc. These strategies are specified in the following strategy module, using the \texttt{matchrew} combinator and some auxiliary functions to obtain the required information from the term:
\begin{lstlisting}[moredelim={[is][]{\#}{\#}}]
smod BLACKBOARD-STRAT is
	protecting BLACKBOARD .

	vars X Y M N : Nat .
	var  B       : Blackboard .

	strats maxmin maxmax minmin @ Blackboard .

	sd maxmin := (matchrew B s.t. X := max(B) /\ Y := min(B)
	               by B using play[M <- X , N <- Y] ) ! .
	sd maxmax := (matchrew B s.t. X := max(B) 
	                           /\ Y := max(remove(X, B))
	               by B using play[M <- X , N <- Y] ) ! .
	sd minmin := (matchrew B s.t. X := min(B)
	                           /\ Y := min(remove(X, B))
	               by B using play[M <- X , N <- Y] ) ! .

	ops max min : Blackboard -> Nat .
	op remove : Nat Blackboard -> Blackboard .

	eq max(N) = N .
	eq max(N B) = #if# N > max(B) then N else max(B) fi .
	eq min(N) = N .
	eq min(N B) = #if# N < min(B) then N else min(B) fi .
	eq remove(X, X B) = B .
endsm
\end{lstlisting}

After executing them, the player can see that \texttt{minmin} is the better option and \texttt{maxmax} the worst (and then prove it mathematically).
\begin{maudexec}
Maude> srew 8 7 4 3 2 1 using maxmax .

Solution 1
rewrites: 106
result NzNat: 2

No more solutions.
rewrites: 106
\end{maudexec}
\begin{maudexec}
Maude> srew 8 7 4 3 2 1 using minmin .

Solution 1
rewrites: 106
result NzNat: 6

No more solutions.
rewrites: 106
\end{maudexec}
\begin{maudexec}
Maude> srew 8 7 4 3 2 1 using maxmin .

Solution 1
rewrites: 117
result NzNat: 3

No more solutions.
rewrites: 117
\end{maudexec} 

	The second example is a generic backtracking algorithm. The specification of the backtracking problem must adhere to the requirements of the strategy theory \texttt{BT-ELEMS}.
\begin{lstlisting}
sth BT-ELEMS is
	protecting BOOL .
	sort State .
	op isSolution : State -> Bool .
	strat expand @ State .
endsth
\end{lstlisting}
It includes a sort \texttt{State} for the problem states, a predicate \texttt{isSolution} to check whether a state is a solution, and a strategy \texttt{expand} to generate the successors of a given state in the search. With these elements the generic algorithm is defined as a strategy \texttt{solve} in the parameterized strategy module \texttt{BACKTRACKING}.
\begin{lstlisting}
smod BACKTRACKING{X :: BT-ELEMS} is
	var S : X$State .

	strat solve @ X$State .
	sd solve := (match S s.t. isSolution(S)) ? idle
		: (expand ; solve) .
endsm
\end{lstlisting}
The strategy \texttt{solve} recursively applies \texttt{expand} to look for a solution and stops when it finds one.

	The generic algorithm can be instantiated with as many instances as desired, for example, the labyrinth problem~\cite{strategies06}, the Hamiltonian cycle problem~\cite{pssm}, the $m$-coloring problem~\cite{stratweb}, etc. Here we will show the 8-queens problem:
\begin{lstlisting}
mod QUEENS is
	protecting LIST{Nat} .
	protecting SET{Nat} .
	protecting EXT-BOOL .

	op isSolution : List{Nat} -> Bool .

	vars N M Diff : Nat .
	var  L        : List{Nat} .
	var  S        : Set{Nat} .

	eq isSolution(L) = size(L) == 8 .

	crl [next] : L => L N if N,S := 1, 2, 3, 4, 5, 6, 7, 8 .

	op isValid : List{Nat} Nat -> Bool .
	op isValid : List{Nat} Nat Nat -> Bool .

	eq isValid(L, M) = isValid(L, M, 1) .
	eq isValid(nil, M, Diff) = true .
	eq isValid(L N, M, Diff) = N =/= M
		and-then N =/= M + Diff and-then M =/= N + Diff
		and-then isValid(L, M, Diff + 1) .
endm
\end{lstlisting}
The states of the 8-queens problem are lists of natural numbers, where a value $m$ in the position $n$ means that there is a queen in the position $(n, m)$ of the board. Such a list is a solution when its size is eight since all the queens have been placed. States are extended by the \texttt{next} rule, which appends a new queen to the board. However, not all possible appends are valid, since the new queen must not share a row or a diagonal with a previous one; and this is what the \texttt{isValid} predicate checks.\footnote{The \texttt{and-then} operator from the \texttt{EXT-BOOL} is short-circuit logical conjunction, since the default \texttt{and} operator always reduces both arguments.} Then, the strategy \texttt{expand} is defined as follows:
\begin{lstlisting}
smod QUEENS-STRAT is
	protecting QUEENS .

	strat expand @ List{Nat} .

	var L : List{Nat} . var N : Nat .
	sd expand := top(next) ; match L N s.t. isValid(L, N) .
endsm
\end{lstlisting}

	Specifying how the module \texttt{QUEENS-STRAT} is an instance of a \texttt{BT-ELEM} problem is achieved by defining a view. Since \texttt{expand} and \texttt{isSolution} in the theory correspond to their homonym elements in the target module, no explicit mapping is required.
\begin{lstlisting}
view QueensBT from BT-ELEMS to QUEENS-STRAT is
	sort State to List{Nat} .
endv
\end{lstlisting}
Finally, we can instantiate the parameterized \texttt{BACKTRACKING} module and execute \texttt{solve} to find solutions for the problem.
\begin{lstlisting}
smod BT-QUEENS is
	protecting BACKTRACKING{QueensBT} .
endsm
\end{lstlisting}
\begin{maudexec}
Maude> dsrew [1] nil using solve .

Solution 1
rewrites: 22104 in 6ms cpu (8ms real) (3340991 rewrites/second)
result NeList{Nat}: 1 5 8 6 3 7 2 4

Maude> srew [1] nil using solve .

Solution 1
rewrites: 404353 in 123ms cpu (122ms real) (3283151 rewrites/second)
result NeList{Nat}: 1 5 8 6 3 7 2 4
\end{maudexec}
The depth-first search of the \texttt{dsrewrite} command finds the first solution after fewer rewrites and less time than the fair search of \texttt{srewrite}, and using less memory. However, with \texttt{srewrite} additional solutions can be obtained faster, since they are already calculated.
\begin{maudexec}
Maude> continue 1 .

Solution 2
rewrites: 0 in 0ms cpu (0ms real) (~ rewrites/second)
result NeList{Nat}: 1 6 8 3 7 4 2 5
\end{maudexec}

\subsection{The 15-puzzle} \label{sec:15puzzle}

	In this section we come back to the 15-puzzle introduced in \cref{sec:language} and show a strategy to solve it. Remember that the game, which dates from the 1870s, consists of fifteen tiles numbered from 1 to 15 lying on a framed square surface of side length 4. The blank left by the absent sixteenth tile can be used to move the numbers from their positions. Given any arrangement of the puzzle, the goal is to put the tiles as in \cref{fig:15puzzle}, in ascending order from left to right and from the top to the bottom, with the blank at the last position. However, only half of the initial settings can be solved, because there exists a relation between the parity of the permutation and the position of the blank, invariant by the allowed moves~\cite[\S 8]{lucas1}. The puzzle can be generalized to any side length $n$ with $n^2-1$ tiles.

	In order to solve the puzzle, a search can be executed with the \texttt{search} command or with the strategy commands \texttt{srewrite} and \texttt{dsrewrite} using the strategy below. However, only puzzles that are near the solution can be expected to be solved like this, since the search space size is impractical for an unguided search, as it has approximately $2.1 \cdot 10^{13}$ states.

\begin{maudexec}
Maude> search [1] puzzle1 =>* P:Puzzle s.t. P:Puzzle = solved .

Solution 1 (state 571)
states: 572
rewrites: 20713 in 20ms cpu (23ms real) (1035650 rewrites/second)
P:Puzzle --> *** solved

Maude> srew [1] puzzle1 using (left | right | up | down) * ;
                       match P:Puzzle s.t. P:Puzzle = solved .

Solution 1
rewrites: 52906 in 72ms cpu (71ms real) (734805 rewrites/second)
result Puzzle: *** solved
\end{maudexec}
where

\begin{center}
\begin{tabular}{c@{\quad}c}
\begin{lstlisting}
eq puzzle1 =  b   6   2   4  ;
              1   5   3   8  ;
              9  10   7  11  ;
             13  14  15  12  .
\end{lstlisting} &
\begin{lstlisting}
eq solved =   1   2   3   4  ;
              5   6   7   8  ;
              9  10  11  12  ;
             13  14  15   b  .
\end{lstlisting}
\end{tabular}
\end{center}

	In fact, this problem is a classic example used to illustrate search heuristics and algorithms like A$^*$. Using this approach and the strategy-parameterized branch-and-bound implementation described in~\cite{pssm}, we have written an instance of this problem that is available in the example collection of the strategy language~\cite{stratweb}. Although finding (even the length of) the shortest sequence of moves that lead to the solution is NP-complete~\cite{npuzzle}, non-optimal solutions can be found in polynomial time with deterministic algorithms not based on search. Next, we describe a strategy that solves the puzzle in $\mathcal O(n^6)$ moves, where $n$ is the side length, and is inspired by the solution method discussed in \cite{lucas1}. The strategy can be clearly improved in several ways, but we want to keep it as simple as possible.

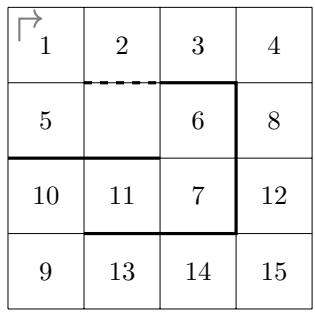
\begin{figure}[h]\centering
\newcommand\innersep{0.025}
\begin{tikzpicture}

	\draw[very thin] (0, 0) grid (4, 4);

	\draw[very thick] (2, 3) -- (3, 3) -- (3, 1) -- (1, 1);
	\draw[very thick] (0, 2) -- (2, 2);
	\draw[very thick, dashed] (1, 3) -- (2, 3);

	\draw[color=gray, thick, ->] (0.15, 3.55) -- (0.15, 3.85) -- (0.45, 3.85);

	\node at (.5, 3.5) {1}; 
	\node at (1.5, 3.5) {2};
	\node at (2.5, 3.5) {3};
	\node at (3.5, 3.5) {4};
	\node at (3.5, 2.5) {8};
	\node at (3.5, 1.5) {12};
	\node at (3.5, .5) {15};
	\node at (2.5, .5) {14};
	\node at (1.5, .5) {13};
	\node at (.5, .5) {9};
	\node at (.5, 1.5) {10};
	\node at (1.5, 1.5) {11};
	\node at (2.5, 1.5) {7};
	\node at (2.5, 2.5) {6};
	\node at (.5, 2.5) {5};
\end{tikzpicture}
\caption{A circuit in the 15-puzzle.} \label{fig:circuit}
\end{figure}

	The key fact is that moving the tiles in a closed circuit through the board does not change the relative position of the numbers within it. \cref{fig:circuit} depicts a possible circuit, where the thick lines represent barriers that cannot be crossed. The act of moving the blank through the circuit, say \emph{rotating}, does not alter the sequence of numbers (1, 2, 3, 4, 8, 12, 15, 14, 13, 9, 10, 11, 7, 6, 5) that can be read clockwise from 1 in the figure. However, all tiles are shifted one position in the direction opposite to the movement of the blank. Using successive rotations, we can place any number above or below the dashed line, and change its relative order in the sequence by slipping it across that line. Like a sorting algorithm, every element could be moved to its correct position in the sequence, except that tiles cannot be swapped but can only jump over pairs.

	These rotations are described in Maude by the strategies \texttt{rotate} and \texttt{reverse}, assuming that the blank is initially below the dashed line, which move the blank in the direction indicated by the arrow and in the reverse one respectively. Remember that numbers are moved in the opposite direction.
\begin{lstlisting}
	strats rotate reverse godown goup goback @ Puzzle .

	sd rotate := left ; up ; right ; right ; right ;
	             down ; down ; down ; left ; left ; left ;
	             up ; right ; right ; up ; left .
\end{lstlisting}
Their definitions are explicit concatenations of rules that follow immediately from~\cref{fig:circuit}. Always without crossing any thick line, the strategies \texttt{goup} and \texttt{godown} move the blank between the positions below and above the dashed line, and \texttt{goback} puts it in the lower-right corner, its desired final position. Their definitions are available in~\cite{stratweb}.

	The solving strategy is called \texttt{solve}. Its first action is moving the blank below the dashed line with \texttt{moveTo}, which uses \texttt{move} from~\cref{sec:modules}. Tile \texttt{1} is then placed above the line by successive rotations, and the \emph{sorting loop} in \texttt{solveLoop} starts. At the end, the strategy will place \texttt{1} above the dashed line again and execute \texttt{goback}, which moves the blank to the lower-right corner, leaving \texttt{1} and the other tiles in their wanted positions.
\begin{lstlisting}
strat  solve                  @ Puzzle .
strat  move         : Nat Nat @ Puzzle .
strats place solveLoop : Tile @ Puzzle .

sd moveTo(X, Y) := matchrew P by P using 
          move(X - blankColumn(P), Y - blankLine(P)) .

sd solve := moveTo(1, 1) ; place(1) ; solveLoop(1) ; place(1) ; goback .

sd place(T) := (match P s.t. T =/= atPos(P, 1, 0) ; rotate) ! .
\end{lstlisting}
The \texttt{solveLoop} strategy iterates on the expected tile sequence, sorting the numbers in the board accordingly. The goal \texttt{sequence} of numbers has been defined explicitly for the circuit. The precondition is that the input parameter \texttt{T} is above the dashed line and preceded by the correct prefix \texttt{LL} in the circuit, and the purpose is to make the next expected tile \texttt{NT} occupy the next position. \texttt{NT} is found in the board by \texttt{findNext} using successive rotations until the expected tile is above the dashed line, as \texttt{place} did. Its second parameter counts the distance, which is finally passed on to \texttt{move} to displace it back as many times as indicated by this number.
\begin{lstlisting}
vars LL LR : Row . vars T NextT Pen Last : Tile .

csd solveLoop(T) := rotate ; findNext(NextT, 0) ; solveLoop(NextT)
 if LL T NextT LR Pen Last := sequence .

csd solveLoop(T) := idle if LL T Pen Last := sequence .

strat findNext : Tile Nat @ Puzzle .
strat move     : Nat      @ Puzzle .

sd findNext(T, N) := match P s.t. T = atPos(P, 1, 0) ? move(N)
		: (rotate ; findNext(T, s(N))) .
\end{lstlisting}
The definitions of \texttt{solveLoop} are conditional to allow obtaining the next tile from the sequence by matching. 

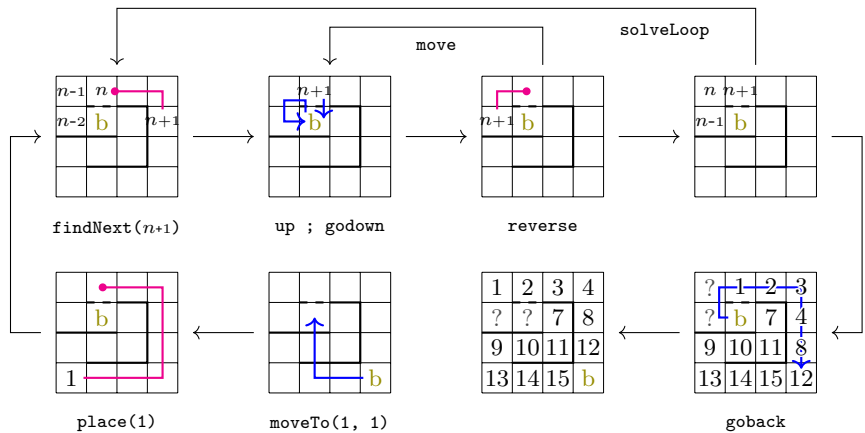
\begin{figure}\centering
\newcommand\ftsquare{
	\draw[very thin] (0, 0) grid (4, 4);

	\draw[thick] (2, 3) -- (3, 3) -- (3, 1) -- (1, 1);
	\draw[thick] (0, 2) -- (2, 2);
	\draw[thick, dashed] (1, 3) -- (2, 3);
}
\newcommand\myscale{.4}
\newcommand\myoffset{7}
\small
\contourlength{.6pt}
\begin{tikzpicture}

	\begin{scope}[scale=\myscale]
		\ftsquare
		\node at (1.5, 2.5) {\color{olive}b};
		\node at (1.5, 3.5) {\scriptsize $n$};
		\node at (0.5, 3.5) {\scriptsize $n$-\tiny$1$};
		\node at (0.5, 2.5) {\scriptsize $n$-\tiny$2$};
		\node at (3.5, 2.5) {\scriptsize $n${\tiny+$1$}};
		\node at (2, -1) {\scriptsize\texttt{findNext($n${\tiny+$1$})}};
		\draw[-{Circle[length=3pt]}, magenta, thick] (3.5, 2.9) -- (3.5, 3.5) -- (1.8, 3.5);
	\end{scope}

	\begin{scope}[scale=\myscale, shift={(\myoffset, 0)}]
		\ftsquare
		\node at (1.5, 2.5) {\color{olive}b};
		\node at (1.5, 3.5) {\scriptsize $n${\tiny+$1$}};
		\node at (2, -1) {\scriptsize\texttt{up ; godown}};
		\draw[->, blue, thick] (1.2, 2.8) -- (1.2, 3.2) -- (0.5, 3.2) -- (0.5, 2.5) -- (1.2, 2.5);
		\draw[->, blue, thick] (1.8, 3.25) -- (1.8, 2.6);
	\end{scope}

	\begin{scope}[scale=\myscale, shift={(2 * \myoffset, 0)}]
		\ftsquare
		\node at (1.5, 2.5) {\color{olive}b};
		\node at (0.5, 2.5) {\scriptsize $n${\tiny+$1$}};
		\node at (2, -1) {\scriptsize\texttt{reverse}};
		\draw[-{Circle[length=3pt]}, magenta, thick] (0.5, 2.9) -- (0.5, 3.5) -- (1.6, 3.5);
	\end{scope}

	\begin{scope}[scale=\myscale, shift={(3 * \myoffset, 0)}]
		\ftsquare
		\node at (1.5, 2.5) {\color{olive}b};
		\node at (1.5, 3.5) {\scriptsize $n${\tiny+$1$}};
		\node at (0.5, 3.5) {\scriptsize $n$};
		\node at (0.5, 2.5) {\scriptsize $n$-\tiny$1$};
\end{scope}

	\begin{scope}[scale=\myscale, shift={(0, -6.5)}]
		\ftsquare
		\node at (1.5, 2.5) {\color{olive}b};
		\node at (0.5, 0.5) {1};
		\draw[-{Circle[length=3pt]}, magenta, thick] (0.9, 0.5) -- (3.5, 0.5) -- (3.5, 3.5) -- (1.4, 3.5);
		\node at (2, -1) {\scriptsize\texttt{place(1)}};
	\end{scope}

	\begin{scope}[scale=\myscale, shift={(\myoffset, -6.5)}]
		\ftsquare
\node at (3.5, 0.5) {\color{olive}b};
		\draw[->, blue, thick] (3.1, 0.5) -- (1.5, 0.5) -- (1.5, 2.5);
		\node at (2, -1) {\scriptsize\texttt{moveTo(1, 1)}};
	\end{scope}

	\begin{scope}[scale=\myscale, shift={(2 * \myoffset, -6.5)}]
		\ftsquare
		\node at (0.5, 3.5) {1};
		\node at (1.5, 3.5) {2};
		\node at (2.5, 3.5) {3};
		\node at (3.5, 3.5) {4};
		\node at (0.5, 2.5) {\color{black!70} ?};
		\node at (1.5, 2.5) {\color{black!70} ?};
		\node at (2.5, 2.5) {7};
		\node at (3.5, 2.5) {8};
		\node at (0.5, 1.5) {9};
		\node at (1.5, 1.5) {10};
		\node at (2.5, 1.5) {11};
		\node at (3.5, 1.5) {12};
		\node at (0.5, 0.5) {13};
		\node at (1.5, 0.5) {14};
		\node at (2.5, 0.5) {15};
		\node at (3.5, 0.5) {\color{olive}b};
	\end{scope}

	\begin{scope}[scale=\myscale, shift={(3 * \myoffset, -6.5)}]
		\ftsquare
		\draw[->, blue, thick] (1.1, 2.5) -- (0.8, 2.5) -- (0.8, 3.5) -- (3.5, 3.5) -- (3.5, 0.8);
\node at (0.5, 3.5) {\color{black!70} ?};
		\node at (1.5, 3.5) {\contour{white}{1}};
		\node at (2.5, 3.5) {\contour{white}{2}};
		\node at (3.5, 3.5) {\contour{white}{3}};
		\node at (0.5, 2.5) {\color{black!70} ?};
		\node at (1.5, 2.5) {\color{olive}b};
		\node at (2.5, 2.5) {7};
		\node at (3.5, 2.5) {\contour{white}{4}};
		\node at (0.5, 1.5) {9};
		\node at (1.5, 1.5) {10};
		\node at (2.5, 1.5) {11};
		\node at (3.5, 1.5) {\contour{white}{8}};
		\node at (0.5, 0.5) {13};
		\node at (1.5, 0.5) {14};
		\node at (2.5, 0.5) {15};
		\node at (3.5, 0.5) {12};
		\node at (2, -1) {\scriptsize\texttt{goback}};
	\end{scope}

\draw[->] (16 * \myscale, 1.75) -- (16 * \myscale, 2.25) -- (9 * \myscale, 2.25) -- (9 * \myscale, 1.75);
\draw[->] (23 * \myscale, 1.75) -- (23 * \myscale, 2.5) -- (2 * \myscale, 2.5) -- (2 * \myscale, 1.75);
\foreach \i in {0, 1, 2} {
		\draw[->] (4.5 * \myscale + 7 * \i * \myscale, .8) -- (6.5 * \myscale + 7 * \i * \myscale, .8);
}
	\foreach \i in {0, 2} {
		\draw[->] (6.5 * \myscale + 7 * \i * \myscale, -1.8) -- (4.5 * \myscale + 7 * \i * \myscale, -1.8);
	}
\draw[->] (-.2, -1.8) -- (-.6, -1.8) -- (-.6, .8) -- (-.2, .8);
	\draw[->] (25 * \myscale + 0.2, .8) -- (25 * \myscale + 0.2 + 0.4, .8) -- (25 * \myscale + 0.2 + 0.4, -1.8) -- (25 * \myscale + 0.2, -1.8);
\node at (12.5 * \myscale, 2) {\scriptsize\texttt{move}};
	\node at (20 * \myscale, 2.2) {\scriptsize\texttt{solveLoop}};
\end{tikzpicture}
\caption{Overview of the \texttt{solve} strategy.} \label{fig:solve}
\end{figure}

	The \texttt{move} strategy is defined recursively. In case the distance is greater than two, the tile is moved down across the dashed line with \texttt{up}, so that it advances two positions against the rotation direction and towards its expected position. Here, an invariant is that the current tile is above the line and the blank is below. To maintain it for the recursive call, \texttt{godown} is called and two reverse rotations make the current tile recover its previous position.
\begin{lstlisting}
sd move(0)       := idle .
sd move(1)       := rotate ; goup ; down ; reverse ; reverse .
sd move(s(s(N))) := up ; godown ; reverse ; reverse ; move(N) .
\end{lstlisting}
This method does not work when the distance is one, but since the tile that occupies the desired position is misplaced, it can be moved clockwise with a similar operation that makes the tile going up across the dashed line instead. This movement is harmless, because it moves the tile to the unordered portion of the sequence. To see that the moved tile does not surpass the first element of the sequence, observe that the last two elements of the list are omitted in \texttt{solveLoop} when the sequence is matched to \texttt{LL T NT LR Pen Last}. Obviously, nothing should be done for the last element, but neither should it be done for the penultimate, which can never be swapped with its neighbor as follows from the parity analysis of the puzzle. If the last two elements are misplaced, the problem is not solvable. A summary of the \texttt{solve} strategy is depicted in~\cref{fig:solve}.

	Now, we can execute the strategy to solve some puzzles:
\begin{maudexec}
Maude> srew (5 1 4 8 ; 2 14 15 3 ; 9 7 6 11 ; 13 10 b 12) using solve .

Solution 1
rewrites: 28868 in 32ms cpu (31ms real) (902937 rewrites/second)
result Puzzle: 1        2       3       4 ;
               5        6       7       8 ;
               9        10      11      12 ;
               13       14      15      b

No more solutions.
rewrites: 28868 in 32ms cpu (31ms real) (902937 rewrites/second)
\end{maudexec}
If the starting puzzle is unsolvable, this is revealed by the position of tiles 5 and 6 being swapped.
\begin{maudexec}
Maude> srew (15	2 1 12 ; 8 5 6 11 ; 4 9	10 7 ; 3 14 13 b) using solve .

Solution 1
rewrites: 40568 in 40ms cpu (41ms real) (1014200 rewrites/second)
result Puzzle: 1        2        3        4 ;
               6        5        7        8 ;
               9        10       11       12 ;
               13       14       15       b

No more solutions.
rewrites: 40568 in 40ms cpu (41ms real) (1014200 rewrites/second)
\end{maudexec}

	When we introduced this example in~\cref{sec:maude}, we admitted that the data representation was not intended to be efficient but just easily readable, to illustrate the strategy language at work. However, we can give the board a more efficient representation as a set of position-to-content pairs:
\begin{lstlisting}
op [_,_,_] : Nat Nat Tile -> Puzzle [ctor] .
op __ : Puzzle Puzzle -> Puzzle [ctor assoc comm id: empty] .
\end{lstlisting}
Like this, the rules \texttt{up} and \texttt{down} can be implemented by simpler unconditional rules:
\begin{lstlisting}
rl [up]   : [X, Y, T] [X, s(Y), b] => [X, Y, b] [X, s(Y), T] .
rl [down] : [X, Y, b] [X, s(Y), T] => [X, Y, T] [X, s(Y), b] .
\end{lstlisting}
And since the strategies above are built solely on the rule names and the function \texttt{atPos}, the strategies do not need to be modified to work with the new representation, in which the problem is solved using fewer rewrites.
\begin{maudexec}
Maude> srew [0,0,5] [0,1,2] ... [2,3,b] ... [3,3,12] using solve .

Solution 1
rewrites: 1630 in 20ms cpu (21ms real) (81500 rewrites/second)
result Puzzle: [0,0,1] [0,1,5] [0,2,9]  [0,3,13]
               [1,0,2] [1,1,6] [1,2,10] [1,3,14]
               [2,0,3] [2,1,7] [2,2,11] [2,3,15]
               [3,0,4] [3,1,8] [3,2,12] [3,3,b]

No more solutions.
rewrites: 1630 in 20ms cpu (21ms real) (81500 rewrites/second)
\end{maudexec}

\subsection{The RIP protocol} \label{sec:RIP}

	This section describes a simple application of the strategy language to the specification of a communication protocol. The \emph{Routing Information Protocol} \cite{rfc2453} is an interior gateway protocol for the interchange of routing information based on distance vectors.

	Internet is composed of different interconnected networks. A message between two hosts in two different networks may travel through a sequence of adjacent networks. The devices that connect them are the \emph{routers}, which decide where to send the data packages so that they arrive to their destinations, preferably by the shortest path. In order to do their job, they need to know the topology of the interconnected networks. This information can be manually established by the system administrator as a table associating a network or arrangement of networks to a physical interface of the router or to the next intermediate router. Such association is called a \emph{routing table}, and this approach, \emph{static routing}. However, routing tables can be constructed dynamically and be aware of network changes, based on data shared by the routers themselves. This is called \emph{dynamic routing}, and RIP is one of its first representative protocols.

	RIPv2 routers periodically exchange their routing tables with their neighbors. When using the IP protocol, each table entry locally describes the next step to reach an aggregation of networks given by an IP address and a mask, here represented in \emph{Classless Inter-Domain Routing} notation.\footnote{IPv4 addresses are 32-bit words. Networks (or aggregations of them) are collections of IP addresses with a common prefix, whose extension is indicated by the mask. CIDR identifies a network by an address followed by the prefix length.} The other relevant fields are the IP address of the next-step router in case the network is not directly reachable, the physical interface that must be used to reach the destination, and the distance, which usually measures the number of router jumps and is called \emph{hop count}.
\begin{lstlisting}
op <_,_,_,_,_> : CIDR IPAddr Interface Nat Nat -> Route [ctor] .
\end{lstlisting}
RIP considers hop-counts above 15 as infinity, so that networks at a greater distance are unreachable. Moreover, the table includes an invalidation timer, to discard entries when no information about them has been received for a significant amount of time.

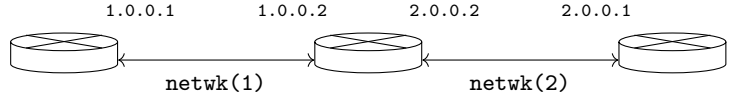
\begin{figure}\centering
\newcommand\router[3]{
	\node (#1) at (#2, #3) [cylinder, shape border rotate=90, draw, minimum height=.5, minimum width=40] {};
	\draw (#2-.5, #3+.16) -- (#2+.5, #3+.33);
	\draw (#2-.5, #3+.33) -- (#2+.5, #3+.16);
}
\begin{tikzpicture}
	\router{A}{-4}{0}
	\router{B}{0}{0}
	\router{C}{4}{0}

	\draw[<->] (A) -- (B);
	\draw[<->] (B) -- (C);

	\node at (-2, -.3) {\small\texttt{netwk(1)}};
	\node at (2, -.3) {\small\texttt{netwk(2)}};
	\node at (3, .66) {\scriptsize \texttt{2.0.0.1}};
	\node at (1, .66) {\scriptsize \texttt{2.0.0.2}};
	\node at (-3, .66) {\scriptsize \texttt{1.0.0.1}};
	\node at (-1, .66) {\scriptsize \texttt{1.0.0.2}};
\end{tikzpicture}
\caption{A simple network topology.} \label{fig:network}
\end{figure}

	In this example, the whole network specification is object-oriented~\cite[\S 8]{maude}.\footnote{Maude traditionally supports object-oriented specifications where \emph{objects} are typically written as \texttt{< $i$ : $c$ | $a_1$ : $v_1$, $\ldots$, $a_n$ : $v_n$ >} with an identifier $i$, a class $c$, and some attributes; they communicate by exchanging \emph{messages} as specified by rewrite rules; and both object and messages live in a soup or multiset called \emph{configuration}. Since Maude 3.3 and previously in Full Maude, syntactic sugar is provided in object-oriented modules (\texttt{omod}s) to define classes along with their attributes, messages, and to simplify the definition of equations and rules. In particular, attributes can be omitted in these statements when their values do not change.}
The objects are the routers, whose attributes are a routing table, and an interface list that enumerates their IP addresses for each directly connected network.
\begin{lstlisting}[escapechar=^]
class Router | table : RouteTable,        *** sets of Route terms
               interfaces : Interfaces .  *** explained below

op none : -> Interfaces [ctor] .
op _|>_ : NetworkId CIDR -> Interfaces [ctor prec 31] .
op __ : Interfaces Interfaces -> Interfaces [ctor assoc comm id: none] .
\end{lstlisting}
Networks are not represented structurally, but by means of unique identifiers of sort \texttt{NetworkId}. Since broadcast and multicast messages used by this protocol are distributed inside the boundaries of a network, this information will be relevant. Every IP message contains a sender and a receiver IP address, and a payload. Multicast and broadcast messages additionally include a network identifier.
\begin{lstlisting}
msg IPMessage : IPAddr IPAddr Payload -> IPMsg .  ** see msg as a synonym of op
msg IPMessage : IPAddr NetworkId IPAddr Payload -> IPMsg .
\end{lstlisting}
In the case of RIP messages, the payload consists of some fields described in the protocol specification, whose main part is the exchanged routing table. The actual definitions of all the previous elements and some operations required to manipulate them are specified in various functional and system modules (\texttt{IP-ADDR}, \texttt{IP-MESSAGES}, \texttt{RIP-ENTRIES}, \texttt{ROUTER}, etc. in the example source).

	The basic operation of the RIP protocol is the interchange of routing tables. This can be triggered as a response to an initial request, or by the detection of changes in the network and the routing tables. However, we will focus on the \emph{gratuitous responses} that routers send approximately every 30 seconds to all their interfaces. These messages are multicast to all the routers, and strategies will be used to deliver them once to all the interested receivers. But first, the router actions are described as rules in a system module \texttt{RIP}. For example, the emission of a (gratuitous) response is specified by the following rule, where the message address 224.0.0.9 is a multicast address recognized by all RIPv2 routers.
\begin{lstlisting}
rl [response] :
   < I : Router | table: RT, interfaces: NId |> A / Mask Ifs >
=> < I : Router | >
   IPMessage(224 . 0 . 0 . 9, NId, A, ripMsg(2, 2, export(RT))) .
\end{lstlisting}
This simple rule will not be admissible in a specification without strategies, because it can be endlessly applied. Its counterpart is the rule in charge of processing the RIP message and updating the routing tables accordingly:
\begin{lstlisting}
crl [readResponse] :
   < I : Router | table: RT, interfaces: NId |> A / Mask Ifs >
   IPMessage(224 . 0 . 0 . 9, NId, O, ripMsg(2, 2, RE))
=> < I : Router | table: import(O, NId, RE, RT) >
   IPMessage(224 . 0 . 0 . 9, NId, O, ripMsg(2, 2, RE))
if O =/= A .
\end{lstlisting}
Notice that the message is not removed from the configuration, since it is a multicast message. The strategy will remove it when it has been received by all the routers, using the \texttt{remove-message} rule. The passage of time is performed by the rule \texttt{update-timer} that increments the internal timers of the routers.

	Over these rules, the protocol operation on a time window of 30 seconds is specified in the \texttt{iteration} strategy. This strategy has two other overloaded versions, in which it delegates more specific tasks.
\begin{lstlisting}
strat iteration                  @ Configuration .
strat iteration : Set{Oid}       @ Configuration .
strat iteration : Oid Interfaces @ Configuration .

var C : Configuration .    var I : Oid .

sd iteration := matchrew C by C using (
                    one(update-tick(allOids(C))) ;
                    iteration(allOids(C))
                ) .

op allOids : Configuration -> Set{Oid} .

eq allOids(none) = empty .
eq allOids(< I : Router | > C) = I, allOids(C) .
eq allOids(M:Msg C) = allOids(C) .
\end{lstlisting}
The parameterless \texttt{iteration} strategy uses the \texttt{allOids} function to collect all the object identifiers of the configuration \texttt{C} captured by \skywd{matchrew}, and passes them to its second overloaded version. Before that, the timers of the routers are updated by the strategy \texttt{update-tick}, which executes the rule \texttt{update-timer} once per router.
\begin{lstlisting}
var RT : RouteTable .   var IS : Set{Oid} .
var D : Configuration . var Ifs : Interfaces .

sd iteration(empty) := handleResponse ! .
sd iteration((I, IS)) := try(
		matchrew C s.t. < I : Router | table: RT, interfaces: Ifs > D := C 
		      by C using one(iteration(I, Ifs)) ;
		handleResponse *
	) ;
	iteration(IS) .
\end{lstlisting}
Using that set of identifiers, the overloaded version \lstinline|iteration(Set{Oid})| iterates over all the objects in the configuration. Each object is matched against a \texttt{Router} object pattern, where the identifier \texttt{I} is the same as the \texttt{I} in the strategy argument. If the matching succeeds, the third overloaded version of \texttt{iteration} is called to send responses from every interface of router, and an undetermined number of them or of those sent by previous routers will be received using the \texttt{handleResponse} strategy. These two strategies will be explained later, but we already see that the emission and reception of responses can be freely interleaved. Otherwise, if the matching does not succeed, the object identifier does not designate a router and the strategy fails inside the \skywd{try}, jumping to the recursive call to \texttt{iteration} for \texttt{IS}. Since the argument is a set, in each strategy call the matches will be multiple, and \texttt{I} will be bound to every element in the set. Hence, objects will be visited nondeterministically in any possible order, and the strategy only enforces that they are visited only once in each rewriting path. This is convenient, because the order in which the routers issue their response is not fixed, and different outcomes could be obtained with different orders. The third overloaded version of \texttt{iteration} sends responses by invoking the \texttt{response} rule for the router \texttt{I} on each interface \texttt{NId}.
\begin{lstlisting}
sd iteration(I, none) := idle .
sd iteration(I, NId |> N Ifs) := response[I <- I, NId <- NId]
                               ; one(iteration(I, Ifs)) .
\end{lstlisting}
Unlike the previous strategies, how \texttt{iteration} visits the interfaces is irrelevant, because the table is not updated in the meanwhile. Thus, every such call is surrounded by a \skywd{one} to avoid unnecessary computation. Alternatively, we could have forced an ordering of the interfaces, either by not making the \texttt{Interfaces} constructor commutative or by sorting them explicitly in the strategy, but we found the usage of \skywd{one} more abstract.

	As we have anticipated, some response messages may remain in the configuration and others may be processed before the next object is addressed due to the \texttt{handleResponse *} expression in the \texttt{iteration} strategy over routers. At the end, where the set of identifiers is \texttt{empty}, all the remaining messages are processed using the normalization operator. The strategy \texttt{handleResponse} takes any multicast message in the soup and delivers it to all its recipients.
\begin{lstlisting}
strat handleResponse            @ Configuration .
strat handleResponse : Set{Oid} @ Configuration .

sd handleResponse := matchrew C s.t.
	IPMessage(A, NId, O, P) D := C
	by C using (one(handleResponse(NId, O, P, allOids(C))) ;
	       remove-message[A <- A, NId <- NId, O <- O, P <- P]) .

sd handleResponse(NId, O, P, empty) := idle .
sd handleResponse(NId, O, P, (I, IS)) :=
	try(readResponse[I <- I, NId <- NId, O <- O, P <- P]) ;
	one(handleResponse(NId, O, P, IS)) .
\end{lstlisting}
It is assumed that all the messages are received simultaneously by all the routers, and so \skywd{one} is used. The definition of this strategy follows the same scheme of the \texttt{iteration} overloaded versions, making each object accept the selected message by fixing the variables in the \texttt{readResponse} rule when applying it.

	As an example, let us apply an iteration on the network of~\cref{fig:network}, where each router has its table filled with the networks to which it is connected, namely 1.0.0.0/8 and 2.0.0.0/8.
\begin{maudexec}
Maude> srew linear using iteration .

Solution 1
rewrites: 136244 in 699ms cpu (699ms real) (194856 rewrites/second)
result Configuration: < r1 : Router | table:
		< 1 . 0 . 0 . 0 / 8, 0 . 0 . 0 . 0, netwk(1), 0, 0 >
		< 2 . 0 . 0 . 0 / 8, 1 . 0 . 0 . 2, netwk(1), 1, 0 >,
		interfaces: netwk(0) |> 1 . 0 . 0 . 1 / 8 >
	< r2 : Router | table:
		< 1 . 0 . 0 . 0 / 8, 0 . 0 . 0 . 0, netwk(1), 0, 0 >
		< 2 . 0 . 0 . 0 / 8, 0 . 0 . 0 . 0, netwk(2), 0, 0 >,
		interfaces: netwk(0) |> 1 . 0 . 0 . 2 / 8
		            netwk(1) |> 2 . 0 . 0 . 2 / 8 >
	< r3 : Router | table:
		< 1 . 0 . 0 . 0 / 8, 2 . 0 . 0 . 2, netwk(2), 1, 0 >
		< 2 . 0 . 0 . 0 / 8, 0 . 0 . 0 . 0, netwk(2), 0, 0 >,
		interfaces: netwk(1) |> 2 . 0 . 0 . 1 / 8 >

No more solutions.
rewrites: 136244 in 699ms cpu (699ms real) (194856 rewrites/second)
\end{maudexec}
Since this network is so small, an iteration is enough to propagate all the routing information, no matter in which order the responses occur. If a similar linear topology is designed with four routers, two iterations are needed to complete the routing information of the routers at both ends, and multiple results are obtained in the first iteration. Other configurations may lead to unwanted situations due to known faults of the protocol like the count to infinity problem. Techniques to mitigate them like the \emph{divided horizon} and \emph{poisoned response}~\cite{rfc2453} have also been specified~\cite{stratweb}.

\subsection{Knuth-Bendix completion} \label{sec:completion}

	In equational logic, given a set $E$ of equations, the \emph{word problem} is deciding whether two given terms $t_1$ and $t_2$ satisfy $t_1 =_E t_2$. Even though the problem is undecidable, incomplete procedures can be constructed for a wide range of instances. A \emph{completion procedure} \cite[\S 7]{allthat} is a method to transform a set of equations into a convergent (confluent and terminating) rewriting system that goes back to Knuth and Bendix~\cite{knuthBendix70}. Whenever it succeeds, provable $E$-equalities can be decided by exhaustively reducing both sides using the calculated rules, and comparing their canonical forms syntactically.

\begin{figure}[t] \centering
	\begin{tabular}{lll}
		Deduce
			& $\begin{array}{c} \langle E, R \rangle \\ \hline \langle E \cup \{ s = t \}, R \rangle \end{array}$
			& if $s \leftarrow_R u \rightarrow_R t$ \\[2.5ex]

		Orient	& $\begin{array}{c} \langle E \cup \{ s = t \}, R \rangle \\ \hline \langle E, R \cup \{ s \to t \} \rangle \end{array}$
			& if $s > t$ \\[2.5ex]

		Delete	& $\begin{array}{c} \langle E \cup \{ s = s \}, R \rangle \\ \hline \langle E, R \rangle \end{array}$
			& \\[2.5ex]
	\end{tabular}
	\hfill
	\begin{tabular}{lll}
		Simplify & $\begin{array}{c} \langle E \cup \{ s = t \}, R \rangle \\ \hline \langle E \cup \{ u = t \}, R \rangle \end{array}$
			 & if $s \to_R^+ u$ \\[2.5ex]

		R-Simplify & $\begin{array}{c} \langle E, R \cup \{ s \to t \} \rangle \\ \hline \langle E, R \cup \{ s \to u \} \rangle \end{array}$
			 & if $t \to_R^+ u$ \\[2.5ex]

		L-Simplify & $\begin{array}{c} \langle E, R \cup \{ s \to t \} \rangle \\ \hline \langle E \cup \{ u = t \}, R \rangle \end{array}$
			 & if $s \to_R^\sqsupset u$
	\end{tabular}

\caption{Bachmair and Dershowitz's inference rules} \label{fig:bachmair}
\end{figure}

\begin{figure}[ht]\centering
	\begin{tikzpicture}
\node[draw] (CP) at (2.5,3) {\tt CRITICAL-PAIRS}; \node[draw] (PA) at (2.5,2) {\tt PARTITION-AUX}; 

\node[draw,dashed] (MO) at (-1.5, 2.5) {\tt MODULE-AND-ORDER};

\node[draw] (C) at (2.5,1) {\tt COMPLETION}; 

\node[draw] (KB) at (-2, -.7) {\tt BASIC-COMPLETION};
		\node[draw] (N) at (1, -.7) {\tt N-COMPLETION};
		\node[draw] (S) at (4, -.7) {\tt S-COMPLETION};
		\node[draw] (ANS) at (7.2, -.7) {\tt ANS-COMPLETION};

		\node at (8.4, 1.8) {\color{black!70}\scriptsize Functional};
		\node at (8.5, 1) {\color{black!70}\scriptsize System};
		\node at (8.5, 0.2) {\color{black!70}\scriptsize Strategy};

		\draw[->] (CP) -- (PA);
		\draw[->] (PA) -- (C);
		\draw[->] (C) -- (KB);
		\draw[->] (C) -- (N);
		\draw[->] (C) -- (S);
		\draw[->] (C) -- (ANS);

		\draw[dotted] (-3.5,1.5) -- (9,1.5);
		\draw[dotted] (-3.5,0.5) -- (9,0.5);
	\end{tikzpicture}
\caption{Module structure of the basic Knuth-Bendix-completion procedures.} \label{fig:completion}
\end{figure}
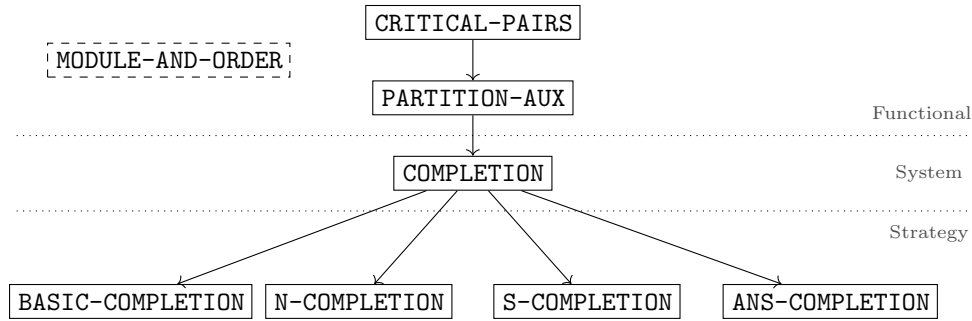

	Many concrete versions of the Knuth-Bendix completion procedure can be expressed as particular ways of applying a set of inference rules (see \cref{fig:bachmair}) proposed by Bachmair and Dershowitz~\cite{bachmair94}, in other words, as specific strategies for that inference system.
Lescanne~\cite{lescanneOrme} described various such procedures as a combination of \emph{transition rules + control}, and implemented them in CAML. Completion procedures have also been implemented in ELAN~\cite{kirchner95}, at the Maude metalevel~\cite{clavel97}, and even using the Maude strategy language~\cite{completion}. However, the separation between rules and control is not clearly enforced in these works, since the rules are adapted to the specific data structure of each method. In this section, we propose an improvement of~\cite{completion} in which a single fixed set of rules is used to express the different specific procedures as stateful strategies. These strategies are specified in separate strategy modules on top of the same system module \texttt{COMPLETION}, as shown in~\cref{fig:completion}.

	The basic state of the Knuth-Bendix-completion procedures is a pair $(E, R)$ of equations and rules. The procedure begins with a set of equations $(E, \emptyset)$, and concludes with a set of rules $(\emptyset, R)$, unless it fails in finite time or it loops forever. Equations are oriented to become rules and critical pairs are calculated during the process. The syntax for equations and rules, the calculation of critical pairs, and the rest of the infrastructure are defined in the functional module \texttt{CRITICAL-PAIRS}. Since each example of equational system to be completed has its own term signature and since an order on the terms is required by the completion procedures, almost every module in~\cref{fig:completion} is parameterized by the \texttt{MODULE} theory or by its extension \texttt{MODULE-AND-ORDER}. Multiple term orders have been proposed in the literature and the algorithms can be instantiated with any of them by means of a view. For facilitating the specification of the order, we provide a parameterized module \texttt{LPO} that derives a \emph{lexicographic path ordering} $>$ on the terms from a strict ordering $\gg$ on the symbols in the signature.
The most basic definitions are those for equations \texttt{Eq}, rules \texttt{Rl}, and sets of these:
\begin{lstlisting}
op _=._ : Term Term -> Eq [comm prec 60] .
op _->_ : Term Term -> Rl [prec 60] .

op __ : EqS EqS -> EqS [assoc comm id: mtEqS prec 70] .
op __ : RlS RlS -> RlS [assoc comm id: mtRlS prec 70] .
\end{lstlisting}

	Bachmair and Dershowitz's inference rules and the state pair on which they operate are defined in the system module \texttt{COMPLETION}:\footnote{Since \texttt{CRITICAL-PAIRS} does not depend on the order, it is parameterized only by \texttt{MODULE}, and the \texttt{ForgetOrder} view from \texttt{MODULE} to \texttt{MODULE-AND-ORDER} allows instantiating it by the parameter \texttt{X}.}
\begin{lstlisting}
mod COMPLETION{X :: MODULE-AND-ORDER} is
	pr CRITICAL-PAIRS{ForgetOrder}{X} .

	sort System .
	op <_,_> : EqS RlS -> System [ctor] .

	var  E     : Eqs .
	vars R QR  : RlS .
	vars s t u : Term .

	*** [...] (the completion inference rules as rewrite rules)
endm
\end{lstlisting}
First, the rule \texttt{Orient} takes an equation $s \approx t$ and orients it as $s \to t$ whenever $s > t$.
\begin{lstlisting}
crl [Orient] : < E s =. t , R >
            => < E, R s -> t > if s > t .
\end{lstlisting}
Since the operator \texttt{=.} is commutative, the equation sides are interchangeable. Trivial equations are removed by \texttt{Delete}, and \texttt{Simplify} reduces any side of an equation by exhaustively rewriting it with the already generated rules. This is implemented, using Maude's metalevel facilities, by the \texttt{reduce} function imported from \texttt{CRITICAL-PAIRS}.
\begin{lstlisting}
rl [Delete] : < E s =. s, R >
           => < E, R > .

crl [Simplify] : < E s =. t, R >
              => < E u =. t, R >
	if u := reduce(s, R) .
\end{lstlisting}
For their part, rules are simplified using the \texttt{R-Simplify} and \texttt{L-Simplify} rules, which simplify the left and righthand side of a rule respectively. Observe that there are two sets in the rule pattern, and the term is only reduced by the rules in \texttt{R}, while the set \texttt{QR} of \emph{quiet rules} is not used. The different completion procedures could select which rules to reduce with by fixing \texttt{R} and \texttt{QR} conveniently.
\begin{lstlisting}
crl [R-Simplify] : < E, R QR s -> t >
                => < E, R QR s -> u >
	if u := reduce(t, R) .

crl [L-Simplify] : < E, R QR s -> t >
                => < E u =. t, R QR >
	if u := reduce>(s -> t, R) .
\end{lstlisting}
Since reductions are decreasing with respect to the term order, the righthand side simplification still satisfies $s > u$. However, the order may not be preserved with \texttt{L-Simplify}, so the rule is removed and reinserted as an equation. Moreover, the lefthand side term $s$ is not reduced exhaustively by all the rules in $R$ as before, instead a single arbitrary rule $l \to r$ in $R$ whose lefthand side $l$ cannot be reduced by $s \to t$ is applied. This is the meaning of the $\to_R^\sqsupset$ arrow in~\cref{fig:bachmair}, and the behavior of the \texttt{reduce>} function.
Finally, \texttt{Deduce} infers new identities by calculating the critical pairs of a rule \texttt{r} with a selected subset of rules \texttt{R}. Here, we have also included a set \texttt{QR} to allow restricting the rules whose critical pairs are calculated.
\begin{lstlisting}
crl [Deduce] : < E, R QR >
            => < E equations(critical-pairs(r, R)), R QR >
if r R' := R QR .
\end{lstlisting}

	Applying the inference rules carelessly does not always lead to a terminating and confluent set of rules, even if one exists. The second entry of the state is always a terminating rewriting system because of the strict order on the terms, but it may not be confluent if some critical pairs have not been calculated. Moreover, the deduction system itself is not terminating, since generating critical pairs leading to known identities may always be possible. Thus, strategies must be used to control the rule execution.

	In the following, we describe a naive completion procedure taken from~\cite[Table 7.7]{terese} and the N-completion procedure of~\cite{lescanneOrme}, both as strategies on Bachmair and Dershowitz's rules. The slightly more complex S-completion and ANS-completion procedures of~\cite{lescanneOrme} are available in the source code of the example~\cite{stratweb}.

\subsubsection{Basic completion}

	The basic completion procedure~\cite[Table 7.7]{terese} follows the script below for a state $(E, R)$:
\begin{enumerate}
	\item Select an equation from $E$. If there is none, stop with success.
	\item Simplify its both sides using the rules in $R$.
	\item If both sides of the equation become identical, remove the equation and go back to step 1. Otherwise, continue to step 4.
	\item Orient the equation and add the resulting rule to $R$. If this is not possible, stop with a failure.
	\item Calculate the critical pairs between the new rule and the whole rule set, and add them as equations to $E$. Go back to step 1.
\end{enumerate}

The procedure can be specified by the following \texttt{compl} strategy, which uses \texttt{deduction} as an auxiliary strategy. It consists of a \skywd{matchrew} combinator, whose content implements the five steps of the script, inside a normalization operator \texttt! that iterates them until they fail.
\begin{lstlisting}
sd compl := (matchrew Sys s.t. < E s =. t, R > := Sys by
               Sys using (try(Simplify[E <- E, s <- s]) ;
                          try(Simplify[E <- E, s <- t]) ;
                          try(Delete[E <- E]) ;
                          (match < E, R > or-else
                             (Orient[E <- E] ;
                              deduction(R))
                          )
                         )
            ) ! .
\end{lstlisting}
First, the \skywd{matchrew} patterns select an equation from the state. If there is none, the matching will fail and the strategy execution will stop due to the normalizing operator. Otherwise, the equation is simplified (step 2) using the \texttt{Simplify} rule. By fixing the variable \texttt{E} in the lefthand side \texttt{< E s =. t, R >} of the rule, the simplification is applied only to the selected equation. Moreover, setting \texttt{s} in the rule to \texttt{s} and \texttt{t} alternatively in the strategy context ensures that both sides are reduced, and since the simplification rule fails when the term is already simplified, these rule applications are surrounded by a \skywd{try} operator.
The simplified version of \texttt{s =. t} may coincide with another equation in \texttt{E}, and so be removed by the idempotence equation of the set. Moreover, the strategy attempts to \texttt{Delete} the equation if it has become a trivial identity (step 3). In either case, the substrategy finishes successfully and another iteration is started because of the normalizing operator. Otherwise, if the equation is already there, the \skywd{or-else} alternative is executed. The rule \texttt{Orient} orients the selected equation unless it is not orientable (step 4). In this latter case, the application fails, and so the iteration loop and the whole procedure stops. As there is still an equation in the state, the failure can be observed in the procedure's output. On success, the \texttt{deduction} strategy is called with the set of rules \texttt{R} from the matching, which contains all rules except the new one.

\begin{lstlisting}
sd deduction(R) := matchrew Sys s.t. < E, R s -> t > := Sys by Sys using
                     Deduce[r <- s -> t, QR <- mtRlS] .
\end{lstlisting}

	In order to generate the critical pairs between the new rule and the whole rule set, the \texttt{deduction} strategy definition applies \texttt{Deduce} with its variable \texttt{r} instantiated to that rule and \texttt{QR} to the empty set. The new rule \texttt{s -> t} is recovered in the \skywd{matchrew} by matching against \texttt{< E, R s -> t >} where \texttt{R} is instantiated to its value from the context, which is the argument of the strategy that contained all the rules except the new one.
Using this procedure, all the critical pairs are generated for every rule, so whenever the strategy terminates with an empty equation set, the resulting set of rules is a convergent rewriting system \cite{huet81}.

	As a simple example to illustrate the procedure, consider a signature with two unary symbols $f$ and $g$, a lexicographic path ordering whose precedence is $f \gg g$, and an equation $f(f(x)) = g(x)$. The algorithm is run by:
\begin{lstlisting}[language={}]
Maude> srew < 'f['f['x:S]] =. 'g['x:S], mtRlS > using compl .

Solution 1
rewrites: 6943 in 6ms cpu (6ms real) (1048000 rewrites/second)
result System: < mtEqS, 'f['f['x:S]] -> 'g['x:S]
                        'f['g['x:S]] -> 'g['f['x:S]] >

No more solutions.
rewrites: 6943 in 6ms cpu (6ms real) (1048000 rewrites/second)
\end{lstlisting}
The algorithm has successfully finished, providing a confluent term rewriting system that solves the word problem for this equational theory.

\subsubsection{N-completion}

	N-completion refines the previous procedure by the simplification of the rules and a more efficient computation of critical pairs. These rules are partitioned into a set $T$ of rules whose critical pairs have not been calculated yet, and its complement $R \setminus T$ (\emph{marked rules} in Huet's terminology~\cite{huet81}). This partition should be part of the procedure state and so, in the original implementation~\cite{completion}, the state pair was extended to a triple and the completion inference rules were changed as a consequence. Here, the information will be held in the strategy parameters:
\begin{lstlisting}
strats N-COMP simplify-eqs @ System .
strats N-COMP success orient deduce simplify-rules : RlS @ System .

sd N-COMP    := N-COMP(mtRlS) .
sd N-COMP(T) := success(T) or-else (match < mtEqS, R >
                 ? deduce(T) : orient(T)) .
\end{lstlisting}
N-completion is implemented by a collection of strategies that take a set of rules as argument, which stands for the set $T$ mentioned before, a subset of rules of the state term. The entry point is the parameterless overloaded version of \texttt{N-COMP} that calls its homonym strategy with $T$ as an empty set. During execution, $T$ is updated and passed on as an argument among the different mutually recursive strategies. According to the \texttt{N-COMP} definition, the procedure tries to execute one of the auxiliary strategies \texttt{success}, \texttt{deduce}, and \texttt{orient}. The first one tests whether the procedure has successfully finished, which is exactly when there are neither pending equations, nor rules whose critical pairs have not yet been calculated.
\begin{lstlisting}
sd success(mtRlS) := match < mtEqS, R > .
\end{lstlisting}
Remember that a call that does not match any definition is a failure, so when $T$ is non-empty this strategy fails. In that case, the procedure continues either with the \texttt{deduce} or with the \texttt{orient} strategy, depending on whether the equation set is empty or not. Hence, equations are greedily oriented and added to $T$ by \texttt{orient}, and only when there are no equations, the critical pairs are calculated by \texttt{deduce}. This strategy nondeterministically takes a rule \texttt{r} from $T$, deduces the identities from its critical pairs with respect to the rest of the rules, and then calls \texttt{simplify-rules}. Since the critical pairs for \texttt{r} have just been calculated, \texttt{r} is no longer included in the strategy argument. The definition is only executed if $T$ is non-empty, but this is an invariant at this point because otherwise \texttt{success} would have succeeded earlier.
\begin{lstlisting}
sd deduce(r T) := Deduce[r <- r, QR <- mtRlS] ;
                  simplify-rules(T) .

sd simplify-rules(T) := matchrew Sys s.t. < E, R > := Sys by Sys using (
				  (L-Simplify[QR <- mtRlS] | R-Simplify[QR <- mtRlS])
				  ? matchrew Sys' s.t. < E', R' > := Sys' by Sys' using
				      simplify-rules(combine(T, R, R'))) :
				  : N-COMP(T)
				) .
\end{lstlisting}
The \texttt{simplify-rules} strategy tries to simplify either the left or righthand side of any rule (inside or outside $T$) using the rest of the rules (for that reason, the quiet rules variable \texttt{QR} is set to the empty set), and calls itself recursively when it succeeds to make the simplification exhaustive. If no more simplifications are possible, a recursive call to \texttt{N-COMP} continues the procedure. \texttt{L-Simplify} and \texttt{R-Simplify} can modify the rule or convert it into an equation, so we should track the changes to update $T$. Using the two nested \skywd{matchrew}, we probe the rule set before and after the simplification, to compare them and find out what changed. The new $T$ is calculated by a function \texttt{combine} defined in \texttt{PARTITION-AUX} by some simple equations.
\[ \mathrm{combine}(T, R, R') = \begin{cases}
		T & \text{ if } r \not\in T \\
		(T \setminus (R \setminus R')) \cup R' \setminus R & \text{ if } r \in T
	\end{cases} \]

	The third auxiliary strategy, \texttt{orient}, is in charge of simplifying and orienting equations. The \skywd{matchrew} is added a condition \texttt{s > t} to know which is the rule that \texttt{Orient} would add, so that it can be included in the set $T$. Incidentally, this condition reduces the number of distinct matches of the subterm operator due to the commutativity of equation symbols.
\begin{lstlisting}
sd orient(T) := simplify-eqs ;
                (match < mtEqS, R >
                  ? N-COMP(T)
                  : matchrew Sys s.t. < s =. t E, R > := Sys
                    /\ s > t by Sys using (
                      Orient[E <- E] ;
                      N-COMP(s -> t T)
                    )
                ) .

sd simplify-eqs := (Delete | Simplify) ! .
\end{lstlisting}

	N-completion is more efficient than the initial procedure, and therefore fewer rewrites are required to compute the same completion of the previous section.
\begin{lstlisting}[language={}]
Maude> srew < 'f['f['x:S]] =. 'g['x:S], mtRlS > using N-COMP .

Solution 1
rewrites: 2279 in 3ms cpu (3ms real) (691654 rewrites/second)
result System: < mtEqS, 'f['f['x:S]] -> 'g['x:S]
                        'f['g['x:S]] -> 'g['f['x:S]] >

No more solutions.
rewrites: 2279 in 3ms cpu (3ms real) (691654 rewrites/second)
\end{lstlisting}

	In order to show a more realistic example and to illustrate the importance of choosing an efficient strategy, we will apply the two procedures to the group axioms $e * x = x$, $I(x) * x = e$ and $(x * y) * z = x * (y * z)$, where $*$ is the binary group operation, $I$ is the inverse, and $e$ the identity element, whose precedence is set to $I \gg * \gg e$ for the lexicographic path ordering. The basic completion algorithm does not terminate in hours for this problem, because of the rapid growth of its search space caused by the inefficient calculation of critical pairs and the lack of rule simplification. On the contrary, N-completion quickly finds a solution using the depth-first search of the \texttt{dsrewrite} command.
\begin{lstlisting}[language={}]
Maude> dsrew [1]
      < '*['e.S ,'x:S] =. 'x:S  '*['I['x:S], 'x:S] =. 'e.S
        '*['*['x:S, 'y:S], 'z:S] =. '*['x:S, '*['y:S, 'z:S]],
        mtRlS > using N-COMP .

Solution 1
rewrites: 222369 in 309ms cpu (311ms real) (718093 rewrites/second)
result System: < mtEqS,
            '*['e.S,'x:S] -> 'x:S
            '*['x3:S,'*['I['x3:S],'z5:S]] -> 'z5:S
            '*['x3:S,'I['x3:S]] -> 'e.S
            '*['z2:S,'e.S] -> 'z2:S
            '*['*['x:S,'y:S],'z:S] -> '*['x:S,'*['y:S,'z:S]]
            '*['I['x:S],'x:S] -> 'e.S
            '*['I['y1:S],'*['y1:S,'z1:S]] -> 'z1:S
            'I['e.S] -> 'e.S
            'I['*['x3:S,'y5:S]] -> '*['I['y5:S],'I['x3:S]]
            'I['I['y1:S]] -> 'y1:S >
\end{lstlisting}
Using the calculated rewriting system (\texttt{result} in the following), we can check whether the term $I(x * (y * z))$ is equal to $(I(z) * I(y)) * I(x)$ by reducing both terms to normal form and comparing the results syntactically.
\begin{lstlisting}[language={}]
Maude> red reduce('I['*['x:S, '*['y:S, 'z:S]]], result) .
rewrites: 21 in 3ms cpu (2ms real) (6300 rewrites/second)
result Term: '*['I['z:S],'*['I['y:S],'I['x:S]]]

Maude> red reduce('*['*['I['z:S], 'I['y:S]], 'I['x:S]], result) .
rewrites: 19 in 0ms cpu (2ms real) (~ rewrites/second)
result Term: '*['I['z:S],'*['I['y:S],'I['x:S]]]
\end{lstlisting}

The strategies for the more complex S-completion and ANS-completion procedures follow the same principles, but these strategies receive more parameters standing for finer rule partitions. All the details can be found in~\cite{stratweb}.

\section{Semantics} \label{sec:semantics}

	An informal description of the semantics of strategy expressions has already been given in the previous sections. Here, we complete and formalize this in two equivalent forms:
\begin{enumerate}
	\item A denotational semantics that describes the results of a strategy execution as a set of terms. It is based on the partial, similar descriptions in~\cite{strategies06,towardsStrategy}.
	\item A small-step operational semantics based on a rewrite theory transformation and expressed in Maude. It is an updated version of the one described in~\cite{rewSemantics}.
\end{enumerate}
The first semantics is more abstract; it emphasizes the results of the strategy, as shown by \texttt{srewrite}, while the second details the evolution of the state as it is rewritten according to the strategy. Intermediate states are also interesting and relevant for some forms of analysis of the controlled systems like model checking. A similar operational semantics has been used to define model checking for such systems in~\cite{smcJournal,smcJournal-btime}.
Another advantage of the rewriting-based semantics is that it is executable and shows that the controlled system can be expressed in rewriting logic itself. This approach has been followed by other works like~\cite{lmcs17}.

\subsection{Set-theoretic semantics}

	The behavior of the strategy language expressions has been described in~\cref{sec:language} by the results they produce for any given initial term. In previous conference papers~\cite{strategies06,rewSemantics}, this description has been partially formalized as a \emph{set-theoretic semantics}, where the denotation of a strategy expression $\alpha$ is a function from terms to sets of terms:
\[ \lBrack \alpha \,\lower1pt\hbox{@}\, \bullet \rBrack : T_\Sigma \to \mathcal P(T_\Sigma) \]
Following this approach, we present here an updated formal semantics of the whole language. To cover all the combinators, some circumstances should be taken into account that complicate the semantic description as follows:
\begin{itemize}
	\item Strategy definitions and \texttt{matchrew} operators can bind variables to values that are accessible within the definition body and the substrategies,  respectively. To pass these values on, variable environments are incorporated as input to the semantic function, and we write $\lBrack \alpha \rBrack(\theta, t)$ instead of $\lBrack \alpha \,\lower1pt\hbox{@}\, t \rBrack$. Variable environments $\theta$ are represented by substitutions $\venv = X \to T_\Sigma$.

	\item With strategy modules, the semantic value of a strategy expression does not only depend on its sole content, but also on the strategy definitions $D$ of the module in which it is evaluated. We describe these definitions as tuples $(sl, \vec p, \delta, C)$, where $\mathit{sl}$ is the strategy name and $\delta$ is the expression to be executed when the input parameters match the lefthand side patterns $\vec p$ and satisfy the condition $C$. We assume that they are numbered from $1$ to $m$, but we will later see that this order is immaterial.

		From the technical point of view, the possibility of defining recursive strategies implies that the semantics cannot be provided by a series of well-founded compositional definitions for each combinator. Hence, we have to resort to more complex tools from domain theory~\cite{abramsky94}. For the presentation of the language semantics, we will assume the existence of a denotation $d_k$ for each definition such that $d_k = \lBrack\delta_k\rBrack$, and express the meaning of a strategy call in these terms. Later, we will justify that  $\Delta = (d_1, \ldots, d_m)$ can be obtained by a fixed point calculation.

	\item Recursive definitions and iterations make nonterminating executions possible. The denotation should indicate whether a computation is terminating or not, and this cannot be expressed by returning only a term set. This fact is valuable even for the compositional definition of the semantics, whose conditional expression is meant to execute its negative branch only when the condition strategy does not provide any result, which must be decided in finite time. Moreover, a nonterminating strategy evaluation can still provide solutions on other rewriting paths, as the \texttt{srewrite} and \texttt{dsrewrite} command do, so any combination of a set of results and a termination status may be possible.

	For those reasons, we extend the output range of the denotation by allowing a symbol $\bot$ representing non-termination to appear in the result sets, now $\mathcal P(T_\Sigma \cup \{\bot\})$. However, to avoid undesired ambiguities, infinite sets will be identified regardless of whether they contain $\bot$ or not, since only nonterminating executions are able to produce infinitely many results. This motivates the following definition for any set $M$:
\[ \mathcal P_\bot(M) \coloneq \mathcal P(M \cup \{ \bot \}) \; / \; {\sim} \]
where
\[A \sim B \iff A = B \vee (A \text{ is not finite and } A \oplus B = \{\bot\}) \]
where $\oplus$ stands for the symmetric difference of sets $A \oplus B = A \cup B \setminus A \cap B$. Thus, $\mathcal P_\bot(T_\Sigma)$ will be the range of the semantic function, and we will also use $\mathcal P_\bot(\venv)$ for some auxiliary operations. In the following, we do not refer to the equivalence classes explicitly but to the sets themselves, taking infinite sets with $\bot$ as representatives of their classes. Hence, we write $\bot \in A$ to express that $A$ contains the symbol $\bot$ or it is infinite.
\end{itemize}

	According to the previous comments, the final form of the denotation for a strategy $\alpha$ is
\[ \lBrack \alpha \rBrack_\Delta : \venv \times T_\Sigma \to \mathcal P_\bot(T_\Sigma) \]
As for a standard denotational definition in the domain theory framework, we have to see the class of denotations $\sfun = \venv \times T_\Sigma \to \mathcal P_\bot(T_\Sigma)$ as a \emph{chain-complete partially ordered set} (ccpo), and prove that for any $\alpha$ the $\sfun^m \to \sfun$ functional that maps $\Delta$ to $\lBrack\alpha\rBrack_\Delta$ is monotonic and continuous to calculate the definitions' semantics using the Kleene fixed-point theorem. Here, we will only highlight the basic ideas, and refer to~\cite{slang} for additional details and proofs. The first step is endowing $\mathcal P_\bot(M)$ with an order to make it a ccpo:\footnote{This order is a particular realization of a flat Plotkin powerdomain~\cite{plotkin76}.}
\[ A \leq B \iff A = B \vee (\bot \in A \wedge A \setminus \{\bot\} \subseteq B) \]
Intuitively, the order expresses how results can be extended by further computation. When approaching $\lBrack\alpha\rBrack(\theta, t)$ by the results $A_n$ of executions with at most $n$ nested recursive calls, $\bot \in A_n$ if some recursive calls have not reached their base cases yet. These results can grow with more solutions as larger depths are allowed, and they can eventually get rid of $\bot$ or keep it forever if the execution does not terminate. On the contrary, sets without $\bot$ are definitive solutions, since all base cases have been reached, and so we call those sets \emph{final}.

\begin{proposition}
	$(\mathcal P_\bot(M), \leq)$ is a chain-complete partially ordered set. Its minimum is the class of $\{\bot\}$, and its maximal elements are the classes of $M$ and the final sets. The union of $\sim$-equivalence classes is well-defined by the union of its representatives, and for any chain $F \subseteq \mathcal P(\mathcal P_\bot(M))$, $\sup F = \bigcup_{A \in F} A$ if $\bot \in A$ for all $A \in F$, and $\sup F = Z$ if there is a $Z \in F$ such that $\bot \not\in Z$. In this case, it is unique.
\end{proposition}

	The denotations $\sfun = \venv \times T_\Sigma \to \mathcal P_\bot(T_\Sigma)$ and $\sfun^m \to \sfun$ are also ccpos by standard results.\footnote{For any ccpos $(D, \leq)$ and $(E, \preceq)$ and any set $N$, $N \to D$ is a ccpo with the order $f \leq g$ iff $f(x) \leq g(x)$ for all $x \in N$, and $E \times D$ is a ccpo with the order $(x, y) \leq (u, v)$ iff $x \leq y \wedge\, u \preceq v$.}
Since various strategy combinators (like concatenation) involve feeding a second strategy with the results of a first one, we will use this operation frequently. In the abstract, we can see it as a function $\mathrm{let} : \mathcal P_\bot(N) \times (N \to \mathcal P_\bot(M)) \to \mathcal P_\bot(M)$ defined as follows for any $A \in \mathcal P_\bot(N)$ and $B : N \to \mathcal P_\bot(M)$:
\[ \mathrm{let}(A, B) \coloneq \{ \bot \mid \bot \in A \} \cup \bigcup_{x \,\in\, A \setminus \{\bot\}} B(x) \]
For readability, instead of $\mathrm{let}(A, B)$ we will use the informal notation $\mathrm{let}\; x \leftarrow A : B(x)$ where $B(x)$ is any set expression depending on $x$.
This functional is monotonic and continuous, and using it we can define a monotonic and continuous composition in $\sfun$. Given two functions $f,g \in \sfun$, we define $g \circ f$ as
\[ (g \circ f)(\theta, t) \coloneq \mathrm{let} \; u \leftarrow f(\varc, t) : g(\varc, u) \]
Then $(\sfun, \circ)$ is a monoid with identity $\mathrm{id}(\varc, t) = \{ t \}$. As usual, we write $f^n$ for the $n$-times composition $f \circ \cdots \circ f$ of $f$ and $f^0$ for the identity.

	Another prerequisite of the semantic infrastructure is matching. For the given rewrite theory $\mathcal R = (\Sigma, E \cup B, R)$, we assume that there is a function $\mathrm{match} : T_\Sigma(X) \times T_\Sigma(X) \to \mathcal P(\venv)$ that provides all the minimal matching substitutions of a pattern $p$ into a term $t$, i.e.\ all $\sigma : X \to T_\Sigma(X)$ that satisfy $\sigma(p) = t$ and $\sigma(y) = y$ for any variable $y \in X$ that does not occur in $p$. For matching anywhere and with extension, the functions $\mathrm{amatch}$ and $\mathrm{xmatch}$ are also considered respectively. Their output is represented by a pair $\venv \times T_\Sigma(X \cup \ominus)$ where the first component is a substitution $\sigma$, and the second is the context $c$ where the match occurs, marked by a distinct variable $\ominus \not\in X$ such that $c[\ominus/\sigma(p)] = t$. However, we often write $c(t)$ for $c[\ominus/t]$.

\subsubsection{Idle and fail}

The semantics of the \skywd{idle} and \skywd{fail} constants follow directly from their descriptions.
\[ 	\ssem\idle(\varc, t) = \{ t \} \qquad
		\ssem\fail(\varc, t) = \emptyset \qquad
	\]

\subsubsection{Rule application}

	\newcommand*\chk{{\mathrm{check}}}

The evaluation of a rule application expression requires finding all matches of each rule with the given label and instantiated with the given initial substitution. In case of rules with rewriting conditions $l_i \,\texttt{=>}\, r_i$, the given strategies must be evaluated in $l_i$, instantiated by the substitutions derived from the previous fragments, and their results matched with $r_i$ to check the conditions and instantiate their variables. We formally describe the application of all rules labeled $rl$ with initial substitution $\rho$ and the mentioned strategies as
\begin{align*}
	\mathrm{ruleApply}&(rl, \rho, \alpha_1 \cdots \alpha_m, \varc, t) = \\
	&\bigcup_{\substack{(rl, l, r, C) \in R \\ \mathrm{nrewf}(C) = m}} \; \bigcup_{(\sigma_0, c) \,\in\, \mathrm{amatch}(\rho(l), t)} \mathrm{let}\; \sigma \leftarrow {\textstyle \chk(C, \sigma_0 \circ \rho, \alpha_1 \cdots \alpha_m, \varc)} : \{ c(\sigma(r)) \}
\end{align*}
where $\mathrm{nrewf}(C)$ is the number of rewriting condition fragments in $C$. Then, the semantics of the application combinator is
	\[ \begin{array}{ll}
		\lBrack\hbox{\lststrat[mathescape,keepspaces]|$rl$[$x_1$ <- $t_1$, $\ldots$, $x_n$ <- $t_n$]{$\alpha_1$, $\ldots$, $\alpha_m$}|}\rBrack_\Delta (\theta, t) = \\[1ex]
		\qquad\mathrm{ruleApply}(rl, \mathrm{id}[x_1 \mapsto \theta(t_1), \ldots, x_n \mapsto \theta(t_n)], \alpha_1 \cdots \alpha_m, \varc, t)
	\end{array} \]
In the definition of $\mathrm{ruleApply}$, the strategies in the expression are passed to the $\chk$ function. Since the strategies should be evaluated in the strategy variable context, the context substitution is also passed to the check function. Its full recursive definition is as follows:
\begin{align*}
	\chk(\hbox{\ttfamily true}, \sigma, \vec\alpha, \varc)
		&= \{ \sigma \} \\
	\chk(\kern.5pt l \;\texttt{=}\; r \wedge C, \sigma, \vec\alpha, \varc)
		&= \chk(C, \sigma, \vec\alpha, \varc) \textbf{ if } \sigma(l) = \sigma(r) \textbf{ else } \emptyset \\
	\chk(\kern.5pt t \;\texttt{:}\; s \wedge C, \sigma, \vec\alpha, \varc)
		&= \chk(C, \sigma, \vec\alpha, \varc) \textbf{ if } \sigma(t) \in T_{\Sigma/E, s}(X) \textbf{ else } \emptyset \\
	\chk(\kern.5pt l \;\texttt{:=}\; r \wedge C, \sigma, \vec\alpha, \varc)
		&= \cup_{\sigma' \in \mathrm{match}(\sigma(l), \sigma(r))} \; \chk(C, \sigma' \circ \sigma, \vec\alpha, \varc) \\
	\chk(\kern.5pt l \;\texttt{=>}\; r \wedge C, \sigma, \alpha \vec\alpha, \varc)
		&= \mathrm{let} \; t \leftarrow  \ssem\alpha(\varc, \sigma(l)) : \cup_{\sigma' \in \mathrm{match}(\sigma(r), t)} \, \chk(C, \sigma' \circ \sigma, \vec\alpha, \varc)
\end{align*}
The range of the $\chk$ function is $\mathcal P_\bot(\venv)$, since the evaluation of rewriting condition fragments may not terminate. However, if the rule does not contain any such fragment, the result is always a plain finite substitution set.

	If the rule application is surrounded by the \lststrat|top| modifier, the matching must only occur at the top, so the $\mathrm{amatch}$ in the $\mathrm{ruleApply}$ definition is replaced by a $\mathrm{xmatch}$. Finally, the \lststrat|all| strategy constant represents a standard rewrite step with the rules in the current module. The rule's rewriting fragments are resolved by an unrestricted search, which is equivalent to using \lststrat|all *| as the controlling strategy of the fragment.

\subsubsection{Tests}

	Tests evaluate to a singleton set with the initial term whenever there is a match of the pattern such that the condition is satisfied. When the pattern does not match the term or no match makes the condition hold, its value is the empty set.
\[ \ssem{\hbox{\lststrat[mathescape]|match $\;P\;$ s.t. $\;C$|}}(\varc, t) = \begin{cases}
		\{ t\} & \exists \sigma \in \mathrm{match}(\varc(P), t) \quad \chk(C, \sigma \circ \varc) \neq \emptyset \\
		\emptyset & \text{otherwise}
	\end{cases} \]
The matching pattern is previously instantiated by the context substitution. Notice that the last two parameters of $\chk$ have been omitted, since $C$ is an equational condition. Other test variants, such as \lststrat|xmatch| and \lststrat|amatch|, use their corresponding matching functions instead.

\subsubsection{Regular expressions}

The semantic value of concatenation is the composition of denotations
\[\ssem{\alpha \seq \beta} = \ssem\beta \circ \ssem\alpha.\]
This means that its results are the collection of the results of $\beta$ for each result of $\alpha$. In case the whole calculation of $\alpha$ does not terminate, the $\bot$ symbol is propagated to the composition. The alternation operator $\alpha \disj \beta$ has a straightforward definition
\[
		\ssem{\alpha \disj \beta}(\varc, t) = \ssem\alpha(\varc, t) \cup \ssem\beta(\varc, t)
	\]
as does the iteration strategy
\[
		\ssem{\alpha\texttt*}(\varc, t) = \bigcup_{n \geq 0} \ssem{\alpha}^n(\varc,t)
	\]
From these definitions and \lststrat[mathescape]|alpha+ $\equiv$ alpha ; alpha*|, it follows that $\ssem{\alpha\texttt+}(\varc, t) = \bigcup_{n \geq 1} \ssem{\alpha}^n(\varc, t)$.

\subsubsection{Conditionals}

	As described in \cref{sec:language}, the condition of the \emph{if-then-else} operator is a strategy itself that is evaluated to decide which branch to take. Any result for the condition $\alpha$ is continued by the positive branch, and discards the execution of the negative one.
\[
		\ssem{\ifthel\alpha\beta\gamma}(\varc, t) = \left\lbrace\begin{array}{ll}
			\ssem\beta \circ \ssem\alpha(\theta, t) 	& \text{ if } \ssem\alpha(\theta, t) \neq \emptyset \\
			\ssem\gamma(\theta, t)				& \text{ if } \ssem\alpha(\theta, t) = \emptyset
		\end{array}\right.
	\]
The negative branch $\gamma$ is only executed if the evaluation of $\alpha$ terminates without obtaining any solution. When $\ssem{\alpha}(\theta, t) = \{\bot\}$, neither $\beta$ nor $\gamma$ are evaluated, and the result of the conditional is $\{\bot\}$ by the first case.

	The semantics of the derived operators can be obtained from the previous definitions. For example,

	 \begin{align*}
		\ssem{\alpha \,\skywd{or-else}\, \beta}(\varc, t) &= \begin{cases}
			\ssem\alpha(\varc, t) & \ssem\alpha(\varc, t) \neq \emptyset \\
			\ssem\beta(\varc, t) & \text{otherwise} \\
		\end{cases} \\
		\ssem{\skywd{test}\texttt{(}\alpha\texttt{)}}(\varc, t) &= \begin{cases}
			\{t\} & \bot \not\in \ssem\alpha(\varc, t) \neq \emptyset \\
			\{\bot\} & \bot \in \ssem\alpha(\varc, t) \\
			\emptyset & \text{otherwise}
		\end{cases} \\
		\ssem{\texttt{not(}\alpha\texttt{)}}(\varc, t) &= \begin{cases}
			\{t\} & \ssem\alpha(\varc, t) = \emptyset \\
			\{\bot\} & \bot \in \ssem\alpha(\varc, t) \\
			\emptyset & \text{otherwise}
\end{cases}
	\end{align*}

\subsubsection{Rewriting of subterms}

	The rewriting of subterms operator is easily defined compositionally as
\newcommand*\stratkwd[1]{\;\texttt{\bfseries\color{darkgray}#1}\;}
\begin{align*}
	\ssem{&\!\stratkwd{matchrew} P \stratkwd{s.t.} C \stratkwd{by} x_1 \stratkwd{using} \alpha_1 \texttt, ...\texttt,\, x_n \stratkwd{using} \alpha_n \,}(\varc, t) \\
		&= \kern-1.5em \bigcup_{\sigma \in \mathrm{mcheck}(t, P, C, \varc)} \kern-1.5em \mathrm{let} \; t_1 \leftarrow \ssem{\alpha_1}(\sigma, \sigma(x_1)), \ldots, t_n \leftarrow \ssem{\alpha_n}(\sigma, \sigma(x_n)) : \sigma[x_1 \mapsto t_1, \ldots, x_n \mapsto t_n](P)
\end{align*}
where $\mathrm{mcheck}(t, P, C, \varc) = \bigcup \{ \chk(C, \sigma_m \circ \varc) : \sigma_m \in \mathrm{match}(\varc(P), t) \}$. The matched subterms $\sigma(x_k)$ are rewritten using the strategy $\alpha_k$ in the variable context $\sigma \circ \varc$, and their results replace them in the subject term, by reinstantiating the pattern with the modified matching substitution. Each combination of subterm results generates a potentially different solution, and if any of the subterm computations contains $\bot$, so does the matchrew.

	The variations of the matchrew combinator, \lststrat|amatchrew| and \lststrat|xmatchrew|, have similar definitions but replacing $\mathrm{match}$ by the appropriate matching function, and rebuilding the matching context.

\subsubsection{Pruning of alternative solutions}

	These semantics do not take the \lststrat|one| operator into account, because that would add another layer of non-determinism and harden the understanding of this exposition. Although the rewriting process controlled by strategies can be nondeterministic, the set of results shown by the \texttt{srewrite} command and described in this section are ideally deterministic in the absence of \texttt{one}. However, \lststrat|one(alpha)| nondeterministically chooses one of results that $\alpha$ produces, passing from the set $\ssem\alpha(\varc, t)$ to any singleton set $\{u\}$ with $u$ in the previous set. In that case, the denotations should produce elements of $\mathcal P(\mathcal P_\bot(T_\Sigma))$.

\subsubsection{Strategy calls} \label{sec:strategycalls}

	Strategy calls are resolved using the definition context $\Delta = (d_1, \ldots, d_m)$,
\begin{align*}
		\ssem{sl&\texttt{(}t_1, \ldots, t_n\texttt{)}}(\varc, t) \\
			&= \bigcup_{(sl, p_1 \cdots p_n, C, \delta_k) \,\in\, D} \; \bigcup_{\sigma \,\in\, \mathrm{mmatch}(\varc(t_1) \cdots \varc(t_n), p_1 \cdots p_n, C)} d_k(\sigma, t)
	\end{align*}
where $D$ is the set of strategy definitions in the module, and $\mathrm{mmatch}(t_1 \cdots t_n, p_1 \cdots p_n, C)$ is defined as the union of all $\chk(C, \sigma)$ for all minimal substitutions satisfying $\sigma(p_k) = t_k$ for all $k$. In other words, all strategy definitions matching the input arguments and satisfying the condition are executed, and their results are gathered in the final one. If no definition can be activated, the result is thus the empty set.

\subsubsection{Correctness and strategy definition calculation}

	The exhaustive and well-founded definition of the semantics on the structure of strategy expressions implies its correctness for any environment $\Delta = (d_1, \ldots, d_m)$. However, it still remains pending how to formally construct such an environment where $d_k = \ssem{\delta_k}$. As anticipated in the introduction, the solution passes through Klenee's fixed point theorem and the monotonicity and continuity of the denotations.

\begin{theorem}
	For any strategy expression $\alpha$, $\lBrack\alpha\rBrack_{(d_1, \ldots, d_m)}$ is monotone and continuous in $d_1, \ldots, d_m$, and so is the operator $F : \sfun^m \to \sfun^m$
\[
		F(d_1, \ldots, d_m) \coloneq \left(\lBrack\delta_1\rBrack_{(d_1, \ldots, d_m)}, \ldots, \lBrack\delta_m\rBrack_{(d_1, \ldots, d_m)}\right)
	\]
Hence, $F$ has a least fixed point $\mathsf{FIX}\, F \in \sfun^m$ which can be calculated as
\[
		\mathsf{FIX} \, F = \sup \,\{ F^n (\{\bot\}, \ldots, \{\bot\}) : n \in \N \}
	\]
\end{theorem}

	Then, the denotation for the definition $k$ is formally defined as \[ d_k \coloneq \left(\mathsf{FIX} \, F\right)_k, \] and satisfies $d_k = F_k\left(\mathsf{FIX} F\right) = \ssem{\delta_k}$.

	If Maude were running on an idealized unbounded-memory machine, the strategy search command \texttt{srewrite in SM : $t$ using $\alpha$} would eventually return any solution in $\ssem\alpha(\mathrm{id}, t)$, and it would finish if and only if $\bot \not \in \ssem\alpha(\mathrm{id}, t)$. In the same situation, \texttt{dsrewrite} will also return the full set of solutions, but if $\bot$ is in the denotation, some solutions may be missed.

\subsection{Rewriting semantics} \label{sec:rewsem}

	This section presents a rewriting-based semantics that transforms a pair $(M, S\!M)$, i.e. a system module $M$ along with a strategy module that defines strategies for $M$, into a rewrite theory $\mathcal S(M, S\!M)$, where strategy expressions can be written and applied to terms. The transformed module implements the syntax of the strategy language and the infrastructure and rules to apply them. For this purpose, some function definitions and rules should be added to the transformed module for each strategy construct. First of all, the signature of the transformed module should include some auxiliary infrastructure for substitutions and matching. Declarations with a type annotation like $S$ are generated for each sort in $M$.
\begin{lstlisting}[mathescape, keepspaces]
sort Substitution .

op _<-_ : Var$S$ $S$ -> Substitution [ctor] .
op none : -> Substitution [ctor] .
op _,_ : Substitution Substitution -> Substitution [ctor assoc id: none] .

op _$\cdot$_ : $S$ Substitution -> $S$ .
\end{lstlisting}
A substitution is defined as a list of variable-to-term bindings, and an infix dot operator represents the application of a substitution to a term.
\begin{lstlisting}[mathescape, keepspaces]
sorts Match MatchSet .
subsort Match < MatchSet .
op <_,_> : Substitution $S$ -> Match [ctor] .
op none : -> MatchSet [ctor] .
op __ : MatchSet MatchSet -> MatchSet [ctor assoc comm id: none] .

op [] : -> $S$ [ctor] .

op getMatch : $S$ $S$ EqCondition -> MatchSet .
op getAmatch : $S$ $S'$ EqCondition -> MatchSet .
op getXmatch : $S$ $S$ EqCondition -> MatchSet .
\end{lstlisting}
Each of the last three operators returns all matches of its second argument into the first, respectively on top, anywhere, or on top with extension. A match is described by a pair containing a substitution and a context. The context \emph{hole} is indicated by means of an overloaded constant \lstinline|[]|.
\begin{lstlisting}[mathescape, keepspaces]
sorts Condition EqCondition .
subsort EqCondition < Condition .

op trueC : -> EqCondition  [ctor] .
op _=_ : $S$ $S$ -> EqCondition [ctor] .
op _:=_ : $S$ $S$ -> EqCondition [ctor] .
op _: $S$ : $S$ -> EqCondition [ctor] .
op _=>_ : $S$ $S$ -> Condition [ctor] .

op _/\_ : Condition Condition -> Condition [ctor assoc id: trueC] .
op _/\_ : EqCondition EqCondition -> EqCondition [ditto] .

op _$\cdot$_ : Condition Substitution -> Condition .
\end{lstlisting}
Equational and rule conditions are defined with the usual syntax, and a substitution can also be recursively applied to them.

	Second, strategy language constructs are expressed as Maude operators. Its signature is similar to the meta-representation of strategies in \cref{sec:metalevel}, but applied at the object level. Here we only include some of them as examples.
\begin{lstlisting}[mathescape, keepspaces]
sorts RuleApp Strat StratCall .
subsorts RuleApp StratCall < Strat < StratList .

op _[_]{_} : Label Substitution StratList -> RuleApp [ctor] .
op match_s.t._ : $S$ EqCondition -> Strat [ctor] .
op _;_ : Strat Strat -> Strat [ctor] .
*** and more

op _$\cdot$_ : Strat Substitution -> Strat .
\end{lstlisting}
Substitutions can be applied to strategy expressions too. The equational definition is straightforward, except for the \skywd{matchrew} case. Pattern variables that designate subterms to be rewritten cannot be replaced syntactically, because the reference would be lost. However, the conflicting substitution assignment can be translated into an equality condition fragment to be added to the strategy expression.
\begin{lstlisting}[mathescape, keepspaces, keywordstyle={[2]}]
ceq matchrew(P:$S$, C, VSL) \cdot Sb =
        matchrew(P:$S$ \cdot SSb, SCond /\ C \cdot Sb, VSL \cdot Sb)
 if { SSb ; SCond } := splitSubs(Sb, VSL) .

eq splitSubs(X:$S$ <- T:$S$ ; Sb, X:$S$ using E) = { Sb ; X:$S$ = T:$S$ } .
eq splitSubs(Sb, X:$S$ using E) = { Sb ; nil } [owise] .
ceq splitSubs(X:$S$ <- T:$S$ ; Sb, (X:$S$ using E, VSL)) =
	{ Sb' ; C /\ X:$S$ = T:$S$) if { Sb' ; C } := splitSubs(Sb, VSL) .
eq splitSubs(Sb, (X:$S$ using E, VSL)) = splitSubs(Sb, VSL) [owise} .
\end{lstlisting}

	Third, the strategy execution infrastructure is based on a series of tasks and continuations.
\begin{lstlisting}[mathescape, keepspaces]
sorts Task Tasks Cont .
subsort Task < Tasks .
op none : -> Tasks [ctor] .
op __ : Tasks Tasks -> Tasks [ctor assoc comm id: none] .
eq T:Task T:Task = T:Task .

op <_@_> : Strat $S$ -> Task [ctor] .
op sol :  $S$ -> Task [ctor] .
op <_;_> : Tasks Cont -> Task [ctor] .

op chkrw : Condition StratList $S$ $S$ -> Cont [ctor] .
op seq : Strat -> Cont [ctor] .
op ifc : Strat Strat $S$ -> Cont [ctor] .
op mrew : $S$ $S'$ Substitution VarStratList -> Cont [ctor] .
op onec : -> Cont [ctor] .
\end{lstlisting}
The application of a strategy $\alpha$ to a term $t$ is represented by a task \lststrat[mathescape, keepspaces]|< alpha @ $t$ >|, and solutions are captured in \lstinline[mathescape]|sol($t$)| tasks. Tasks can be rewritten and fork new tasks, which represent different search states. They are all gathered in an associative and commutative soup of sort \texttt{Tasks}. Nested searches are represented by the \lstinline|<_;_>| constructor, which additionally contains a \emph{continuation} that the results from the inner search must execute to be a solution for the outer execution level. Continuations are specified as terms of sort \texttt{Cont}.

\subsubsection{Idle and fail}

	The \skywd{idle} and \skywd{fail} meaning is given by the following rules that convert the \texttt{idle} task to a solution, and remove the \texttt{fail} task.
\begin{lstlisting}[mathescape, keepspaces, keywordstyle={[2]}]
rl < idle @ T:$S$ > => sol(T:$S$) .
rl < fail @ T:$S$ > => none .
\end{lstlisting}

\subsubsection{Rule application}

	For each unconditional rule or each conditional rule without rewriting fragments $l \;\texttt{=>}\; r$ with label $\mathit{label}$ in $M$, a rule as below is appended to the transformed module:
\begin{lstlisting}[mathescape, keepspaces]
crl < $\mathit{label}$[Sb]{empty} @ T:$S$ > => gen-sols(MAT, $r$ \cdot Sb)
if MAT := getAmatch($l$ \cdot Sb, T:$S$, C) .

eq gen-sols(none, T:$S'$) = none .
eq gen-sols(< Sb, Cx:$S$ > MAT, T:$S'$) =
	sol(replace(Cx:$S$, T:$S'$ \cdot Sb)) gen-sols(MAT, T:$S'$) .
\end{lstlisting}
The possibly-empty substitution \texttt{Sb} from the application expression is applied to both sides of the rule. Then the partially instantiated lefthand side is matched against the subject term, and the resulting matches are passed to the \texttt{gen-sols} function. This function traverses the set generating a solution task for each match, by instantiating the righthand side of the rule with the matching substitution, and building up the context with \texttt{replace}.

	The treatment of rewriting conditions is much more involved, because they must be rewritten according to the given strategies. To handle this situation, we make use of a continuation. Consider a rule
\begin{lstlisting}[mathescape, keepspaces]
crl [$label$] : $l$ => $r$ if $C_0$ /\ $u_1$ => $v_1$ /\ $C_1$ /\ $\ldots$ /\ $C_{n-1}$ /\ $u_n$ => $v_n$ /\ $C_n$ .
\end{lstlisting}
where $C_k$ are equational conditions, which may be empty. Let $RC$ be the condition fragments from $C_1$ to $C_n$. For each such rule, we generate
\begin{lstlisting}[mathescape, keepspaces]
var C  : EqCondition .
var RC : Condition .

crl < $label$[Sb]{E1, $\ldots$, En} @ T:$S$ >
   => gen-rw-tks(MAT, $u_1$ \cdot Sb, ($u_1$ => $v_1$ /\ $RC$) \cdot Sb,
                 E1 $\ldots$ En, $r$ \cdot Sb)
   if MAT := getAmatch($l$ \cdot Sb, T:$S$, $C_0$) .

eq gen-rw-tks(none, U:$S'$, RC, EL, Rhs:$S''$) = none .
eq gen-rw-tks(< Sb, Cx:$S$ > MAT, T:$S'$, RC, (E, EL), Rhs:$S''$) =
	< < E @ T:$S'$ \cdot Sb > ; chkrw(RC \cdot Sb, (E, EL), Rhs:$S''$ \cdot Sb, Cx:$S$) >
	gen-rw-tks(MAT, T:$S'$, RC, (E, EL), Rhs:$S''$) .
\end{lstlisting}
The function \texttt{gen-rw-tks} traverses the set of matches like \texttt{gen-sols}, but a continuation task is generated for each match. Its nested computation applies the first given strategy to the lefthand side of the first rewriting fragment instantiated by the matching substitution. Its \texttt{chkrw} continuation stores the condition, the pending controlling strategies, the rule righthand side and the subterm context. This allows checking the condition recursively and stepwise. When a solution is obtained in the nested computation, it must be matched against the righthand side of the fragment and all the matches must be continued as potentially different condition solutions.
\begin{lstlisting}[mathescape, keepspaces]
crl < sol(R:$S$) TS ; 
      chkrw(U:$S$ => V:$S$ /\ C /\ U':$S'$ => V':$S'$ /\ RC,
            (E, E', EL), Rhs:$S''$, Cx:$S''$) >
 => < TS ; chkrw(U:$S$ => V:$S$ /\ C /\ U':$S'$ => V':$S'$ /\ RC,
                 (E, E', EL), Rhs:$S''$, Cx:$S'''$) >
      gen-rw-tks2(MAT, U':$S'$, (U':$S'$ => V':$S'$ /\ RC),
                  (E', EL), Rhs:$S''$, Cx:$S'''$)
   if MAT := getMatch(V:$S$, R:$S$, C) .

eq gen-rw-tks2(none, T:$S'$, RC, EL, Rhs:$S''$, Cx:$S'''$) = none .
eq gen-rw-tks2(< Sb, Cx:$S$ > MAT, T:$S'$, RC, (E, EL), Rhs:$S''$,
               Cx:$S'''$) = < < E @ T:$S'$ \cdot Sb > ;
       chkrw(RC \cdot Sb, (E, EL), Rhs:$S''$ \cdot Sb, Cx:$S'''$) >
     gen-rw-tks2(MAT, T:$S'$, RC, (E, EL), Rhs:$S''$, Cx:$S'''$) .
\end{lstlisting}
Here, \texttt{gen-rw-tks2} walks over the matches for the righthand side of the previous condition fragment, and generates continuation tasks that evaluate the next condition fragment as already done for the initial fragment. Clearly, the base case of this process is reached when no rewriting fragment remains.
\begin{lstlisting}[mathescape, keepspaces]
crl < sol(R:$S$) TS ; chkrw(U:$S$ => V:$S$ /\ C, E, Rhs:$S'$, Cx:$S''$) >
 => < TS ; chkrw(U:$S$ => V:$S$ /\ C, E, Rhs:$S'$, Cx:$S''$) >
    gen-sols2(MAT, Rhs:$S'$, Cx:$S''$)
 if MAT := getMatch(V:$S$, R:$S$, C) .

eq gen-sols2(none, Rhs:$S$, Cx:$S'$) = none .
eq gen-sols2(< Sb, Cx':$S''$ > MAT, Rhs:$S$, Cx:$S'$)
 = sol(replace(Cx:$S'$, Rhs:$S$ \cdot Sb))
   gen-sols2(MAT, Rhs:$S$, Cx:$S'$) .
\end{lstlisting}
The function \texttt{gen-sols2} finally composes the solutions of the rule application by rebuilding the term using the successively instantiated righthand side of the rule. In the case that any of the nested strategy evaluations fails, the whole rule application fails.
\begin{lstlisting}[mathescape, keepspaces]
rl < none ; chkrw(RC, EL, Rhs:$S'$, Cx:$S$) > => none .
\end{lstlisting}

\subsubsection{Tests}

	As described before, tests behave like \texttt{idle} if there is a match satisfying the condition, and like a \texttt{fail} otherwise.
\begin{lstlisting}[mathescape, keepspaces, keywordstyle={[2]}]
crl < match P:$S$ s.t. C @ T:$S$ > => sol(T:$S$)
 if < Sb, Cx:$S$ > MAT := getMatch(P, T:$S$, C) .

crl < match P:$S$ s.t. C @ T:$S$ > => none
 if getMatch(P:$S$, T:$S$, C) = none .
\end{lstlisting}
The \lststrat|amatch| and \lststrat|xmatch| variants are defined by similar pairs of rules. The only difference is the search function, \texttt{getAmatch} and \texttt{getXmatch} respectively, which can be implemented in Maude by means of the family of \texttt{metaMatch} functions.

\subsubsection{Regular expressions}

Regular expressions can be handled by a series of simple rules:
\begin{lstlisting}[mathescape, keepspaces]
rl < E | E' @ T:$S$ > => < E @ T:$S$ > < E' @ T:$S$ > .
rl < E ; E' @ T:$S$ > => < < E @ T:$S$ > ; seq(E') > .
rl < sol(R:$S$) TS ; seq(E') > => < E' @ R:$S$ >
                                < TS ; seq(E') > .
rl < none ; seq(E') > => none .
rl < E * @ T:$S$ > => sol(T:$S$) < E ; (E *) @ T:$S$ > .
eq E + = E ; E *  .
\end{lstlisting}
The rule for alternation splits the task into two subtasks, where each of them continues with one of the alternatives. The rule for concatenation creates a nested task to evaluate the first of the concatenated strategies and leaves the second strategy pending using the \texttt{seq} continuation. Each solution found in the nested search is then continued using the strategy in the continuation. When the subsearch runs out of tasks, the task is discarded. The iteration rule, following its recursive definition, produces both a solution for the empty iteration, and a task that evaluates the iteration body concatenated with the iteration itself. The non-empty iteration is equationally reduced to this equivalent expression.

\subsubsection{Conditionals}

The semantics of conditionals is also expressed by a continuation and a subsearch for the strategy condition. The \texttt{ifc} continuation maintains the strategies for both branches of the conditional and the initial term, which will be used if the negative branch has to be evaluated.
\begin{lstlisting}[mathescape, keepspaces]
rl < E ? E' : E'' @ T:$S$ > => < < E @ T:$S$ > ;
                               ifc(E', E'', T:$S$) > .
rl < sol(R:$S$) TS ; ifc(E', E'', T:$S$) > => < E' @ R:$S$ >
                                          < TS ; seq(E') > .
rl < none ; ifc(E', E'', T:$S$) > => < E'' @ T:$S$ > .
\end{lstlisting}
When a solution is found for the condition, the result is given a task to be continued by the positive branch strategy. Moreover, the conditional \texttt{ifc} continuation is transformed in a \texttt{seq} continuation, since the execution of the negative branch is already discarded. On the other hand, if the tasks in the subcomputation get exhausted, the negative branch is evaluated in the initial term by means of a new task.

	The semantics of the derived operators is implicitly given by equationally translating them into their equivalent expressions:
\begin{lstlisting}[keywordstyle={[2]}]
eq E or-else E' = E ? idle : E' .
eq not(E) = E ? fail : idle .
eq try(E) = E ? idle : idle .
eq test(E) = not(not(E)) .
\end{lstlisting}

\subsubsection{Rewriting of subterms}

	The rewriting of subterms operator requires rewriting each subterm found by the given strategy. Like for rewriting conditions, this is handled using a continuation \texttt{mrew(P, Sb, Cx, X, VSL)} that holds the main pattern \texttt{P}, the substitution \texttt{Sb} and context \texttt{Cx} of its occurrence in the subject term, the variable whose subterm is currently being rewritten, and the list of pending $\overline t$ \skywd{using} $\overline \alpha$ pairs.
\begin{lstlisting}[mathescape, keepspaces, keywordstyle={[2]}]
crl < amatchrew(P:$S$, C, VSL) @ T:$S_0$ > => gen-mrew(MAT, P:$S$, VSL)
 if MAT := getAmatch(P:$S$, T:$S_0$, C) .

eq gen-mrew(none, P:$S$, VSL) = none .
ceq gen-mrew(< Sb, Cx:$S_0$ > MAT, P:$S$, VSL) =
    < < E \cdot Sb @ X:$S'$ \cdot Sb > ; mrew(P:$S$, Cx:$S_0$, Sb, VSL) >
    gen-mrew(MAT, P:$S$, VSL)
 if X:$S'$ using E := firstPair(VSL) .
\end{lstlisting}
For each match of the main pattern in the subject term, a continuation task is created. It starts to evaluate the first strategy in the matched subterm, which is recovered by \lstinline[mathescape]|X:$S'$ \cdot Sb|. The substitution \texttt{Sb} is also applied to the strategy, since it is allowed to contain free occurrences of the pattern and condition variables.

When the evaluation of a subterm gives a solution, the \texttt{mrew} task is split into two subtasks: the first one keeps looking for other solutions for the same subterm, and another one continues with the evaluation of the next subterm. The creation of the last task is similar to the initial case, but the information is instead obtained from the continuation. The result of the subterm rewriting is substituted in the copy of the main pattern carried by the continuation. This way, when all the subterms are processed the copy of the pattern will have the initial subterms replaced by some results, so that the rest of the variables can be instantiated with the initial substitution, and the initial term is rebuilt with the new subterms by means of the context stored in the continuation.
\begin{lstlisting}[mathescape, keepspaces, keywordstyle={[2]}]
crl < sol(T:$S'$) TS ; mrew(P:$S$, Cx:$S_0$, Sb, (X:$S'$ using E', VSL)) > =>
   < TS ; mrew(P:$S$, Cx:$S_0$, Sb, (X:$S'$ using E', VSL)) >
   < < E \cdot Sb @ Y:$S''$ \cdot Sb > ; mrew(P:$S$ \cdot (X:$S'$ <- T:$S'$), Cx:$S_0$, Sb, VSL) > 
 if Y:$S''$ using E := firstPair(VSL) .

rl < sol(T:$S'$) TS ; mrew(P:$S$, Cx:$S_0$, Sb, X:$S'$ using E) > =>
   < TS ; mrew(P:$S$, Cx:$S_0$, Sb, X:$S'$ using E) >
   sol(replace(Cx:$S_0$, P \cdot (X:$S'$ <- T:$S'$)) \cdot Sb)  .

rl < none ; mrew(P:$S$, Cx:$S_0$, Sb, VSL) > => none .
\end{lstlisting}
When the subterm search tasks are exhausted, the whole \lststrat|amatchrew| execution is discarded. Identical rules are used for the other variants, \skywd{matchrew} and \skywd{xmatchrew}, except for the first rule, where \texttt{genAmatch} should be replaced by the appropriate function.

\subsubsection{Pruning of solutions}

	The semantics of the \texttt{one} combinator can be expressed using a trivial continuation \texttt{onec}:
\begin{lstlisting}[mathescape, keepspaces, keywordstyle={[2]}]
rl < one(E) @ T:$S$ > => < < E @ T:$S$ > ; onec > .
rl < sol(T:$S$) TS ; onec > => sol(T:$S$) .
rl < none ; onec > => none .
\end{lstlisting}
Which solution is selected depends on the internal strategy of the rewriting engine for applying rules and ordering matches. Using the \texttt{search} command, every possible solution will be selected in some rewriting branch. Better performance is obtained if the second rule above is run just after the first solution appears inside the task, so that no unnecessary work is done. This is a situation were strategies are valuable ``at the meta-level'', that is for the rewrite theory $\mathcal S(M, S\!M)$.

\subsubsection{Strategy modules and calls}

	Strategy modules, their declarations and definitions, can be represented as Maude terms, like we did for the metalevel in \cref{sec:metalevel}. To simplify this presentation, we assume that the strategy definitions are collected in a definition set \texttt{DEFS}.
\begin{lstlisting}[mathescape, keepspaces]
eq DEFS = ($slabel$($p_1$, $\ldots$, $p_n$), $\delta$, $C$) , $\ldots$ .

rl < SC:StratCall @ T:$S$ > => find-defs(DEFS, SC:StratCall,
                                       T:$S$) .

eq find-defs(none, SC, T:$S$) = none .
ceq find-defs((Slhs, Def, C) Defs, SC, T:$S$) =
    find-defs2(MAT, T:$S$, Def) find-defs(Defs, SC, T:$S$)
 if MAT := getMatch(Slhs, SC, C) .

eq find-defs2(none, T:$S$, Def) = none .
eq find-defs2(< Sb, Cx:$S$ > MAT, T:$S$, Def) =
   < Def \cdot Sb @ T:$S$ > find-defs2(MAT, T:$S$, Def) .
\end{lstlisting}
The function \texttt{find-defs} traverses all the strategy definitions in \texttt{DEFS} and tries to match the strategy call term with their lefthand sides, and check their equational conditions. The strategy \texttt{find-defs2} takes these matches and produces a task \lstinline[mathescape]|< Def \cdot Sb @ T:$S$ >| for each of them, to continue rewriting \texttt{T} with the definition strategy, whose free variables are bound according to the matching substitution.

\subsection{Relating both semantics}

The previous semantics are equivalent in the sense specified in the following proposition, i.e., they produce the same solutions and terminate for the same input data. 

\begin{proposition}[\cite{slang}]
	In any module $(M, S\!M)$, for any term $t \in T_{\Sigma/E}$, and for any strategy expression $\alpha$, $t' \in \ssem\alpha(\varc, t)$ iff {\normalfont $\texttt< \; \alpha\; \texttt{@} \; t \; \texttt> \to^*_{\mathcal S(M, S\!M)} \texttt{sol(}t'\texttt) \; T\!S$} for some $T\!S$ of sort {\normalfont\ttfamily Tasks}. Moreover, $\bot \in \ssem\alpha(\varc, t)$ iff there is an infinite derivation from {\normalfont $\texttt< \; \alpha\; \texttt{@} \; t \; \texttt>$} in $\mathcal S(M, S\!M)$.
\end{proposition}

\begingroup
\renewcommand*\proofname{Proof sketch}
\begin{proof}
The proof proceeds by generalizing the statement to $t' \in \mathrm{dsem}(T\!S)$ iff $T\!S \to^*_{\mathcal S(M, S\!M)} \texttt{sol(}t'\texttt) \; T\!S'$, where $\mathrm{dsem} : T_{\mathcal S(M, S\!M)} \to \mathcal P_\bot(T_\Sigma)$ is an extended denotation satisfying $\mathrm{dsem}(\texttt< \; \alpha\; \texttt{@} \; t \; \texttt>) = \ssem{\alpha}(\mathrm{id}, t)$. Induction is carried out, first on the order of the approximants of the denotational semantics, and then structurally on the semantic terms.
\end{proof}
\endgroup

\section{Implementation} \label{sec:implementation}

	The first prototype for the language was a reflective implementation built as an extension of Full Maude, similar to the rewriting semantics described in \cref{sec:rewsem}. After some experimentation with that prototype, the strategy language started to be incorporated into the builtin Maude functionality, programmed in C++. This implementation provides better performance and integration with the rest of the Maude system, the module hierarchy, and the universal reflective theory.

	Implementing a nondeterministic language is a challenging task that involves several nontrivial design decisions. The commands \texttt{srewrite} and \texttt{dsrewrite} must explore the rewriting graph permitted by the strategy, and this graph may contain loops and other non-terminating rewriting sequences. Moreover, recursion and the existence of rewriting conditions in rules allow a state to have infinitely many successors, i.e., the rewriting graph need not be finitary. Hence, the best we can do, on a both potential infinitely branching and infinitely deep tree, is to support \emph{fairness}: whenever $t$ can be rewritten to $t'$ using a strategy, the solution $t'$ should be found in a finite computation. Nevertheless, this computation may not be practical in terms of memory and time.

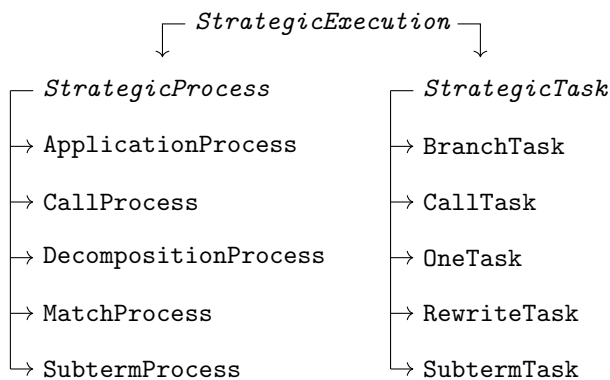
\begin{figure}[h]\centering
{\ttfamily\begin{tikzpicture}[every node/.style={anchor=west}, y=2em]
	\node (SE) at (2, .2) {\itshape StrategicExecution};

	\node (SP) at (0, -1) {\itshape StrategicProcess};
	\node (AP) at (0, -2) {ApplicationProcess};
	\node (CP) at (0, -3) {CallProcess};
	\node (DP) at (0, -4) {DecompositionProcess};
	\node (MP) at (0, -5) {MatchProcess};
	\node (StP) at (0, -6) {SubtermProcess};

	\node (ST) at (5, -1) {\itshape StrategicTask};
	\node (BT) at (5, -2) {BranchTask};
	\node (CT) at (5, -3) {CallTask};
	\node (OT) at (5, -4) {OneTask};
	\node (RT) at (5, -5) {RewriteTask};
	\node (StT) at (5, -6) {SubtermTask};

	\draw[->] (SE.west) -- +(-.3, 0) -- +(-.3, -.6);
	\draw[->] (SE.east) -- +(.3, 0) -- +(.3, -.6);

	\draw     (-.3, -1) -- (-.3, -6);
	\draw     (-.3, -1) -- (0, -1);
	\draw[->] (-.3, -2) -- (0, -2);
	\draw[->] (-.3, -3) -- (0, -3);
	\draw[->] (-.3, -4) -- (0, -4);
	\draw[->] (-.3, -5) -- (0, -5);
	\draw[->] (-.3, -6) -- (0, -6);

	\draw     (4.7, -1) -- (4.7, -6);
	\draw     (4.7, -1) -- (5, -1);
	\draw[->] (4.7, -2) -- (5, -2);
	\draw[->] (4.7, -3) -- (5, -3);
	\draw[->] (4.7, -4) -- (5, -4);
	\draw[->] (4.7, -5) -- (5, -5);
	\draw[->] (4.7, -6) -- (5, -6);
\end{tikzpicture}}
\caption{Class diagram of the main processes and tasks.} \label{fig:classdiag}
\end{figure}

	The fair implementation consists on a series of \emph{processes} and \emph{tasks} (see \cref{fig:classdiag}). Processes make the tree grow by processing expressions, applying rules, and calling strategies. Tasks are used to confine subsearches and continuations, and delimit variable contexts. Roughly speaking, processes can be understood as the \lststrat|<_@_>| tasks of the rewriting semantics in \cref{sec:rewsem}, and tasks as \lststrat|<_;_>|-terms. The same infrastructure is used for the depth-first search \texttt{dsrewrite} command, but processes are run differently.

	There are other alternatives to implement a strategy language, which can also be compiled instead of interpreted. Specifications in ELAN, where rules and strategies are mixed together, can be compiled to an executable binary through a C code generator~\cite{vittek96,remElan}. However, this will perhaps be simpler in that language where strategies are by default deterministic, unless some combinators are used. Nondeterminism is handled by a custom library supporting choice points and backtracking. In a recent work~\cite{qmaude}, an alternative implementation of a probabilistic extension of Maude strategies has been written in Python to generate probabilistic graphs. Strategies are compiled into an assembly-like language that is then executed by a virtual machine without explicit processes and tasks. Some aspects of these other approaches could be taken into consideration to improve the Maude implementation in the future.

\subsection{Processes}

	We contemplate the growing search tree as a pool of processes, each with a subject term to rewrite and a stack of strategy expressions to use. For achieving fairness, they are executed in round-robin by the \texttt{srewrite} command, while \texttt{dsrewrite} command follows a FIFO policy. Processes are similar to the \lststrat|<_@_>| tasks of the rewriting semantics, where the strategy continuation is replaced by a stack of strategy expressions. This stack essentially corresponds to a concatenation of strategies, and arise naturally from the accumulation of pending strategies during the execution of complex expressions. All processes are kept in a circular double-linked list with a moving pointer to the currently active process, which does a small amount of search computation and hands over the baton to the next one.

	There are several types of processes specialized in different tasks, each a subclass of an abstract class \texttt{StrategyProcess} in the implementation. The process in charge of the actual rewriting is the \emph{application process}. Generated by an application expression, in each turn it tries to make a new rewrite by finding the next position and rule that can be applied, considering the given label and initial substitution. Another essential process is the \emph{decomposition process}, since it pops strategy expressions from the pending stack, and acts according to their types. Each kind of strategy expression is also a subclass of an abstract class \texttt{StrategyExpression} with a method that specifies how each one should be \emph{decomposed} by the decomposition process. For example, \skywd{idle} is handled by doing nothing, so that the strategy is simply removed from the stack. On the other hand, \skywd{fail} causes the process to terminate, aborting its search path. Concatenations \lststrat|alpha ; beta| simply push their arguments to the stack in the right order, and the application strategy inserts an instance of the aformentioned application process into the list. Nondeterministic strategies like the alternation and the iteration spawn a copy of the decomposition process for each alternative.
When this process reaches the bottom of the stack, it informs its parent that the current term is a solution.

	Strategy calls are handled by \emph{call processes}, which try to match the strategy call term with the lefthand side of the definitions in the module, and create a decomposition process for each match in any of the definitions, one at a time.  Finally, \emph{subterm processes} find all matches of a \lststrat|matchrew| main pattern in the subject term, and create a decomposition process for each selected subterm with its associated strategy. These are controlled and gathered under a \emph{subterm task}.

\subsection{Tasks}

	Executing process independently is not enough, as we have seen with the rewriting semantics. Sometimes we want to treat a chunk of the search tree as an entity in its own right, as a subsearch or as a call frame. For example, the execution of the conditional operator $\ifthel\alpha\beta\gamma$ needs to know whether $\alpha$ has reached a solution to discard or activate the evaluation of the negative branch. This was achieved in the rewriting semantics of \cref{sec:rewsem} using \lststrat|<_;_>| terms, which gather a collection of processes along with a continuation. The various tasks of the C++ implementation, each a subclass of an abstract class \texttt{StrategicTask}, essentially reproduce the continuations of the semantics.

	Since searches and contexts can be nested, tasks form a hierarchy. Each process and each task (except the root task) belongs to some task and lives in a task-local double-linked list, different from the global list of processes. New processes created by processes in a task are usually attached to that same task and sometimes to the parent task, if they do not logically belong to the subsearch. Tasks and processes notify their parents whether they have succeeded or terminated, and unlink themselves from their owner's list.

	One of the most important tasks is the \emph{branch task}, charged of dealing with conditionals $\ifthel\alpha\beta\gamma$ and its derived operators. When informed by one of its processes that a solution has been reached, it creates a decomposition process on the parent task with the original pending stack but with $\beta$ on top. If the branch task runs out of processes without finding a solution, the negative branch is executed by a decomposition process from the initial term. Similar actions are performed for the combinators \lststrat|not| and \lststrat|test|. The execution of the \lststrat|one(alpha)| strategy is achieved by a task that executes $\alpha$ and interrupts the subcomputation as the first solution is reported.

	Other nested computations are the evaluation of rewriting condition fragments, which is done by \emph{rewrite tasks} and \emph{match processes}, and the rewriting of subterms. In the latter case, a \emph{subterm task} is used to delimit the variable context inside the \lststrat|matchrew| and manage the term reconstruction with the results of the evaluation of the subterms. Each subterm is processed in a separate child task, and its results are stored in a table from which all the combinations of subterm results are generated. They are searched in parallel, since their processes are part of the same process list. Finally, \emph{call tasks} house strategy call evaluations and separate their variable contexts.

\subsection{Modules}

	Strategy modules and theories are implemented like their functional and system counterparts. In addition to extending the Maude grammar with the statements of strategy modules and the strategy-specific commands, the data structure for modules is extended with slots for two additional kinds of module items: strategy declaration and strategy definitions. When processing the module, strategy declaration are checked to be well-formed and free of unbound variables, like other statements are. All the machinery of importation, renaming, and parameterization has also been extended to work with strategy modules, theories, and strategy mappings in views.

	One of the challenges when implementing strategy modules is how to efficiently handle the execution of strategy definitions. This involves matching the strategy call term with all definitions in the module to find which should be executed. We wanted to reuse the existing infrastructure for pattern matching. However, strategy calls are not terms in any signature, only their arguments are, and matching each argument separately is not reasonable. In effect, combining their respective matching substitution is far from being straightforward, and the optimized matching algorithms of Maude rely on a global view of the pattern for efficiency. Our solution is creating a hidden sort for strategy calls in each module and hidden tuple symbols for each strategy signature, which are also used in other parts of the Maude implementation. Each strategy definition and each call expression holds a term constructed with such a tuple for the lefthand side pattern or the call arguments, which can then be matched efficiently.

\subsubsection{Metalevel}

	The reflection of the strategy language, strategy modules and theories at the metalevel consists of the declaration of their metarepresentation in the modules \texttt{META-STRATEGY}, \texttt{META-MODULE}, and \texttt{META-THEORY} of the Maude prelude. Moreover, some more routine changes are needed in the C++ implementation to translate back and forth between metarepresentation and the usual representation of strategy expressions and statements. The \texttt{metaSrewrite} descent function does this conversion and then executes the given strategy just like the \texttt{srewrite} and \texttt{dsrewrite} commands.

\subsection{Efficiency considerations}

In order to avoid repeating work, tasks maintain a hash set of those states seen in previous executions, identified by their subject term and the pending strategies stack. When the decomposition process is executed, this set is checked and redundant search paths are aborted. This optimization is sound since all processes in the same task belong to the same subsearch and share the same variable environment. Checking the equivalence of two states involves comparing the terms and strategy stacks. The comparison of the latter is straightforward, since stacks are represented as shared nodes in a persistent global tree, whose branches are built by alternative push operations on the same stack.

Regarding strategy calls, we keep a list of definitions for each named strategy, so that they can be quickly checked on a strategy call. For optimizing a common case, call processes are not generated for strategies that are defined by a single unconditional definition without input parameters, whose defining expression is directly pushed on top of the stack.

\subsection{Performance comparison}

	As explained in the introduction, the strategy language was first implemented in Maude as an extension of Full Maude. We have compared the performance of the new C++ implementation with respect to that prototype using the original examples written for it. Since we have fixed some bugs and made some improvements when updating these examples to the current version of the language, instead of using their original code for the benchmarks, we have slightly adapted the renewed versions to run in the prototype. Additionally, we have included the 15-puzzle in the comparison, since it is the running example of this article. Moreover, the strategy language prototype was based on an old version of Full Maude that is incompatible with the current Maude releases, so we have minimally adapted it to run in Full Maude 3.1. This version of the extended interpreter comes with the strategy language as standard through the Core Maude implementation, but we have disabled what may interfere with the prototype. Notice that the meaning of the strategy search command differs from the prototype to the current version: the former's \texttt{srew} command looks for a single solution of the strategy, and we identify it with the new \texttt{dsrew [1]} command; its \texttt{srewall} command is the current \texttt{srew} that looks for all solutions of the strategy.

	\Cref{table:proto} collects some cost measures of the execution of both implementations on each example. The last three columns show the quotient of these measures with the C++ implementation in the denominator, so that the improvement is greater as these numbers are. The total time and number of rewrites are calculated as the sum of those printed by Maude for each command, thus excluding module parsing time. The value of the memory peak refers to the whole example execution, and the memory used by the interpreter and the prototype have been subtracted before calculating the coefficients. Otherwise, the prototype always uses much more memory due to Full Maude.

\begin{table}\centering
\begin{tabular}{lrrrrr}
	\toprule
			& \multicolumn2c{Time (ms)}	& \multicolumn3c{Measure proportion (old/new)} 		\\  \cmidrule(lr){2-3} \cmidrule(lr){4-6}
	Example		& New		& Old		& Time		& Rewrites	& Memory peak	\\
	\midrule
	Sudoku
	\cite{sudoku}
	& 908
	& 90639
	& 99.82
	& 19.9
	& 20.39
	\\
	Neural networks
	\cite{neuralNetworks}
	& 1061
	& 26496
	& 24.97
	& 8.16
	& 1.64
	\\
	Eden
	\cite{eden}
	& 12
	& 1360
	& 113.33
	& 417.61
	& 2.74
	\\
	Ambient calculus
	\cite{ambientCalculus}
	& 16
	& 125735
	& 7858.44
	& 85759.7
	& 180.47
	\\
	Basic completion
	(\cref{sec:completion})
	& 91
	& 7322
	& 80.46
	& 66.56
	& 2.2
	\\
	Blackboard
	\cite{towardsStrategy}
	& 1373
	& 47027
	& 34.25
	& 5451.67
	& 0.27
	\\
	15-puzzle
	(\cref{sec:15puzzle})
	& 182
	& 2232
	& 12.26
	& 4.69
	& 3.02
	\\
	\bottomrule
\end{tabular}
\caption{Performance comparison between the original prototype and the C++ implementation.} \label{table:proto}
\end{table}

	While the improvement is not uniform in all examples, the C++ unsurprisingly outperforms the Maude-based one in all of them. For the semantics of the ambient calculus, which is the example with the highest absolute execution time, the speedup is very significant. The memory usage is also drastically reduced. In the specification of Eden, a parallel Haskell-like programming language, the difference is not bigger because we have not included examples that finish in few milliseconds in the C++ implementation but take too much time to be waited for in the prototype. The same happens with the basic completion procedures of \cref{sec:completion}.
The higher memory usage peak of the C++ implementation in the blackboard example is caused by the allocation of nodes of the directed acyclic graph in which terms are represented, and it could be explained by the way Maude reserves and reuse memory for them. In absolute terms, the memory peak of the Core Maude implementation is 26.71 Mb while that of the prototype is 119.17 Mb, and the total memory even after discounting the total memory used by Full Maude alone is 22.77 times greater in the Maude-based implementation.

	This comparison exhibits some other advantages of the new implementation in addition to performance, since translating some of these examples back to the old prototype was not an easy task in some cases. Its strategy modules do not admit parameterization, module importation, or even selecting which module is being controlled. Hence, modular and parametric strategy specifications must be flattened in a single module of the prototype. This is specially visible in the basic completion example. Moreover, the syntactical analysis of strategy modules is less robust in the prototype, errors are less informative, and strategies with more than two arguments cannot be declared.

\section{Strategy language comparison} \label{sec:comparison}

	In the Introduction, we mentioned that the Maude strategy language has been influenced by other existing strategic rewriting systems like ELAN (1993) \cite{elan}, Stratego (1998) \cite{stratego}, and TOM (2001) \cite{tom}. While ELAN and Maude are standalone specification languages, the last two have specialized goals. Stratego is aimed at \emph{program transformation} and it is now distributed as part of the Spoofax Language Workbench~\cite{spoofax}, a platform for the development of textual, usually domain-specific, programming languages. TOM is a rewriting extension of Java that allows defining signatures, rules, and strategies inside the Java code and interoperate with them. There are other more recent strategy languages, like ρLog (2004) \cite{rholog} integrated into the Mathematica computing system as a plugin, and Porgy (2009) \cite{porgyJournal} for strategic graph (instead of term) rewriting.

	All these languages have similar foundations and repertories of strategy combinators, as shown in~\cref{tab:syntax}. Most combinators of any of the languages are either available in the others or can otherwise be expressed with more elaborate expressions. However, we highlight the main differences regarding Maude:

\begin{table}\centering\small
\begin{tabular}{ccccc} \toprule
	Maude 			&  ELAN					& Stratego			& TOM 			& ρLog			\\
	\midrule
	\ttfamily idle		& \ttfamily id				& \ttfamily id			& \ttfamily Identity	& \ttfamily Id		\\
	\ttfamily fail		& \ttfamily fail			& \ttfamily fail		& \ttfamily Fail 	& 			\\
	$\mathit{label}$	& $\mathit{label}$			& $\mathit{label}$		& Inline rules 		& $\mathit{label}$	\\
	\ttfamily top		& By default				& By default			& By default 		& By default 		\\
	\ttfamily ;		& \ttfamily ;				& \ttfamily ;			& \ttfamily Sequence 	& $\circ$		\\
	\ttfamily |		& \ttfamily dk				& \ttfamily +			& \ttfamily Choice 	& \ttfamily | 		\\
	\ttfamily or-else	& \ttfamily first			& \ttfamily <+			& Using Java		& \ttfamily Fst 	\\
	\ttfamily match $P$	& Using rules or library		& \ttfamily ?$P$		& {\ttfamily \%match}  	& Using rules		\\
\ttfamily $\alpha$ ? $\beta$ : $\gamma$	& \ttfamily\scriptsize if $\alpha$ then $\beta$ orelse $\gamma$ fi			& \ttfamily $\alpha$ < $\beta$ + $\gamma$	& Using Java	& Using rules \\
	\ttfamily *		& \ttfamily iterate*			& 	 			& 			& \ttfamily * 		\\	\ttfamily +		& \ttfamily iterate+			& 	 			& 			&			\\	\ttfamily !		& \ttfamily repeat*			& \ttfamily repeat		& \ttfamily Repeat 	& \ttfamily NF 		\\
	\ttfamily one($\alpha$)	& \ttfamily dc{\normalfont/}first one($\alpha$)	& Implicit		& Implicit 		& 			\\
	\ttfamily try		& \ttfamily first($\alpha$, id)		& \ttfamily try			& \ttfamily Try 	& \ttfamily Fst[$\alpha$, Id]			\\
\ttfamily test		& Using rules				& \ttfamily where		& Using Java		& Using rules		\\
	\ttfamily $\alpha \equiv \mathit{sl}$($t_1$, $\ldots$, $t_n$)	& \ttfamily call($\alpha$) & \ttfamily call($\mathit{sl}\!$ | \kern-5pt | $\!t_1$, $\ldots$, $t_n$)	& \ttfamily $\alpha$.apply & Using rules \\
\bottomrule
\end{tabular}
\caption{Strategy language syntax comparison} \label{tab:syntax}
\end{table}
\begin{enumerate}
	\item Maude enforces a clear separation between rules and strategies, while in ELAN and TOM strategies can be used in the definition of rules and they are sometimes required to cooperate. This dependency is more explicit in ρLog, where strategies are defined using an extension of the syntax for defining rules. Stratego allows and promotes this separation, but also lets strategies directly set the subject term, define new rules, and apply inline ones. The separation between rules and strategies in Maude is a conscious design decision to ease the understanding, analysis, and verification of the specification, respecting the \emph{separation of concerns} principle. 

	\item Strategies are potentially nondeterministic in all of these languages. However, ELAN, ρLog, and Maude compute the whole set of results for a given term, while Stratego and TOM explore a single rewriting path by resolving the nondeterminism arbitrarily.

	\item The application of a rule is the common basic element of these languages. However, all except Maude apply rules at the top by default. Instead, Maude applies them on any subterm within the subject term, unless explicitly indicated by \skywd{top}.

\item Parameterization is available by means of modules in ELAN and Maude, or at the level of strategy definitions in ρLog and Stratego, whose approach is more succinct. In fact, strategy expressions in Stratego are usually simpler, since the variety of strategy combinators is wider, some combinators are generic, and the signatures of strategies are not declared (types are not checked).

	\item Stratego, TOM, and ρLog allow programming generic traversals of terms, which need to be made explicit for each known operator in the signature in Maude. The generic possibilities of ρLog follow from the untyped Wolfram language of Mathematica on which it is based, since variables can be used not only for terms but also for function symbols and argument lists. 

	\item Maude and Stratego allow binding variables in strategy expressions. In Maude their scopes are delimited nested regions like the \skywd{matchrew} substrategies, while in Stratego explicit scopes can be declared and variables are defined from left to right within the same expression level.

	\item ELAN, Stratego, and TOM include a library of reusable strategy definitions, while no such library currently exists for the Maude strategy language.
\end{enumerate}

	Some features, like the generic term traversals of Stratego (item 5), have not been incorporated into the Maude language for the sake of simplicity. This is also the case for \emph{congruence operators} that make each symbol $f$ in the signature a strategy combinator such that \texttt{$f$($\alpha_1$, \ldots, $\alpha_n$)} applies the given strategies to the corresponding subterms. In~\cite{metatrans-jlamp}, these combinators have been implemented in Maude at the metalevel using \skywd{matchrew} operators, which can be seen as their counterparts in this language. Generic traversals have been addressed in~\cite{metatrans-jlamp} using a similar reflective strategy transformation, but we briefly describe here how they can be written by hand. Specifically, the three primitives or one-step descent operators of Stratego are \texttt{all}, to apply a strategy to all the direct subterms of the subject term; \texttt{some}, to apply a strategy to all the children in which it does not fail, and as long as it succeeds in at least one; and \texttt{one}, to apply the strategy to the first subterm in which it succeeds from left to right. These can be defined in the Maude strategy language for a fixed $\alpha$ with the following definition for each $n$-ary symbol $f$:
\begin{lstlisting}[mathescape]
sd st_all := matchrew $f$(x1, ..., xn) by x1 using $\alpha$, ...,
                                       xn using $\alpha$ .
sd st_one := matchrew $f$(x1, ..., xn) by x1 using $\alpha$ or-else
              ... or-else
             matchrew $f$(x1, ..., xn) by xn using $\alpha$ .
\end{lstlisting}
The \texttt{some} operator can be defined as \lstinline[mathescape]{test(st_one($\alpha$)) ; st_all(try($\alpha$))} in pseudocode according to its definition. It can also be given its own definition using a sequence of \skywd{matchrew} that apply the strategy $\alpha$ on every argument, letting it fail in all but one argument each.
\begin{lstlisting}[mathescape]
sd st_some := (matchrew $f$(x1, ..., xn) by x1 using $\alpha$ ;
               matchrew $f$(x1, ..., xn) by x2 using try($\alpha$),
	                                   ..., xn using try($\alpha$)) 
	            or-else ... or-else
              matchrew $f$(x1, ..., xn) by xn using $\alpha$ .
\end{lstlisting}
These definitions could be parametric on $\alpha$ by using parameterized strategy modules.

Similarly, the choosing operators from ELAN have not been included in the Maude language either, but almost all of them can be defined easily using Maude's strategy language constructs:
\[
	  \begin{array}{r@{\;\;\equiv\;\;}l}
		\texttt{first($\alpha_1$, \ldots, $\alpha_n$)} &\alpha_1 \;\skywd{or-else}\; \cdots \;\skywd{or-else}\; \alpha_n \\
		\texttt{dk($\alpha_1$, \ldots, $\alpha_n$)} &\alpha_1 \;\texttt|\; \cdots \;\texttt|\; \alpha_n \\
		\texttt{first one($\alpha_1$, \ldots, $\alpha_n$)} &\texttt{\skywd{one}($\alpha_1$)} \;\skywd{or-else}\; \cdots \;\skywd{or-else}\; \texttt{\skywd{one}($\alpha_n$)} \\
		\texttt{dc one($\alpha_1$, \ldots, $\alpha_n$)} &\texttt{\skywd{one}(}\alpha_1 \;\texttt|\; \cdots \;\texttt|\; \alpha_n\texttt{)} \\
	  \end{array}
	\]
The exception is \texttt{dc($\alpha_1$, $\ldots$, $\alpha_n$)} that returns all the results of only one of the $\alpha_i$ chosen nondeterministically. With the \skywd{one} operator we are only able to take either one or all of the results of any strategy expression. In the opposite direction, there are Maude constructs that are not available in ELAN, but they can also be expressed by more involved strategy expressions, in the worst case resorting to rules.

\section{Conclusion}

	The Maude strategy language is a useful resource to write and execute clear and natural strategy specifications where the description of rules is decoupled from the specification of how they should be applied, following the \emph{separation of concerns} principle. While the Maude strategy language was proposed almost twenty years ago and its first prototype appeared soon after, these goals have not been completely achieved until the recent introduction of the language into the official Maude interpreter with support for first-class strategy modules and reflection. Moreover, the efficient implementation of the language at the C++ level makes more interesting and complex applications possible.

	This article includes a comprehensive description of the current design of the strategy language, its complete formal semantics, and the relevant details about its implementation. We have shown that some challenges are involved in both the formal semantics and the implementation, and the benefit of this approach have been shown by various examples. Strategies have been used to specify Knuth-Bendix completion procedures on top of a well-established inference system, which helps reasoning about the correctness of the proposed algorithms. They have also been applied to solve games and to specify communication protocols. Additional examples presented in previous works include applications to the semantics of programming languages, process algebras, membrane systems, neural networks, linear programming, and a Sudoku solver, among others.

	Once systems have been described in rewriting logic, verification and analysis become relevant. Testing and comparing specifications executed with alternative controlling strategies could provide interesting information, and verification techniques like model checking \cite{fscd,smcJournal,smcJournal-btime} have been developed for these systems. Quantitative extensions of the strategy language have also been implemented and applied for probabilistic and statistical model checking~\cite{qmaude}. Moreover, support for strategies can be added to the Maude-based theorem provers and to other related formal tools~\cite{upvt23}.

\paragraph{Acknowledgements} 
Martí-Oliet, Rubio, and Verdejo are partially supported by Spanish AEI project ProCode (PID2019-108528RB-C22/AEI/10.13039/50110001103). Rubén Rubio has been partially supported by the Spanish Ministry of Universities under grant FPU17/02319.

\bibliographystyle{ACM-Reference-Format}

\begin{thebibliography}{61}



\ifx \showCODEN    \undefined \def \showCODEN     #1{\unskip}     \fi
\ifx \showDOI      \undefined \def \showDOI       #1{#1}\fi
\ifx \showISBNx    \undefined \def \showISBNx     #1{\unskip}     \fi
\ifx \showISBNxiii \undefined \def \showISBNxiii  #1{\unskip}     \fi
\ifx \showISSN     \undefined \def \showISSN      #1{\unskip}     \fi
\ifx \showLCCN     \undefined \def \showLCCN      #1{\unskip}     \fi
\ifx \shownote     \undefined \def \shownote      #1{#1}          \fi
\ifx \showarticletitle \undefined \def \showarticletitle #1{#1}   \fi
\ifx \showURL      \undefined \def \showURL       {\relax}        \fi
\providecommand\bibfield[2]{#2}
\providecommand\bibinfo[2]{#2}
\providecommand\natexlab[1]{#1}
\providecommand\showeprint[2][]{arXiv:#2}

\bibitem[Abramsky(1994)]{abramsky94}
\bibfield{author}{\bibinfo{person}{Samson Abramsky}.}
  \bibinfo{year}{1994}\natexlab{}.
\newblock \bibinfo{booktitle}{\emph{Domain theory}}. Vol.~\bibinfo{volume}{3}.
\newblock \bibinfo{publisher}{Clarendon Press}, \bibinfo{pages}{1--168}.
\newblock
\showISBNx{978-0-19-853762-5}


\bibitem[Alpuente et~al\mbox{.}(2023)]{upvt23}
\bibfield{author}{\bibinfo{person}{Mar{\'{\i}}a Alpuente},
  \bibinfo{person}{Demis Ballis}, \bibinfo{person}{Santiago Escobar},
  \bibinfo{person}{D. Gal{\'{a}}n}, {and} \bibinfo{person}{Julia
  Sapi{\~{n}}a}.} \bibinfo{year}{2023}\natexlab{}.
\newblock \showarticletitle{Safety enforcement via programmable strategies in
  Maude}.
\newblock \bibinfo{journal}{\emph{J. Log. Algebraic Methods Program.}}
  \bibinfo{volume}{132} (\bibinfo{year}{2023}), \bibinfo{pages}{100849}.
\newblock
\urldef\tempurl \url{https://doi.org/10.1016/j.jlamp.2023.100849}
\showDOI{\tempurl}


\bibitem[Andrei and Lucanu(2009)]{membrane}
\bibfield{author}{\bibinfo{person}{Oana Andrei} {and} \bibinfo{person}{Dorel
  Lucanu}.} \bibinfo{year}{2009}\natexlab{}.
\newblock \showarticletitle{Strategy-Based Proof Calculus for Membrane
  Systems}. In \bibinfo{booktitle}{\emph{Proceedings of the Seventh
  International Workshop on Rewriting Logic and its Applications, WRLA 2008,
  Budapest, Hungary, March 29-30, 2008}} \emph{(\bibinfo{series}{Electronic
  Notes in Theoretical Computer Science}, Vol.~\bibinfo{volume}{238(3)})},
  \bibfield{editor}{\bibinfo{person}{Grigore Roşu}} (Ed.).
  \bibinfo{publisher}{Elsevier}, \bibinfo{pages}{23--43}.
\newblock
\urldef\tempurl \url{https://doi.org/10.1016/j.entcs.2009.05.011}
\showDOI{\tempurl}


\bibitem[Baader and Nipkow(1998)]{allthat}
\bibfield{author}{\bibinfo{person}{Franz Baader} {and} \bibinfo{person}{Tobias
  Nipkow}.} \bibinfo{year}{1998}\natexlab{}.
\newblock \bibinfo{booktitle}{\emph{Term Rewriting and All That}}.
\newblock \bibinfo{publisher}{Cambridge University Press}.
\newblock
\showISBNx{978-1-139-17275-2}
\urldef\tempurl \url{https://doi.org/10.1017/CBO9781139172752}
\showDOI{\tempurl}


\bibitem[Bachmair and Dershowitz(1994)]{bachmair94}
\bibfield{author}{\bibinfo{person}{Leo Bachmair} {and} \bibinfo{person}{Nachum
  Dershowitz}.} \bibinfo{year}{1994}\natexlab{}.
\newblock \showarticletitle{Equational Inference, Canonical Proofs, and Proof
  Orderings}.
\newblock \bibinfo{journal}{\emph{J. {ACM}}} \bibinfo{volume}{41},
  \bibinfo{number}{2} (\bibinfo{year}{1994}), \bibinfo{pages}{236--276}.
\newblock
\urldef\tempurl \url{https://doi.org/10.1145/174652.174655}
\showDOI{\tempurl}


\bibitem[Balland et~al\mbox{.}(2007)]{tom}
\bibfield{author}{\bibinfo{person}{Emilie Balland}, \bibinfo{person}{Paul
  Brauner}, \bibinfo{person}{Radu Kopetz}, \bibinfo{person}{Pierre{-}Etienne
  Moreau}, {and} \bibinfo{person}{Antoine Reilles}.}
  \bibinfo{year}{2007}\natexlab{}.
\newblock \showarticletitle{Tom: {P}iggybacking Rewriting on {Java}}. In
  \bibinfo{booktitle}{\emph{Term Rewriting and Applications, 18th International
  Conference, {RTA} 2007, Paris, France, June 26-28, 2007, Proceedings}}
  \emph{(\bibinfo{series}{Lecture Notes in Computer Science},
  Vol.~\bibinfo{volume}{4533})}, \bibfield{editor}{\bibinfo{person}{Franz
  Baader}} (Ed.). \bibinfo{publisher}{Springer}, \bibinfo{pages}{36--47}.
\newblock
\showISBNx{978-3-540-73447-5}
\urldef\tempurl \url{https://doi.org/10.1007/978-3-540-73449-9_5}
\showDOI{\tempurl}


\bibitem[Barendregt(2014)]{barendregt}
\bibfield{author}{\bibinfo{person}{H.P. Barendregt}.}
  \bibinfo{year}{2014}\natexlab{}.
\newblock \bibinfo{booktitle}{\emph{The Lambda Calculus: Its Syntax and
  Semantics} (\bibinfo{edition}{2} ed.)}. \bibinfo{series}{Studies in Logic and
  the Foundations of Mathematics}, Vol.~\bibinfo{volume}{131}.
\newblock \bibinfo{publisher}{North Holland}.
\newblock
\showISBNx{978-0-444-87508-2}


\bibitem[Borovanský et~al\mbox{.}(2001)]{elan}
\bibfield{author}{\bibinfo{person}{Peter Borovanský}, \bibinfo{person}{Claude
  Kirchner}, \bibinfo{person}{Hélène Kirchner}, {and}
  \bibinfo{person}{Christophe Ringeissen}.} \bibinfo{year}{2001}\natexlab{}.
\newblock \showarticletitle{Rewriting with Strategies in {ELAN:} {A} Functional
  Semantics}.
\newblock \bibinfo{journal}{\emph{Int. J. Found. Comput. Sci.}}
  \bibinfo{volume}{12}, \bibinfo{number}{1} (\bibinfo{year}{2001}),
  \bibinfo{pages}{69--95}.
\newblock
\urldef\tempurl \url{https://doi.org/10.1142/S0129054101000412}
\showDOI{\tempurl}


\bibitem[Bouhoula et~al\mbox{.}(1997)]{spmel}
\bibfield{author}{\bibinfo{person}{Adel Bouhoula}, \bibinfo{person}{Jean-Pierre
  Jouannaud}, {and} \bibinfo{person}{José Meseguer}.}
  \bibinfo{year}{1997}\natexlab{}.
\newblock \showarticletitle{Specification and Proof in Membership Equational
  Logic}. In \bibinfo{booktitle}{\emph{TAPSOFT'97: Theory and Practice of
  Software Development, 7th International Joint Conference CAAP/FASE, Lille,
  France, April 14-18, 1997, Proceedings}} \emph{(\bibinfo{series}{Lecture
  Notes in Computer Science}, Vol.~\bibinfo{volume}{1214})},
  \bibfield{editor}{\bibinfo{person}{Michel Bidoit} {and} \bibinfo{person}{Max
  Dauchet}} (Eds.). \bibinfo{publisher}{Springer}, \bibinfo{pages}{67--92}.
\newblock
\showISBNx{3-540-62781-2}
\urldef\tempurl \url{https://doi.org/10.1007/BFb0030589}
\showDOI{\tempurl}


\bibitem[Bourdier et~al\mbox{.}(2009)]{extstrat}
\bibfield{author}{\bibinfo{person}{Tony Bourdier}, \bibinfo{person}{Horatiu
  Cirstea}, \bibinfo{person}{Daniel~J. Dougherty}, {and}
  \bibinfo{person}{Hélène Kirchner}.} \bibinfo{year}{2009}\natexlab{}.
\newblock \showarticletitle{Extensional and Intensional Strategies}. In
  \bibinfo{booktitle}{\emph{Proceedings Ninth International Workshop on
  Reduction Strategies in Rewriting and Programming, {WRS} 2009, Brasilia,
  Brazil, 28th June 2009}} \emph{(\bibinfo{series}{Electronic Proceedings in
  Theoretical Computer Science}, Vol.~\bibinfo{volume}{15})},
  \bibfield{editor}{\bibinfo{person}{Maribel Fernández}} (Ed.).
  \bibinfo{pages}{1--19}.
\newblock
\urldef\tempurl \url{https://doi.org/10.4204/EPTCS.15.1}
\showDOI{\tempurl}


\bibitem[Braga and Verdejo(2007)]{operational}
\bibfield{author}{\bibinfo{person}{Christiano Braga} {and}
  \bibinfo{person}{Alberto Verdejo}.} \bibinfo{year}{2007}\natexlab{}.
\newblock \showarticletitle{Modular Structural Operational Semantics with
  Strategies}. In \bibinfo{booktitle}{\emph{Proceedings of the Third Workshop
  on Structural Operational Semantics, SOS 2006, Bonn, Germany, August 26,
  2006}} \emph{(\bibinfo{series}{Electronic Notes in Theoretical Computer
  Science}, Vol.~\bibinfo{volume}{175(1)})},
  \bibfield{editor}{\bibinfo{person}{Rob van Glabbeek} {and}
  \bibinfo{person}{Peter~D. Mosses}} (Eds.). \bibinfo{publisher}{Elsevier},
  \bibinfo{pages}{3--17}.
\newblock
\urldef\tempurl \url{https://doi.org/10.1016/j.entcs.2006.10.024}
\showDOI{\tempurl}


\bibitem[Bravenboer et~al\mbox{.}(2008)]{stratego}
\bibfield{author}{\bibinfo{person}{Martin Bravenboer},
  \bibinfo{person}{Karl~Trygve Kalleberg}, \bibinfo{person}{Rob Vermaas}, {and}
  \bibinfo{person}{Eelco Visser}.} \bibinfo{year}{2008}\natexlab{}.
\newblock \showarticletitle{Stratego/{XT} 0.17. {A} language and toolset for
  program transformation}.
\newblock \bibinfo{journal}{\emph{Sci. Comput. Program.}} \bibinfo{volume}{72},
  \bibinfo{number}{1-2} (\bibinfo{year}{2008}), \bibinfo{pages}{52--70}.
\newblock
\urldef\tempurl \url{https://doi.org/10.1016/j.scico.2007.11.003}
\showDOI{\tempurl}


\bibitem[Cirstea et~al\mbox{.}(2017)]{lmcs17}
\bibfield{author}{\bibinfo{person}{Horatiu Cirstea}, \bibinfo{person}{Sergueï
  Lenglet}, {and} \bibinfo{person}{Pierre{-}Etienne Moreau}.}
  \bibinfo{year}{2017}\natexlab{}.
\newblock \showarticletitle{Faithful (meta-)encodings of programmable
  strategies into term rewriting systems}.
\newblock \bibinfo{journal}{\emph{Log. Methods Comput. Sci.}}
  \bibinfo{volume}{13}, \bibinfo{number}{4} (\bibinfo{year}{2017}).
\newblock


\bibitem[Clavel(2003)]{strategiesClavel}
\bibfield{author}{\bibinfo{person}{Manuel Clavel}.}
  \bibinfo{year}{2003}\natexlab{}.
\newblock \showarticletitle{Strategies and user interfaces in {Maude} at work}.
  In \bibinfo{booktitle}{\emph{Proceedings of the 3rd International Workshop on
  Reduction Strategies in Rewriting and Programming, WRS 2003, Valencia, Spain,
  June 8, 2003}} \emph{(\bibinfo{series}{Electronic Notes in Theoretical
  Computer Science}, Vol.~\bibinfo{volume}{86(4)})},
  \bibfield{editor}{\bibinfo{person}{Bernhard Gramlich} {and}
  \bibinfo{person}{Salvador Lucas}} (Eds.). \bibinfo{publisher}{Elsevier},
  \bibinfo{pages}{570--592}.
\newblock
\urldef\tempurl \url{https://doi.org/10.1016/S1571-0661(05)82612-X}
\showDOI{\tempurl}


\bibitem[Clavel et~al\mbox{.}(2023)]{maude}
\bibfield{author}{\bibinfo{person}{Manuel Clavel}, \bibinfo{person}{Francisco
  Durán}, \bibinfo{person}{Steven Eker}, \bibinfo{person}{Santiago Escobar},
  \bibinfo{person}{Patrick Lincoln}, \bibinfo{person}{Narciso Martí-Oliet},
  \bibinfo{person}{José Meseguer}, \bibinfo{person}{Rubén Rubio}, {and}
  \bibinfo{person}{Carolyn Talcott}.} \bibinfo{year}{2023}\natexlab{}.
\newblock \bibinfo{booktitle}{\emph{Maude Manual v3.3.1}}.
\newblock
\urldef\tempurl \url{https://maude.lcc.uma.es/maude-manual}
\showURL{\tempurl}


\bibitem[Clavel et~al\mbox{.}(2007)]{allmaude}
\bibfield{author}{\bibinfo{person}{Manuel Clavel}, \bibinfo{person}{Francisco
  Durán}, \bibinfo{person}{Steven Eker}, \bibinfo{person}{Patrick Lincoln},
  \bibinfo{person}{Narciso Martí-Oliet}, \bibinfo{person}{José Meseguer},
  {and} \bibinfo{person}{Carolyn~L. Talcott}.} \bibinfo{year}{2007}\natexlab{}.
\newblock \bibinfo{booktitle}{\emph{All About {Maude} - {A} High-Performance
  Logical Framework, How to Specify, Program and Verify Systems in Rewriting
  Logic}}. \bibinfo{series}{Lecture Notes in Computer Science},
  Vol.~\bibinfo{volume}{4350}.
\newblock \bibinfo{publisher}{Springer}.
\newblock
\showISBNx{978-3-540-71940-3}
\urldef\tempurl \url{https://doi.org/10.1007/978-3-540-71999-1}
\showDOI{\tempurl}


\bibitem[Clavel and Meseguer(1996)]{clavel96}
\bibfield{author}{\bibinfo{person}{Manuel Clavel} {and} \bibinfo{person}{José
  Meseguer}.} \bibinfo{year}{1996}\natexlab{}.
\newblock \showarticletitle{Reflection and Strategies in Rewriting Logic}. In
  \bibinfo{booktitle}{\emph{Proceedings of the First International Workshop on
  Rewriting Logic and its Applications, WRLA'96, Asilomar, California,
  September 3-6, 1996}} \emph{(\bibinfo{series}{Electronic Notes in Theoretical
  Computer Science}, Vol.~\bibinfo{volume}{4})},
  \bibfield{editor}{\bibinfo{person}{José Meseguer}} (Ed.).
  \bibinfo{publisher}{Elsevier}, \bibinfo{pages}{126--148}.
\newblock
\urldef\tempurl \url{https://doi.org/10.1016/S1571-0661(04)00037-4}
\showDOI{\tempurl}


\bibitem[Clavel and Meseguer(1997)]{clavel97}
\bibfield{author}{\bibinfo{person}{Manuel Clavel} {and} \bibinfo{person}{José
  Meseguer}.} \bibinfo{year}{1997}\natexlab{}.
\newblock \showarticletitle{Internal Strategies in a Reflective Logic}. In
  \bibinfo{booktitle}{\emph{Proceedings of the CADE-14 Workshop on Strategies
  in Automated Deduction}}, \bibfield{editor}{\bibinfo{person}{Bernhard
  Gramlich} {and} \bibinfo{person}{Hélène Kirchner}} (Eds.).
  \bibinfo{address}{Townsville, Australia}, \bibinfo{pages}{1--12}.
\newblock


\bibitem[Durán et~al\mbox{.}(2020)]{maude30}
\bibfield{author}{\bibinfo{person}{Francisco Durán}, \bibinfo{person}{Steven
  Eker}, \bibinfo{person}{Santiago Escobar}, \bibinfo{person}{Narciso
  Martí-Oliet}, \bibinfo{person}{José Meseguer}, \bibinfo{person}{Rubén
  Rubio}, {and} \bibinfo{person}{Carolyn Talcott}.}
  \bibinfo{year}{2020}\natexlab{}.
\newblock \showarticletitle{Programming and Symbolic Computation in {Maude}}.
\newblock \bibinfo{journal}{\emph{J. Log. Algebraic Methods Program.}}
  \bibinfo{volume}{110} (\bibinfo{year}{2020}), \bibinfo{numpages}{58}~pages.
\newblock
\urldef\tempurl \url{https://doi.org/10.1016/j.jlamp.2019.100497}
\showDOI{\tempurl}


\bibitem[Durán and Meseguer(2012)]{crc-jlap}
\bibfield{author}{\bibinfo{person}{Francisco Durán} {and}
  \bibinfo{person}{Jos{\'{e}} Meseguer}.} \bibinfo{year}{2012}\natexlab{}.
\newblock \showarticletitle{On the {C}hurch-{R}osser and coherence properties
  of conditional order-sorted rewrite theories}.
\newblock \bibinfo{journal}{\emph{J. Log. Algebraic Methods Program.}}
  \bibinfo{volume}{81}, \bibinfo{number}{7-8} (\bibinfo{year}{2012}),
  \bibinfo{pages}{816--850}.
\newblock
\urldef\tempurl \url{https://doi.org/10.1016/j.jlap.2011.12.004}
\showDOI{\tempurl}


\bibitem[Durán et~al\mbox{.}(2011)]{mfe}
\bibfield{author}{\bibinfo{person}{Francisco Durán}, \bibinfo{person}{Camilo
  Rocha}, {and} \bibinfo{person}{José~María Álvarez}.}
  \bibinfo{year}{2011}\natexlab{}.
\newblock \showarticletitle{Towards a {Maude} {F}ormal {E}nvironment}. In
  \bibinfo{booktitle}{\emph{Formal Modeling: Actors, Open Systems, Biological
  Systems - Essays Dedicated to Carolyn Talcott on the Occasion of Her 70th
  Birthday}} \emph{(\bibinfo{series}{Lecture Notes in Computer Science},
  Vol.~\bibinfo{volume}{7000})}, \bibfield{editor}{\bibinfo{person}{Gul Agha},
  \bibinfo{person}{Olivier Danvy}, {and} \bibinfo{person}{José Meseguer}}
  (Eds.). \bibinfo{publisher}{Springer}, \bibinfo{pages}{329--351}.
\newblock
\showISBNx{978-3-642-24932-7}
\urldef\tempurl \url{https://doi.org/10.1007/978-3-642-24933-4_17}
\showDOI{\tempurl}


\bibitem[Eker et~al\mbox{.}(2021)]{stratweb}
\bibfield{author}{\bibinfo{person}{Steven Eker}, \bibinfo{person}{Narciso
  Martí-Oliet}, \bibinfo{person}{José Meseguer}, \bibinfo{person}{Isabel
  Pita}, \bibinfo{person}{Rubén Rubio}, {and} \bibinfo{person}{Alberto
  Verdejo}.} \bibinfo{year}{2021}\natexlab{}.
\newblock \bibinfo{booktitle}{\emph{Strategy language for {Maude}}}.
\newblock
\urldef\tempurl \url{https://maude.ucm.es/strategies}
\showURL{\tempurl}


\bibitem[Eker et~al\mbox{.}(2007)]{strategies06}
\bibfield{author}{\bibinfo{person}{Steven Eker}, \bibinfo{person}{Narciso
  Martí-Oliet}, \bibinfo{person}{José Meseguer}, {and}
  \bibinfo{person}{Alberto Verdejo}.} \bibinfo{year}{2007}\natexlab{}.
\newblock \showarticletitle{Deduction, Strategies, and Rewriting}. In
  \bibinfo{booktitle}{\emph{Proceedings of the 6th International Workshop on
  Strategies in Automated Deduction, STRATEGIES 2006, Seattle, WA, USA, August
  16, 2006}} \emph{(\bibinfo{series}{Electronic Notes in Theoretical Computer
  Science}, Vol.~\bibinfo{volume}{174(11)})},
  \bibfield{editor}{\bibinfo{person}{Myla Archer}, \bibinfo{person}{Thierry~Boy
  de~la Tour}, {and} \bibinfo{person}{César Muñoz}} (Eds.).
  \bibinfo{publisher}{Elsevier}, \bibinfo{pages}{3--25}.
\newblock
\urldef\tempurl \url{https://doi.org/10.1016/j.entcs.2006.03.017}
\showDOI{\tempurl}


\bibitem[Fernández et~al\mbox{.}(2019)]{porgyJournal}
\bibfield{author}{\bibinfo{person}{Maribel Fernández},
  \bibinfo{person}{Hélène Kirchner}, {and} \bibinfo{person}{Bruno Pinaud}.}
  \bibinfo{year}{2019}\natexlab{}.
\newblock \showarticletitle{Strategic port graph rewriting: an interactive
  modelling framework}.
\newblock \bibinfo{journal}{\emph{Math. Struct. Comput. Sci.}}
  \bibinfo{volume}{29}, \bibinfo{number}{5} (\bibinfo{year}{2019}),
  \bibinfo{pages}{615--662}.
\newblock
\urldef\tempurl \url{https://doi.org/10.1017/S0960129518000270}
\showDOI{\tempurl}


\bibitem[Goguen(1984)]{parameterized}
\bibfield{author}{\bibinfo{person}{Joseph~A. Goguen}.}
  \bibinfo{year}{1984}\natexlab{}.
\newblock \showarticletitle{Parameterized Programming}.
\newblock \bibinfo{journal}{\emph{{IEEE} Trans. Software Eng.}}
  \bibinfo{volume}{10}, \bibinfo{number}{5} (\bibinfo{year}{1984}),
  \bibinfo{pages}{528--544}.
\newblock
\urldef\tempurl \url{https://doi.org/10.1109/TSE.1984.5010277}
\showDOI{\tempurl}


\bibitem[Hidalgo-Herrero et~al\mbox{.}(2007)]{eden}
\bibfield{author}{\bibinfo{person}{Mercedes Hidalgo-Herrero},
  \bibinfo{person}{Alberto Verdejo}, {and} \bibinfo{person}{Yolanda
  Ortega-Mallén}.} \bibinfo{year}{2007}\natexlab{}.
\newblock \showarticletitle{Using {Maude} and Its Strategies for Defining a
  Framework for Analyzing {Eden} Semantics}. In
  \bibinfo{booktitle}{\emph{Proceedings of the Sixth International Workshop on
  Reduction Strategies in Rewriting and Programming, WRS 2006, Seattle, WA,
  USA, August 11, 2006}} \emph{(\bibinfo{series}{Electronic Notes in
  Theoretical Computer Science}, Vol.~\bibinfo{volume}{174(10)})},
  \bibfield{editor}{\bibinfo{person}{Sergio Antoy}} (Ed.).
  \bibinfo{publisher}{Elsevier}, \bibinfo{pages}{119--137}.
\newblock
\urldef\tempurl \url{https://doi.org/10.1016/j.entcs.2007.02.051}
\showDOI{\tempurl}


\bibitem[Huet(1981)]{huet81}
\bibfield{author}{\bibinfo{person}{Gérard~P. Huet}.}
  \bibinfo{year}{1981}\natexlab{}.
\newblock \showarticletitle{A Complete Proof of Correctness of the
  {Knuth-Bendix} Completion Algorithm}.
\newblock \bibinfo{journal}{\emph{J. Comput. Syst. Sci.}} \bibinfo{volume}{23},
  \bibinfo{number}{1} (\bibinfo{year}{1981}), \bibinfo{pages}{11--21}.
\newblock
\urldef\tempurl \url{https://doi.org/10.1016/0022-0000(81)90002-7}
\showDOI{\tempurl}


\bibitem[Kirchner(2015)]{rewritingStrategies}
\bibfield{author}{\bibinfo{person}{Hélène Kirchner}.}
  \bibinfo{year}{2015}\natexlab{}.
\newblock \showarticletitle{Rewriting Strategies and Strategic Rewrite
  Programs}. In \bibinfo{booktitle}{\emph{Logic, Rewriting, and Concurrency -
  Essays dedicated to José Meseguer on the Occasion of His 65th Birthday}}
  \emph{(\bibinfo{series}{Lecture Notes in Computer Science},
  Vol.~\bibinfo{volume}{9200})}, \bibfield{editor}{\bibinfo{person}{Narciso
  Martí-Oliet}, \bibinfo{person}{Peter~Csaba Ölveczky}, {and}
  \bibinfo{person}{Carolyn~L. Talcott}} (Eds.). \bibinfo{publisher}{Springer},
  \bibinfo{pages}{380--403}.
\newblock
\showISBNx{978-3-319-23164-8}
\urldef\tempurl \url{https://doi.org/10.1007/978-3-319-23165-5_18}
\showDOI{\tempurl}


\bibitem[Kirchner and Moreau(1995)]{kirchner95}
\bibfield{author}{\bibinfo{person}{Hélène Kirchner} {and}
  \bibinfo{person}{Pierre{-}Etienne Moreau}.} \bibinfo{year}{1995}\natexlab{}.
\newblock \showarticletitle{Prototyping Completion with Constraints Using
  Computational Systems}. In \bibinfo{booktitle}{\emph{Rewriting Techniques and
  Applications, 6th International Conference, RTA-95, Kaiserslautern, Germany,
  April 5-7, 1995, Proceedings}} \emph{(\bibinfo{series}{Lecture Notes in
  Computer Science}, Vol.~\bibinfo{volume}{914})},
  \bibfield{editor}{\bibinfo{person}{Jieh Hsiang}} (Ed.).
  \bibinfo{publisher}{Springer}, \bibinfo{pages}{438--443}.
\newblock
\showISBNx{3-540-59200-8}
\urldef\tempurl \url{https://doi.org/10.1007/3-540-59200-8_79}
\showDOI{\tempurl}


\bibitem[Knuth and Bendix(1970)]{knuthBendix70}
\bibfield{author}{\bibinfo{person}{Donald~E. Knuth} {and}
  \bibinfo{person}{Peter~B. Bendix}.} \bibinfo{year}{1970}\natexlab{}.
\newblock \showarticletitle{Simple Word Problems in Universal Algebras}. In
  \bibinfo{booktitle}{\emph{Computational Problems in Abstract Algebra.
  Proceedings of a Conference Held at Oxford Under the Auspices of the Science
  Research Council Atlas Computer Laboratory, 29th August to 2nd September
  1967}}, \bibfield{editor}{\bibinfo{person}{John Leech}} (Ed.).
  \bibinfo{publisher}{Pergamon Press}, \bibinfo{pages}{263--297}.
\newblock
\showISBNx{978-0-08-012975-4}
\urldef\tempurl \url{https://doi.org/10.1016/B978-0-08-012975-4.50028-X}
\showDOI{\tempurl}


\bibitem[Kowalski(1979)]{kowalski}
\bibfield{author}{\bibinfo{person}{Robert~A. Kowalski}.}
  \bibinfo{year}{1979}\natexlab{}.
\newblock \showarticletitle{Algorithm = Logic + Control}.
\newblock \bibinfo{journal}{\emph{Commun. {ACM}}} \bibinfo{volume}{22},
  \bibinfo{number}{7} (\bibinfo{year}{1979}), \bibinfo{pages}{424--436}.
\newblock
\urldef\tempurl \url{https://doi.org/10.1145/359131.359136}
\showDOI{\tempurl}


\bibitem[Lescanne(1990)]{lescanneOrme}
\bibfield{author}{\bibinfo{person}{Pierre Lescanne}.}
  \bibinfo{year}{1990}\natexlab{}.
\newblock \showarticletitle{Implementations of Completion by Transition Rules +
  Control: {ORME}}. In \bibinfo{booktitle}{\emph{Algebraic and Logic
  Programming, Second International Conference, Nancy, France, October 1-3,
  1990, Proceedings}} \emph{(\bibinfo{series}{Lecture Notes in Computer
  Science}, Vol.~\bibinfo{volume}{463})},
  \bibfield{editor}{\bibinfo{person}{Hélène Kirchner} {and}
  \bibinfo{person}{Wolfgang Wechler}} (Eds.). \bibinfo{publisher}{Springer},
  \bibinfo{pages}{262--269}.
\newblock
\showISBNx{3-540-53162-9}
\urldef\tempurl \url{https://doi.org/10.1007/3-540-53162-9_44}
\showDOI{\tempurl}


\bibitem[Malkin(1998)]{rfc2453}
\bibfield{author}{\bibinfo{person}{G. Malkin}.}
  \bibinfo{year}{1998}\natexlab{}.
\newblock \bibinfo{booktitle}{\emph{RIP Version 2}}.
\newblock \bibinfo{type}{{RFC}} 2453. \bibinfo{institution}{Internet
  Engineering Task Force}.
\newblock
\urldef\tempurl \url{https://tools.ietf.org/html/rfc2453}
\showURL{\tempurl}


\bibitem[Marin and Kutsia(2006)]{rholog}
\bibfield{author}{\bibinfo{person}{Mircea Marin} {and} \bibinfo{person}{Temur
  Kutsia}.} \bibinfo{year}{2006}\natexlab{}.
\newblock \showarticletitle{Foundations of the rule-based system ρLog}.
\newblock \bibinfo{journal}{\emph{J. Appl. Non Class. Logics}}
  \bibinfo{volume}{16}, \bibinfo{number}{1-2} (\bibinfo{year}{2006}),
  \bibinfo{pages}{151--168}.
\newblock
\urldef\tempurl \url{https://doi.org/10.3166/jancl.16.151-168}
\showDOI{\tempurl}


\bibitem[Martí-Oliet et~al\mbox{.}(2004)]{towardsStrategy}
\bibfield{author}{\bibinfo{person}{Narciso Martí-Oliet},
  \bibinfo{person}{José Meseguer}, {and} \bibinfo{person}{Alberto Verdejo}.}
  \bibinfo{year}{2004}\natexlab{}.
\newblock \showarticletitle{Towards a Strategy Language for {Maude}}. In
  \bibinfo{booktitle}{\emph{Proceedings of the Fifth International Workshop on
  Rewriting Logic and its Applications, WRLA 2004, Barcelona, Spain, March
  27-April 4, 2004}} \emph{(\bibinfo{series}{Electronic Notes in Theoretical
  Computer Science}, Vol.~\bibinfo{volume}{117})},
  \bibfield{editor}{\bibinfo{person}{Narciso Martí-Oliet}} (Ed.).
  \bibinfo{publisher}{Elsevier}, \bibinfo{pages}{417--441}.
\newblock
\urldef\tempurl \url{https://doi.org/10.1016/j.entcs.2004.06.020}
\showDOI{\tempurl}


\bibitem[Martí-Oliet et~al\mbox{.}(2009)]{rewSemantics}
\bibfield{author}{\bibinfo{person}{Narciso Martí-Oliet},
  \bibinfo{person}{José Meseguer}, {and} \bibinfo{person}{Alberto Verdejo}.}
  \bibinfo{year}{2009}\natexlab{}.
\newblock \showarticletitle{A Rewriting Semantics for {Maude} Strategies}. In
  \bibinfo{booktitle}{\emph{Proceedings of the Seventh International Workshop
  on Rewriting Logic and its Applications, WRLA 2008, Budapest, Hungary, March
  29-30, 2008}} \emph{(\bibinfo{series}{Electronic Notes in Theoretical
  Computer Science}, Vol.~\bibinfo{volume}{238(3)})},
  \bibfield{editor}{\bibinfo{person}{Grigore Roşu}} (Ed.).
  \bibinfo{publisher}{Elsevier}, \bibinfo{pages}{227--247}.
\newblock
\urldef\tempurl \url{https://doi.org/10.1016/j.entcs.2009.05.022}
\showDOI{\tempurl}


\bibitem[Martí{-}Oliet et~al\mbox{.}(2007)]{ccs}
\bibfield{author}{\bibinfo{person}{Narciso Martí{-}Oliet},
  \bibinfo{person}{Miguel Palomino}, {and} \bibinfo{person}{Alberto Verdejo}.}
  \bibinfo{year}{2007}\natexlab{}.
\newblock \showarticletitle{Strategies and simulations in a semantic
  framework}.
\newblock \bibinfo{journal}{\emph{J. Algorithms}} \bibinfo{volume}{62},
  \bibinfo{number}{3-4} (\bibinfo{year}{2007}), \bibinfo{pages}{95--116}.
\newblock
\urldef\tempurl \url{https://doi.org/10.1016/j.jalgor.2007.04.002}
\showDOI{\tempurl}


\bibitem[Meseguer(1992)]{rewritingLogic}
\bibfield{author}{\bibinfo{person}{José Meseguer}.}
  \bibinfo{year}{1992}\natexlab{}.
\newblock \showarticletitle{Conditional rewriting logic as a unified model of
  concurrency}.
\newblock \bibinfo{journal}{\emph{Theor. Comput. Sci.}} \bibinfo{volume}{96},
  \bibinfo{number}{1} (\bibinfo{year}{1992}), \bibinfo{pages}{73--155}.
\newblock
\urldef\tempurl \url{https://doi.org/10.1016/0304-3975(92)90182-F}
\showDOI{\tempurl}


\bibitem[Meseguer(2012)]{20years}
\bibfield{author}{\bibinfo{person}{José Meseguer}.}
  \bibinfo{year}{2012}\natexlab{}.
\newblock \showarticletitle{Twenty years of rewriting logic}.
\newblock \bibinfo{journal}{\emph{J. Log. Algebr. Program.}}
  \bibinfo{volume}{81}, \bibinfo{number}{7-8} (\bibinfo{year}{2012}),
  \bibinfo{pages}{721--781}.
\newblock
\urldef\tempurl \url{https://doi.org/10.1016/j.jlap.2012.06.003}
\showDOI{\tempurl}


\bibitem[Moreau(2000)]{remElan}
\bibfield{author}{\bibinfo{person}{Pierre{-}Etienne Moreau}.}
  \bibinfo{year}{2000}\natexlab{}.
\newblock \showarticletitle{{REM} (Reduce Elan Machine): Core of the New {ELAN}
  Compiler}. In \bibinfo{booktitle}{\emph{Rewriting Techniques and
  Applications, 11th International Conference, {RTA} 2000, Norwich, UK, July
  10-12, 2000, Proceedings}} \emph{(\bibinfo{series}{Lecture Notes in Computer
  Science}, Vol.~\bibinfo{volume}{1833})},
  \bibfield{editor}{\bibinfo{person}{Leo Bachmair}} (Ed.).
  \bibinfo{publisher}{Springer}, \bibinfo{pages}{265--269}.
\newblock
\urldef\tempurl \url{https://doi.org/10.1007/10721975\_19}
\showDOI{\tempurl}


\bibitem[Pettorossi and Proietti(2002)]{pettorossi}
\bibfield{author}{\bibinfo{person}{Alberto Pettorossi} {and}
  \bibinfo{person}{Maurizio Proietti}.} \bibinfo{year}{2002}\natexlab{}.
\newblock \showarticletitle{Program Derivation = Rules + Strategies}. In
  \bibinfo{booktitle}{\emph{Computational Logic: Logic Programming and Beyond,
  Essays in Honour of Robert A. Kowalski, Part {I}}}
  \emph{(\bibinfo{series}{Lecture Notes in Computer Science},
  Vol.~\bibinfo{volume}{2407})}, \bibfield{editor}{\bibinfo{person}{Antonis~C.
  Kakas} {and} \bibinfo{person}{Fariba Sadri}} (Eds.).
  \bibinfo{publisher}{Springer}, \bibinfo{pages}{273--309}.
\newblock
\showISBNx{3-540-43959-5}
\urldef\tempurl \url{https://doi.org/10.1007/3-540-45628-7_12}
\showDOI{\tempurl}


\bibitem[Plotkin(1976)]{plotkin76}
\bibfield{author}{\bibinfo{person}{Gordon~D. Plotkin}.}
  \bibinfo{year}{1976}\natexlab{}.
\newblock \showarticletitle{A Powerdomain Construction}.
\newblock \bibinfo{journal}{\emph{{SIAM} J. Comput.}} \bibinfo{volume}{5},
  \bibinfo{number}{3} (\bibinfo{year}{1976}), \bibinfo{pages}{452--487}.
\newblock
\urldef\tempurl \url{https://doi.org/10.1137/0205035}
\showDOI{\tempurl}


\bibitem[Ratner and Warmuth(1990)]{npuzzle}
\bibfield{author}{\bibinfo{person}{Daniel Ratner} {and}
  \bibinfo{person}{Manfred~K. Warmuth}.} \bibinfo{year}{1990}\natexlab{}.
\newblock \showarticletitle{The $(n^2 -1)$-Puzzle and Related Relocation
  Problems}.
\newblock \bibinfo{journal}{\emph{J. Symb. Comput.}} \bibinfo{volume}{10},
  \bibinfo{number}{2} (\bibinfo{year}{1990}), \bibinfo{pages}{111--138}.
\newblock
\urldef\tempurl \url{https://doi.org/10.1016/S0747-7171(08)80001-6}
\showDOI{\tempurl}


\bibitem[Rosa-Velardo et~al\mbox{.}(2006)]{ambientCalculus}
\bibfield{author}{\bibinfo{person}{Fernando Rosa-Velardo},
  \bibinfo{person}{Clara Segura}, {and} \bibinfo{person}{Alberto Verdejo}.}
  \bibinfo{year}{2006}\natexlab{}.
\newblock \showarticletitle{Typed Mobile Ambients in {Maude}}. In
  \bibinfo{booktitle}{\emph{Proceedings of the 6th International Workshop on
  Rule-Based Programming, RULE 2005, Nara, Japan, April 23, 2005}}
  \emph{(\bibinfo{series}{Electronic Notes in Theoretical Computer Science},
  Vol.~\bibinfo{volume}{147(1)})}, \bibfield{editor}{\bibinfo{person}{Horatiu
  Cirstea} {and} \bibinfo{person}{Narciso Martí-Oliet}} (Eds.).
  \bibinfo{publisher}{Elsevier}, \bibinfo{pages}{135--161}.
\newblock
\urldef\tempurl \url{https://doi.org/10.1016/j.entcs.2005.06.041}
\showDOI{\tempurl}


\bibitem[Rubio(2022)]{mitesis}
\bibfield{author}{\bibinfo{person}{Rubén Rubio}.}
  \bibinfo{year}{2022}\natexlab{}.
\newblock \emph{\bibinfo{title}{Model checking of strategy-controlled systems
  in rewriting logic}}.
\newblock \bibinfo{thesistype}{Ph.\,D. Dissertation}.
  \bibinfo{school}{Universidad Complutense de Madrid}.
\newblock
\urldef\tempurl \url{https://eprints.ucm.es/71531}
\showURL{\tempurl}


\bibitem[Rubio et~al\mbox{.}(2023)]{qmaude}
\bibfield{author}{\bibinfo{person}{Rub{\'{e}}n Rubio}, \bibinfo{person}{Narciso
  Mart{\'{\i}}{-}Oliet}, \bibinfo{person}{Isabel Pita}, {and}
  \bibinfo{person}{Alberto Verdejo}.} \bibinfo{year}{2023}\natexlab{}.
\newblock \showarticletitle{QMaude: Quantitative Specification and Verification
  in Rewriting Logic}. In \bibinfo{booktitle}{\emph{Formal Methods - 25th
  International Symposium, {FM} 2023, L{\"{u}}beck, Germany, March 6-10, 2023,
  Proceedings}} \emph{(\bibinfo{series}{Lecture Notes in Computer Science},
  Vol.~\bibinfo{volume}{14000})}, \bibfield{editor}{\bibinfo{person}{Marsha
  Chechik}, \bibinfo{person}{Joost{-}Pieter Katoen}, {and}
  \bibinfo{person}{Martin Leucker}} (Eds.). \bibinfo{publisher}{Springer},
  \bibinfo{pages}{240--259}.
\newblock
\urldef\tempurl \url{https://doi.org/10.1007/978-3-031-27481-7\_15}
\showDOI{\tempurl}


\bibitem[Rubio et~al\mbox{.}(2019a)]{fscd}
\bibfield{author}{\bibinfo{person}{Rubén Rubio}, \bibinfo{person}{Narciso
  Martí-Oliet}, \bibinfo{person}{Isabel Pita}, {and} \bibinfo{person}{Alberto
  Verdejo}.} \bibinfo{year}{2019}\natexlab{a}.
\newblock \showarticletitle{Model checking strategy-controlled rewriting
  systems}. In \bibinfo{booktitle}{\emph{4th International Conference on Formal
  Structures for Computation and Deduction, {FSCD} 2019, June 24-30, 2019,
  Dortmund, Germany}} \emph{(\bibinfo{series}{LIPIcs},
  Vol.~\bibinfo{volume}{131})}, \bibfield{editor}{\bibinfo{person}{Herman
  Geuvers}} (Ed.). \bibinfo{publisher}{Schloss Dagstuhl - Leibniz-Zentrum für
  Informatik}, \bibinfo{pages}{34:1--34:18}.
\newblock
\urldef\tempurl \url{https://doi.org/10.4230/LIPIcs.FSCD.2019.34}
\showDOI{\tempurl}


\bibitem[Rubio et~al\mbox{.}(2019b)]{pssm}
\bibfield{author}{\bibinfo{person}{Rubén Rubio}, \bibinfo{person}{Narciso
  Martí-Oliet}, \bibinfo{person}{Isabel Pita}, {and} \bibinfo{person}{Alberto
  Verdejo}.} \bibinfo{year}{2019}\natexlab{b}.
\newblock \showarticletitle{Parameterized strategies specification in {Maude}}.
  In \bibinfo{booktitle}{\emph{Recent Trends in Algebraic Development
  Techniques}} \emph{(\bibinfo{series}{Lecture Notes in Computer Science},
  Vol.~\bibinfo{volume}{11563})}, \bibfield{editor}{\bibinfo{person}{José
  Fiadeiro} {and} \bibinfo{person}{Ionuț Țuțu}} (Eds.).
  \bibinfo{publisher}{Springer}, \bibinfo{pages}{27--44}.
\newblock
\urldef\tempurl \url{https://doi.org/10.1007/978-3-030-23220-7_2}
\showDOI{\tempurl}


\bibitem[Rubio et~al\mbox{.}(2021a)]{slang}
\bibfield{author}{\bibinfo{person}{Rubén Rubio}, \bibinfo{person}{Narciso
  Martí-Oliet}, \bibinfo{person}{Isabel Pita}, {and} \bibinfo{person}{Alberto
  Verdejo}.} \bibinfo{year}{2021}\natexlab{a}.
\newblock \bibinfo{booktitle}{\emph{The semantics of the {Maude} strategy
  language}}.
\newblock \bibinfo{type}{{T}echnical {R}eport} 01/21.
  \bibinfo{institution}{Departamento de Sistemas Informáticos y Computación,
  Universidad Complutense de Madrid}.
\newblock
\urldef\tempurl \url{https://eprints.ucm.es/67449/}
\showURL{\tempurl}


\bibitem[Rubio et~al\mbox{.}(2021b)]{smcJournal-btime}
\bibfield{author}{\bibinfo{person}{Rubén Rubio}, \bibinfo{person}{Narciso
  Martí-Oliet}, \bibinfo{person}{Isabel Pita}, {and} \bibinfo{person}{Alberto
  Verdejo}.} \bibinfo{year}{2021}\natexlab{b}.
\newblock \showarticletitle{Strategies, model checking and branching-time
  properties in {Maude}}.
\newblock \bibinfo{journal}{\emph{J. Log. Algebr. Methods Program.}}
  \bibinfo{volume}{123} (\bibinfo{year}{2021}), \bibinfo{numpages}{28}~pages.
\newblock
\urldef\tempurl \url{https://doi.org/10.1016/j.jlamp.2021.100700}
\showDOI{\tempurl}


\bibitem[Rubio et~al\mbox{.}(2022a)]{metatrans-jlamp}
\bibfield{author}{\bibinfo{person}{Rubén Rubio}, \bibinfo{person}{Narciso
  Martí-Oliet}, \bibinfo{person}{Isabel Pita}, {and} \bibinfo{person}{Alberto
  Verdejo}.} \bibinfo{year}{2022}\natexlab{a}.
\newblock \showarticletitle{Metalevel transformation of strategies}.
\newblock \bibinfo{journal}{\emph{J. Log. Algebr. Methods Program.}}
  \bibinfo{volume}{124} (\bibinfo{year}{2022}), \bibinfo{numpages}{21}~pages.
\newblock
\urldef\tempurl \url{https://doi.org/10.1016/j.jlamp.2021.100728}
\showDOI{\tempurl}


\bibitem[Rubio et~al\mbox{.}(2022b)]{smcJournal}
\bibfield{author}{\bibinfo{person}{Rubén Rubio}, \bibinfo{person}{Narciso
  Martí-Oliet}, \bibinfo{person}{Isabel Pita}, {and} \bibinfo{person}{Alberto
  Verdejo}.} \bibinfo{year}{2022}\natexlab{b}.
\newblock \showarticletitle{Model checking strategy-controlled systems in
  rewriting logic}.
\newblock \bibinfo{journal}{\emph{Automat. Softw. Eng.}} \bibinfo{volume}{29},
  \bibinfo{number}{1} (\bibinfo{year}{2022}), \bibinfo{numpages}{57}~pages.
\newblock
\urldef\tempurl \url{https://doi.org/10.1007/s10515-021-00307-9}
\showDOI{\tempurl}


\bibitem[Rubio et~al\mbox{.}(2022c)]{memstratmc-jlamp}
\bibfield{author}{\bibinfo{person}{Rubén Rubio}, \bibinfo{person}{Narciso
  Martí-Oliet}, \bibinfo{person}{Isabel Pita}, {and} \bibinfo{person}{Alberto
  Verdejo}.} \bibinfo{year}{2022}\natexlab{c}.
\newblock \showarticletitle{Simulating and model checking membrane systems
  using strategies in {Maude}}.
\newblock \bibinfo{journal}{\emph{J. Log. Algebr. Methods Program.}}
  \bibinfo{volume}{124} (\bibinfo{year}{2022}), \bibinfo{numpages}{25}~pages.
\newblock
\urldef\tempurl \url{https://doi.org/10.1016/j.jlamp.2021.100727}
\showDOI{\tempurl}


\bibitem[Santos{-}García and Palomino(2007)]{sudoku}
\bibfield{author}{\bibinfo{person}{Gustavo Santos{-}García} {and}
  \bibinfo{person}{Miguel Palomino}.} \bibinfo{year}{2007}\natexlab{}.
\newblock \showarticletitle{Solving {Sudoku} Puzzles with Rewriting Rules}. In
  \bibinfo{booktitle}{\emph{Proceedings of the 6th International Workshop on
  Rewriting Logic and its Applications, WRLA 2006, Vienna, Austria, April 1-2,
  2006}} \emph{(\bibinfo{series}{Electronic Notes in Theoretical Computer
  Science}, Vol.~\bibinfo{volume}{176(4)})},
  \bibfield{editor}{\bibinfo{person}{Grit Denker} {and}
  \bibinfo{person}{Carolyn Talcott}} (Eds.). \bibinfo{publisher}{Elsevier},
  \bibinfo{pages}{79--93}.
\newblock
\urldef\tempurl \url{https://doi.org/10.1016/j.entcs.2007.06.009}
\showDOI{\tempurl}


\bibitem[Santos-García et~al\mbox{.}(2009)]{neuralNetworks}
\bibfield{author}{\bibinfo{person}{Gustavo Santos-García},
  \bibinfo{person}{Miguel Palomino}, {and} \bibinfo{person}{Alberto Verdejo}.}
  \bibinfo{year}{2009}\natexlab{}.
\newblock \showarticletitle{Rewriting Logic Using Strategies for Neural
  Networks: An Implementation in {Maude}}. In
  \bibinfo{booktitle}{\emph{International Symposium on Distributed Computing
  and Artificial Intelligence, {DCAI} 2008, University of Salamanca, Spain,
  22th-24th October 2008}} \emph{(\bibinfo{series}{Advances in Soft Computing},
  Vol.~\bibinfo{volume}{50})}, \bibfield{editor}{\bibinfo{person}{Juan~M.
  Corchado}, \bibinfo{person}{Sara Rodríguez}, \bibinfo{person}{James Llinas},
  {and} \bibinfo{person}{José~M. Molina}} (Eds.).
  \bibinfo{publisher}{Springer}, \bibinfo{pages}{424--433}.
\newblock
\showISBNx{978-3-540-85862-1}
\urldef\tempurl \url{https://doi.org/10.1007/978-3-540-85863-8_50}
\showDOI{\tempurl}


\bibitem[Terese(2003)]{terese}
\bibfield{author}{\bibinfo{person}{Terese}.} \bibinfo{year}{2003}\natexlab{}.
\newblock \bibinfo{booktitle}{\emph{Term Rewriting Systems}}.
\newblock \bibinfo{publisher}{Cambridge University Press}.
\newblock
\showISBNx{978-0-521-39115-3}


\bibitem[Verdejo and Martí-Oliet(2011)]{completion}
\bibfield{author}{\bibinfo{person}{Alberto Verdejo} {and}
  \bibinfo{person}{Narciso Martí-Oliet}.} \bibinfo{year}{2011}\natexlab{}.
\newblock \showarticletitle{Basic completion strategies as another application
  of the {Maude} strategy language}. In \bibinfo{booktitle}{\emph{Proceedings
  10th International Workshop on Reduction Strategies in Rewriting and
  Programming, {WRS} 2011, Novi Sad, Serbia, 29 May 2011}}
  \emph{(\bibinfo{series}{Electronic Proceedings in Theoretical Computer
  Science}, Vol.~\bibinfo{volume}{82})},
  \bibfield{editor}{\bibinfo{person}{Santiago Escobar}} (Ed.).
  \bibinfo{pages}{17--36}.
\newblock
\urldef\tempurl \url{https://doi.org/10.4204/EPTCS.82.2}
\showDOI{\tempurl}


\bibitem[Visser(2005)]{visser05}
\bibfield{author}{\bibinfo{person}{Eelco Visser}.}
  \bibinfo{year}{2005}\natexlab{}.
\newblock \showarticletitle{A survey of strategies in rule-based program
  transformation systems}.
\newblock \bibinfo{journal}{\emph{J. Symb. Comput.}} \bibinfo{volume}{40},
  \bibinfo{number}{1} (\bibinfo{year}{2005}), \bibinfo{pages}{831--873}.
\newblock
\urldef\tempurl \url{https://doi.org/10.1016/j.jsc.2004.12.011}
\showDOI{\tempurl}


\bibitem[Vittek(1996)]{vittek96}
\bibfield{author}{\bibinfo{person}{Marian Vittek}.}
  \bibinfo{year}{1996}\natexlab{}.
\newblock \showarticletitle{A Compiler for Nondeterministic Term Rewriting
  Systems}. In \bibinfo{booktitle}{\emph{Rewriting Techniques and Applications,
  7th International Conference, RTA-96, New Brunswick, NJ, USA, July 27-30,
  1996, Proceedings}} \emph{(\bibinfo{series}{Lecture Notes in Computer
  Science}, Vol.~\bibinfo{volume}{1103})},
  \bibfield{editor}{\bibinfo{person}{Harald Ganzinger}} (Ed.).
  \bibinfo{publisher}{Springer}, \bibinfo{pages}{154--167}.
\newblock
\urldef\tempurl \url{https://doi.org/10.1007/3-540-61464-8\_50}
\showDOI{\tempurl}


\bibitem[Wachsmuth et~al\mbox{.}(2014)]{spoofax}
\bibfield{author}{\bibinfo{person}{Guido Wachsmuth}, \bibinfo{person}{Gabriël
  D.~P. Konat}, {and} \bibinfo{person}{Eelco Visser}.}
  \bibinfo{year}{2014}\natexlab{}.
\newblock \showarticletitle{Language Design with the {Spoofax} {Language}
  {Workbench}}.
\newblock \bibinfo{journal}{\emph{{IEEE} Softw.}} \bibinfo{volume}{31},
  \bibinfo{number}{5} (\bibinfo{year}{2014}), \bibinfo{pages}{35--43}.
\newblock
\urldef\tempurl \url{https://doi.org/10.1109/MS.2014.100}
\showDOI{\tempurl}


\bibitem[Éduard Lucas(1992)]{lucas1}
\bibfield{author}{\bibinfo{person}{Éduard Lucas}.}
  \bibinfo{year}{1992}\natexlab{}.
\newblock \bibinfo{booktitle}{\emph{Recréations mathématiques}
  (\bibinfo{edition}{2} ed.)}.
\newblock \bibinfo{publisher}{Albert Blanchard}, \bibinfo{address}{Paris}.
\newblock


\end{thebibliography}

\end{document}